\documentclass[%
 preprint,
 amsmath,amssymb,
 aps,
 prb,
]{revtex4-2}

\usepackage{graphicx}
\usepackage{subcaption}
\usepackage{dcolumn}
\usepackage{bm}

\newcommand{\mean}[1]{\langle #1 \rangle}

\begin{document}

\title{Stability and distortion of fcc-LaH$_{10}$ with path-integral molecular dynamics}

\author{Kevin K. Ly}
\email{kkly2@illinois.edu}
\author{David M. Ceperley}
\email{ceperley@illinois.edu}
\affiliation{Department of Physics, University of Illinois at Urbana-Champaign}

\date{\today}

\begin{abstract}
    The synthesis of the high temperature superconductor LaH$_{10}$ requires pressures in excess of 100 GPa, wherein it adopts a face-centered cubic structure.
    Upon decompression, this structure undergoes a distortion which still supports superconductivity, but with a much lower critical temperature.
    Previous calculations have shown that quantum and anharmonic effects are necessary to stabilize the cubic structure, but have not resolved the low pressure distortion.
    Using large scale path-integral molecular dynamics enabled by a machine learned potential, we show that a rhombohedral distortion appears at sufficiently low pressures, even with quantum and anharmonic effects.
    We also highlight the importance of quantum zero point motion in stabilizing the cubic structure.
\end{abstract}

\maketitle


\section{\label{sec:intro}Introduction}

In recent years, the search for room temperature superconductors has been revitalized by advances in high-throughput first principles calculations and subsequent successes in experiments.
These high superconducting temperatures are enabled by metallic hydrogen, which is expected to support phonon mediated superconductivity at room temperature \cite{ashcroft_metallic_H}.
In order to metallize hydrogen, megabar pressures must be applied to dissociate the molecular H$_2$ units, a feat which has yet to be unambiguously achieved for the solid phase.
To circumvent these prohibitively high pressures, it was suggested that one could make a hydrogen-dominant alloy instead \cite{ashcroft_alloy}.
The idea is to provide a scaffolding that exerts ``chemical pressure'' on the hydrogen sublattice, forcing the molecular units to dissociate and form an atomic structure at lower pressures.

Inspired by this idea, initial high-throughput calculations based on density functional theory (DFT) predicted the possible formation of several metallic rare-earth superhydrides at around 200 GPa \cite{structure_search_PRL, structure_search_PNAS}.
These calculations combined structure searching algorithms with first principles electron-phonon coupling calculations to evaluate the superconducting temperatures for candidate structures; for a review, see \cite{review_high_pressure}.
Subsequently, independent experiments successfully synthesized the candidate lanthanum superhydrides and showed superconductivity at temperatures up to $T_c \approx 250$ K \cite{LaH_PRL, LaH_Nature}.
By combining the available experimental results and DFT calculations, it is inferred that the superconducting structure is a face-centered cubic structure, hereafter referred to as fcc-LaH$_{10}$.

By necessity, high-throughput calculations forgo a more careful treatment of the hydrogen sublattice, for example by ignoring anharmonic effects.
Within the harmonic approximation, fcc-LaH$_{10}$ is unstable to distortions below 250 GPa.
Initial \emph{classical} ab-initio molecular dynamics (AIMD) simulations showed the fcc structure to be stable above 140 GPa \cite{LaH_AIMD}.
By including quantum effects with the stochastic self-consistent harmonic approximation (SSCHA), Errea et al. found stability down to 120 GPa \cite{LaH_SSCHA}.
These are consistent with experiments, which find an fcc lanthanum sublattice down to 150 GPa and below.
However, upon further decompression, the experiments have also observed that the sublattice undergoes a distortion \cite{LaH_synthesis, LaH_structural_instability}, which is reversible and coincides with a significant decrease in $T_c$.
The SSCHA calculations did not find such a distortion in their studied pressure range and the authors suggested that the observed distortion could be due to anisotropy in the diamond anvil cell (DAC).
More recent quantum AIMD simulations found signs of instability at sufficiently low pressures \cite{AIMD_2022}, but the scale of these calculations limits the resolution of the distortion.
As we will show, simulations in small cells lead to fluctuations that obscure the distortion, especially near the phase boundary.

MD based methods are more versatile than the SSCHA, which is designed primarily to calculate phonons.
For example, in \cite{LaH_AIMD} it was found that the protons in LaH$_{10}$ can undergo noticeable diffusion beginning at around 800 K, a phenomenon that cannot be described by vibrations.
Recently, it has been shown that the anharmonicity of metallic hydrogen is not fully captured by SSCHA, as compared against path-integral molecular dynamics (PIMD) \cite{metallic_H_anharmonicity}, particularly at low temperatures.

AIMD calculations can be very expensive, so they are of limited scale.
At each timestep, the total energy and forces acting on the atoms are computed using DFT.
Such computations can only be performed a limited number of times, on systems of limited size.
A workaround which has exploded in popularity in the past decade is to construct interatomic potentials that are fitted to DFT calculations.
These potentials can be evaluated by a computer much more cheaply, and enable dynamical simulations at much larger scales.
With the incorporation of machine learning techniques, these potentials are now able to match DFT energies and forces with significantly greater accuracy than classical potentials; see \cite{MLP_review_JCP, NNP_review, MLP_benchmarks, representations} for some reviews.

We constructed a machine learned potential (MLP) for LaH$_{10}$ in order to perform both classical and path integral molecular dynamics.
Varying the temperature and pressure, and turning quantum effects on and off, we investigated the stability and distortion of the high pressure cubic structure.
Consistent with previous studies, our PIMD simulations favor the cubic structure over a large pressure range.
At sufficiently low pressures, a distortion is observed, which we are able to resolve as rhombohedral.
This resolution is enabled by large scale simulations accessible with the MLP.
Furthermore, while lowering the temperature in classical MD destabilizes the cubic structure at all of the pressures studied here, the same is \emph{not} true in PIMD, down to 100 K.
This may mean that zero point motion (ZPM) plays a significant role in stabilizing the cubic structure at lower pressures, and should be considered in future hydride structure studies.

\section{\label{sec:methods}Methods}

\subsection{\label{sec:dft}DFT}

All DFT calculations were performed with \texttt{QUANTUM ESPRESSO} \cite{QE_2009, QE_2017} using ultrasoft pseudopotentials from \texttt{pslibrary} \cite{pslibrary}.
An energy cutoff of 50 Ry and a shifted $3^3$ k-point grid for Brillouin zone integration are used for all supercell ($N = 352 \text{ and } 297$) calculations.
We checked that this corresponded to convergence in energies and forces to better than 5 meV/atom and 20 meV/\AA, respectively.
All results shown are based on the Perdew-Burke-Ernzerhof (PBE) functional \cite{PBE}, though we have also constructed models based on the Perdew-Zunger local density approximation (LDA) \cite{PZ}.
Our PBE DFT calculations are consistent with those of previous calculations: for example, within the harmonic approximation, we find fcc-LaH$_{10}$ to be dynamically unstable below about 250 GPa [supplemental material].
The choice of density functional significantly affects pressure estimates, which we will discuss at the end.

\subsection{\label{sec:model}Model}

We used the Deep Potential (DP) method to construct a MLP for LaH$_{10}$ \cite{DP, DP_SE}.
In this approach, a deep neural network is trained to predict the energy $U$ (and forces and pressures, by appropriate differentiation) of a given structure.
Instead of calculating structural fingerprints such as Behler-Parrinello symmetry functions \cite{BP} or Smooth Overlap of Atomic Positions (SOAP) \cite{SOAP}, the requisite physical symmetries for the model are satisfied by a symmetry-preserving embedding network, which itself is also trainable.
DP models have been used to study Al-Ce alloys \cite{AlCe}, the liquid-liquid transition in phosphorus \cite{phosphorus}, the phase diagram of water \cite{DP_water}, supersolidity in deuterium \cite{supersolid_D}, and the tetragonal distortion of strontium titanate \cite{STO}, to name a few examples.

In the DP model we used a cutoff of $4.0$ \AA.
For the densities studied here, this cutoff corresponds to local environments containing between 60 and 110 atoms.
The deep neural network consists of an embedding and fitting network, with sizes (25, 50, 100) and (240, 240, 240), respectively.
A feature of DP which is not shared by all current MLPs is the usage of virial information, which we take advantage of.
We found that the inclusion of virial information in training improved performance, likely because our data covers a pressure range of about 90 GPa.

Our final model is actually the sum of two parts, a pair potential $f(\{ \boldsymbol{R} \})$ plus the DP model $g(\{ \boldsymbol{R} \})$.
The DP model is trained to learn the difference between the DFT energy and the pair potential $U - f$.
This pair potential enforces repulsion between the atoms at very short ranges, ensuring that atoms never get closer than they should.
Importantly, subtracting from the training data a reasonable pair potential reduces the range of energies, since the energy can increase dramatically as atoms get close to each other.
This also reduces the variance of the training data.
The resultant dataset of differences is easier for the model to learn from.
By ``reasonable'' pair potential, we mean that that it only needs be accurate for very short distances, as the DP model can learn the remainder.
To construct these pair potentials, we calculated the energy of the isolated La-La, La-H, and H-H dimers as a function of bond length with \texttt{PySCF} \cite{libcint, pyscf_2018, pyscf_2020}, shown in the supplementary material.

\subsection{\label{sec:data}Training data}

\begin{figure}
    \begin{subfigure}{0.49\linewidth}
        \includegraphics[width=\textwidth]{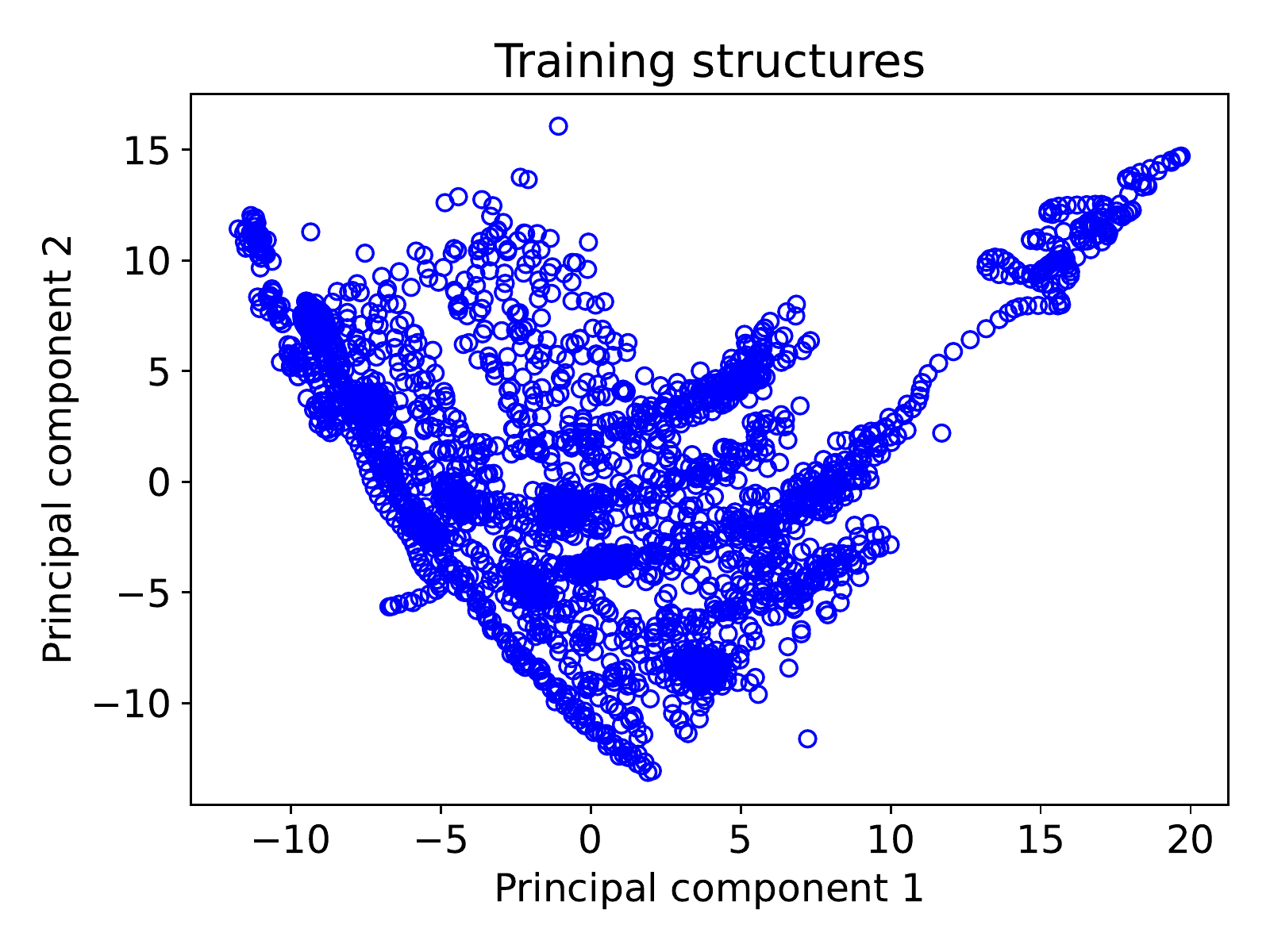}
        \caption{\label{fig:pca}}
    \end{subfigure}
    \begin{subfigure}{0.49\linewidth}
        \includegraphics[width=\textwidth]{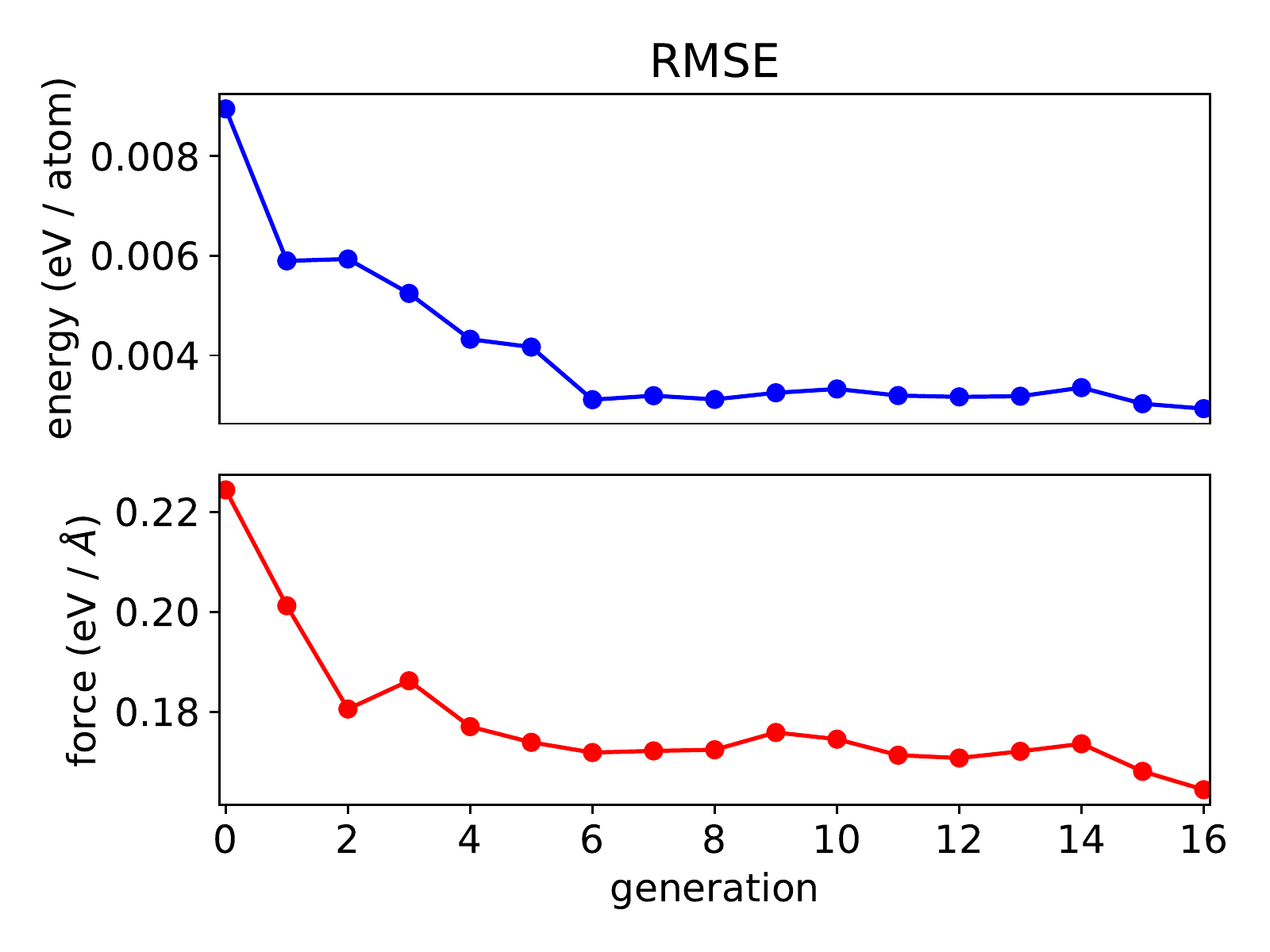}
        \caption{\label{fig:convergence}}
    \end{subfigure}
    \caption{\label{fig:data}%
        (a) 2D projection of the structures used for training.
        Each point is an averaged SOAP representation of a structure.
        A simple way to read this is that points which are close to each other are similar, as determined by SOAP.
        (b) Root mean square errors (RMSE) in the energies and forces of our models at each generation in our iterative procedure.
    }
\end{figure}

The structures used for training were generated in an iterative procedure.
Following the DPGEN protocol \cite{DPGEN}, we began with a relatively small initial pool of structures from AIMD.
A committee of 3 models was trained on this data.
The models are identical in architecture and differ only in the random seeds used for training.
As a result, the models do not make identical predictions, yielding generally different forces for a given structure.
One of these models is then used to sample more structures with MD at various thermodynamic conditions.
By performing MD using the model, rather than performing AIMD, a large quantity of structures may be generated quickly.
From this large pool of structures a subset is selected for labelling, meaning their energies and forces will be calculated with DFT.
Given a structure, each model from the committee evaluates the forces $\boldsymbol{F}_{i}$ on every atom $i$.
Consider the maximum deviation among the committee
\begin{equation}
    \label{eq:model_deviation}
    \mathcal{E} \equiv \max_i \sqrt{\langle | \boldsymbol{F}_i - \langle \boldsymbol{F}_i \rangle |^2 \rangle}
\end{equation}
where $\langle \dots \rangle$ represents an average over the committee.
In other words, the standard deviation among the committee is calculated for each force in the structure, and $\mathcal{E}$ is chosen to be the largest deviation.
This is meant to represent the committee's uncertainty on that structure.
We then selected the structures for which the uncertainties are greatest.
Once these structures are labelled, they are added to the training dataset, and the procedure is repeated: train a new committee, use one of the models to generate more structures, then poll the new committee on these samples to determine which to include for labelling.

This procedure allows us to generate structures efficiently across the phase space of interest, and also includes a failsafe: it is possible for early iterations to perform poorly under certain thermodynamic conditions, especially those beyond the scope of the training data.
Consequently, sampling based on these models may yield highly unusual structures which should not be included.
If this is so, we expect the uncertainty $\mathcal{E}$ on such structures to be anomalously high.
Such cases are excluded from consideration.

If the models are sufficiently accurate, the committee uncertainty may not vary enough to distinguish structures.
More importantly, our goal is to have a diverse set of structures spanning the phase space of interest, so that resultant models would be uniformly accurate over the desired thermodynamic conditions.
To this end, in the final iterations the subset selection criteria is changed to approximately maximize structural diversity.
Specifically, given a structure $i$, a set of SOAP descriptors $\boldsymbol{x}_i$ is computed using \texttt{ASAP} \cite{ASAP}.
A subset is chosen such that
\begin{equation}
    \label{eq:distance}
    \mathcal{S} \equiv \sum_{i < j} | \boldsymbol{x}_i - \boldsymbol{x}_j |^2
\end{equation}
the sum of distances over all pairs of structures is maximized.

In total, we generated 2779 structures, 259 of which were set aside for testing, with the rest used for training.
A visualization of these structures is shown in Figure \ref{fig:pca}, where each point represents a different structure.
Each structure is represented by a vector of SOAP descriptors, and a principal component analysis (PCA) is performed in order to make a 2D projection of the data.
In Figure \ref{fig:convergence} we show how the accuracy of the models varies during our iterative procedure.
Each model is tested on the same dataset, the final set of test structures.
We see that, for example, even though the model at generation 12 has never seen the types of structures generated in subsequent generations, it still performs well on them.

\subsection{\label{sec:structure}Structure}

\begin{figure}
    \begin{subfigure}{0.49\linewidth}
        \includegraphics[width=\textwidth]{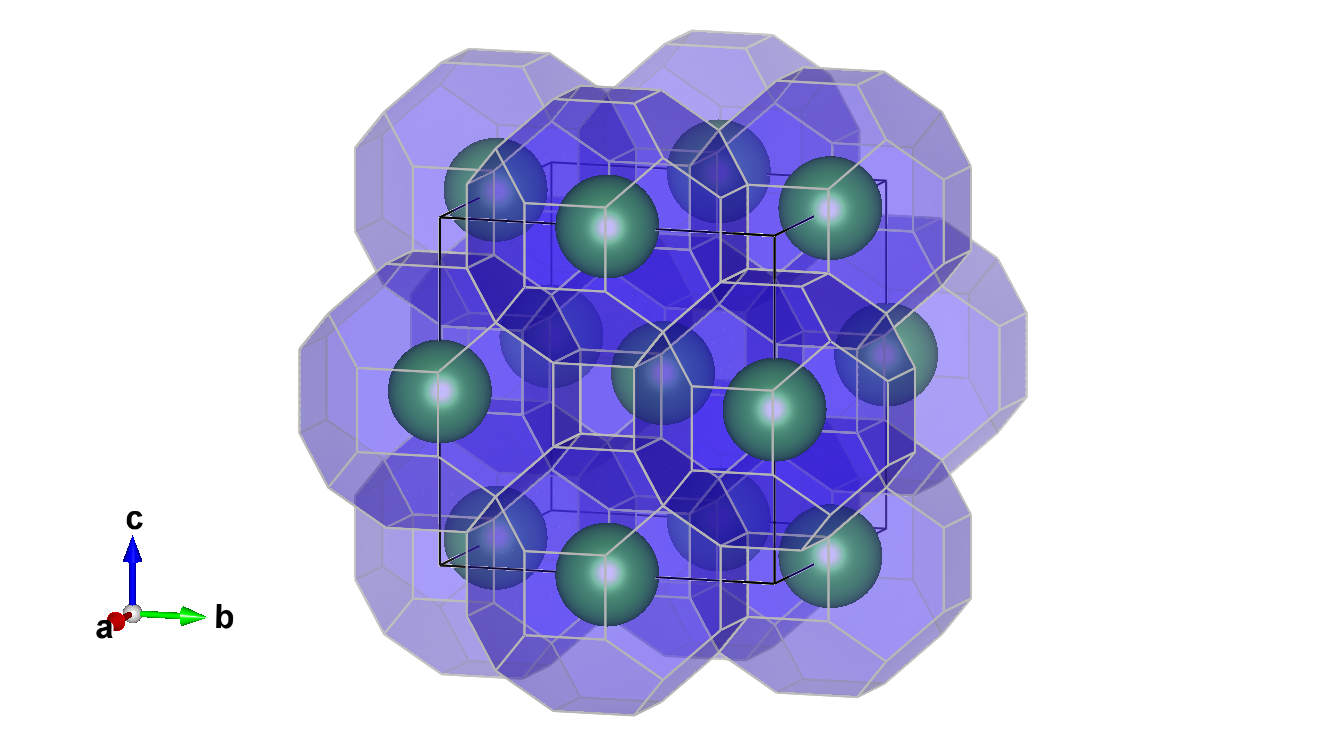}
        \caption{\label{fig:fm3m}}
    \end{subfigure}
    \begin{subfigure}{0.49\linewidth}
        \includegraphics[width=\textwidth]{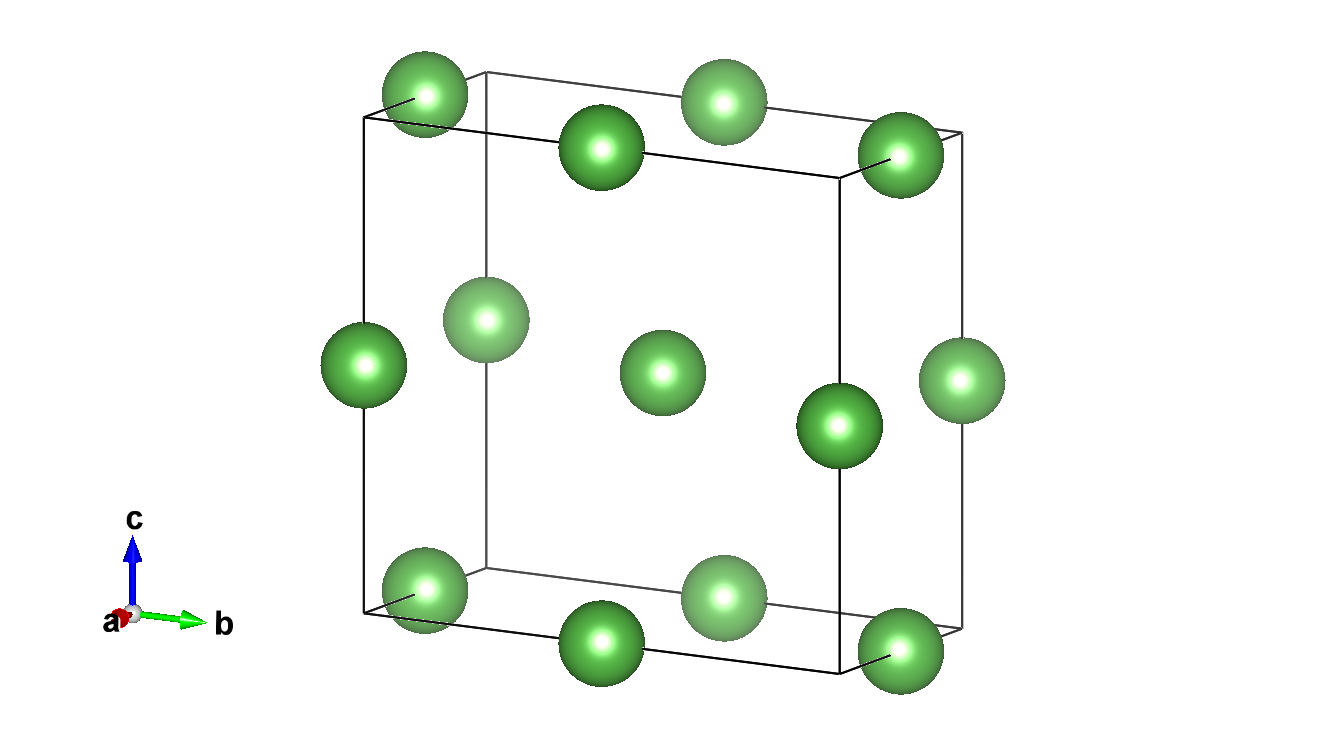}
        \caption{\label{fig:r3m}}
    \end{subfigure}
    \caption{\label{fig:structures}
        (a) Conventional cell of fcc-LaH$_{10}$.
        The lanthanum atoms are shown in green, and the hydrogem atoms are not explicitly shown but reside on the corners of the blue polyhedra, forming a clathrate structure.
        (b) Rhombohedral distortion of the conventional fcc unit cell.
        The distortion is exaggerated for clarity.
    }
\end{figure}

We focus entirely on fcc-LaH$_{10}$ and distortions thereof.
Shown in Figure \ref{fig:fm3m} is the conventional cubic cell for the fcc structure, with the hydrogen atoms occupying the corners of the polyhedra.
As we will show, a rhombohedral distortion appears at lower pressure, shown in Figure \ref{fig:r3m}.
The hydrogen atoms have been removed and the lanthanum atoms have been shrunk, while the distortion has been exaggerated, for visual clarity.

For the fcc structure with a lattice constant of $a = 5.1$ \AA, the corresponding hydrogen sublattice has a nearest neighbor spacing of 1.1 \AA.
For reference, atomic hydrogen in the candidate Cs-IV structure \cite{cs_iv_structure} has a nearest neighbor spacing of 1.06 \AA \ at a density of $r_s = 1.37$.
This density is well below where one would expect to be able to metallize pure hydrogen.
Similarly, the  lanthanum sublattice is significantly larger than that of pure fcc-lanthanum, which at 50 GPa has a lattice parameter $a = 4.25$ \AA \ \cite{La_EOS}.
Note that pure fcc-lanthanum also undergoes a rhombohedral distortion below 50 GPa.

To study potential distortions of the cubic structure, we performed molecular dynamics simulations targeting the isothermal-isobaric (NPT) ensemble.
We used \texttt{LAMMPS} \cite{LAMMPS} and \texttt{i-PI} \cite{i-pi} to perform classical and path-integral simulations.
The former is used to calculate energies, forces, and pressures from our final model, and serves as a driver for the latter.
To allow arbitrary distortions of the simulation cell, we used a flexible barostat \cite{barostat} in which all cell parameters are allowed to fluctuate.
In other words, the side lengths $a, b, c$ and angles $\alpha, \beta, \gamma$ between them can and will vary over the course of a given simulation.
For reference, the cubic cell shown in Figure \ref{fig:fm3m} has $a = b = c$ and $\alpha = \beta = \gamma = 90^\circ$, while the distortion shown in Figure \ref{fig:r3m} has $a = b = c$ and $\alpha = 94^\circ, \beta = \gamma = 86^\circ$.

\section{\label{sec:results}Results}

\begin{figure}
%
    \begin{subfigure}{0.49\linewidth}
        \includegraphics[width=\textwidth]{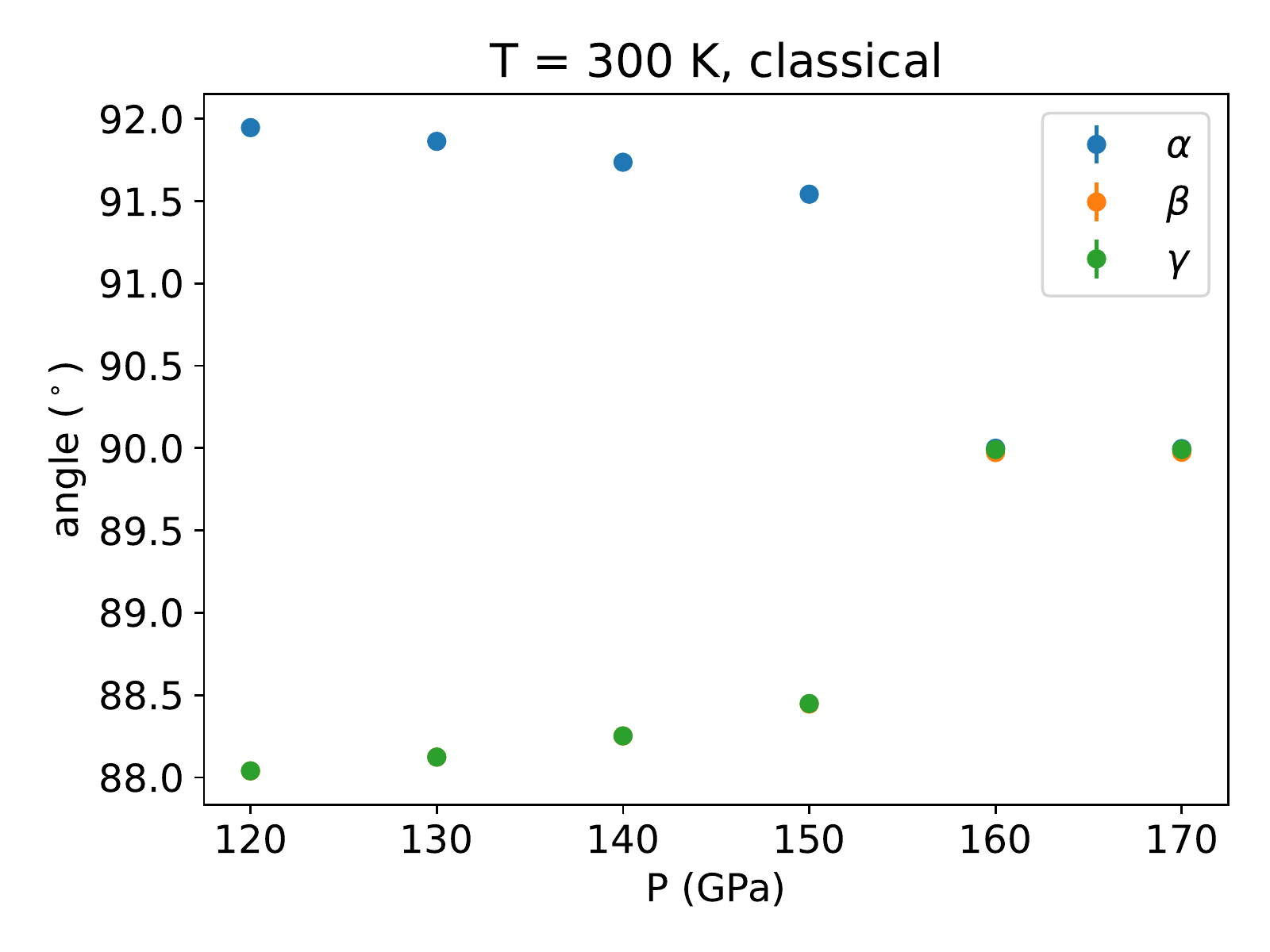}
        \caption{\label{fig:classical_300}}
    \end{subfigure}
    \begin{subfigure}{0.49\linewidth}
        \includegraphics[width=\textwidth]{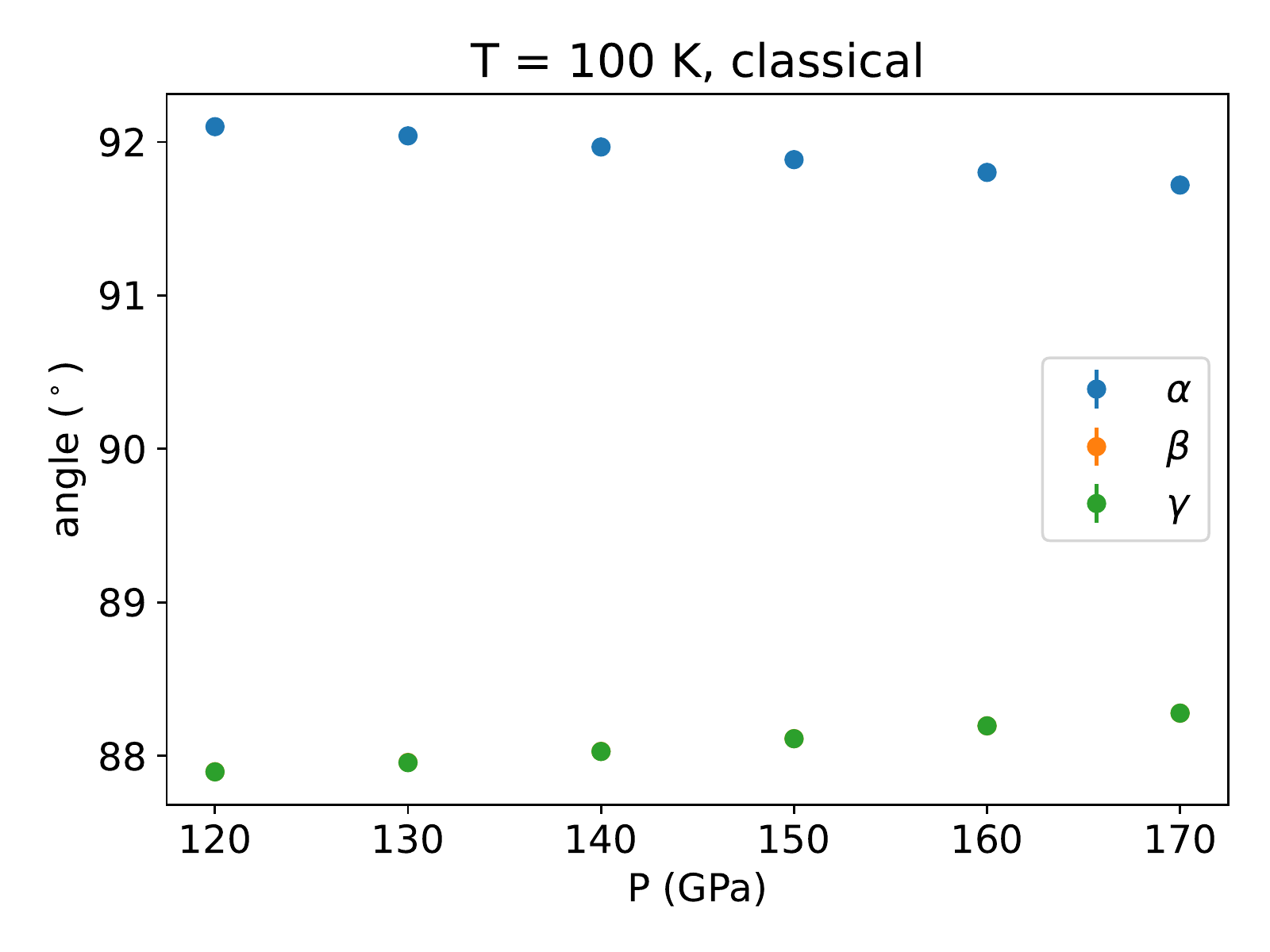}
        \caption{\label{fig:classical_100}}
    \end{subfigure}
    \caption{\label{fig:classical}
        The mean angles as the pressure is varied at (b) 300 K and (c) 100 K.
    }
\end{figure}

We first show our classical simulations in Figure \ref{fig:classical}.
Varying the temperature and pressure, we measured the three side lengths $a(t), b(t), c(t)$ and three angles $\alpha(t), \beta(t), \gamma(t)$, where $t$ represents the simulation time.
In all cases we found that the side lengths $\mean{a} = \mean{b} = \mean{c}$; as such, we show here only the angles.
Shown in Figures \ref{fig:classical_300} and \ref{fig:classical_100} are the mean angles at every pressure; the error bars are not visible at this scale.
Since $a = b = c$, the order of the angles $\alpha, \beta, \gamma$ is arbitrary, by symmetry.
The reason that $\alpha$ always appears to be the largest is because we initialized the simulations in a rhombohedral cell, with $\alpha$ chosen to be the largest.

At 300 K and above 150 GPa, the simulation cell is overall cubic.
Recall that within the harmonic approximation the cubic structure is not stable at these pressures, meaning that classical anharmonic effects are stabilizing it here.
When the pressure is lowered, all three angles deviate from $90^\circ$ in a rhombohedral manner $|\alpha - 90^\circ| = |\beta - 90^\circ| = |\gamma - 90^\circ|$.
When the temperature is lowered to 100 K, the cubic structure no longer appears in this range, and requires much higher pressures.
This is consistent with what we expect, since as $T \to 0$ we should effectively recover the harmonic approximation.

\begin{figure}
    \begin{subfigure}{0.49\linewidth}
        \includegraphics[width=\textwidth]{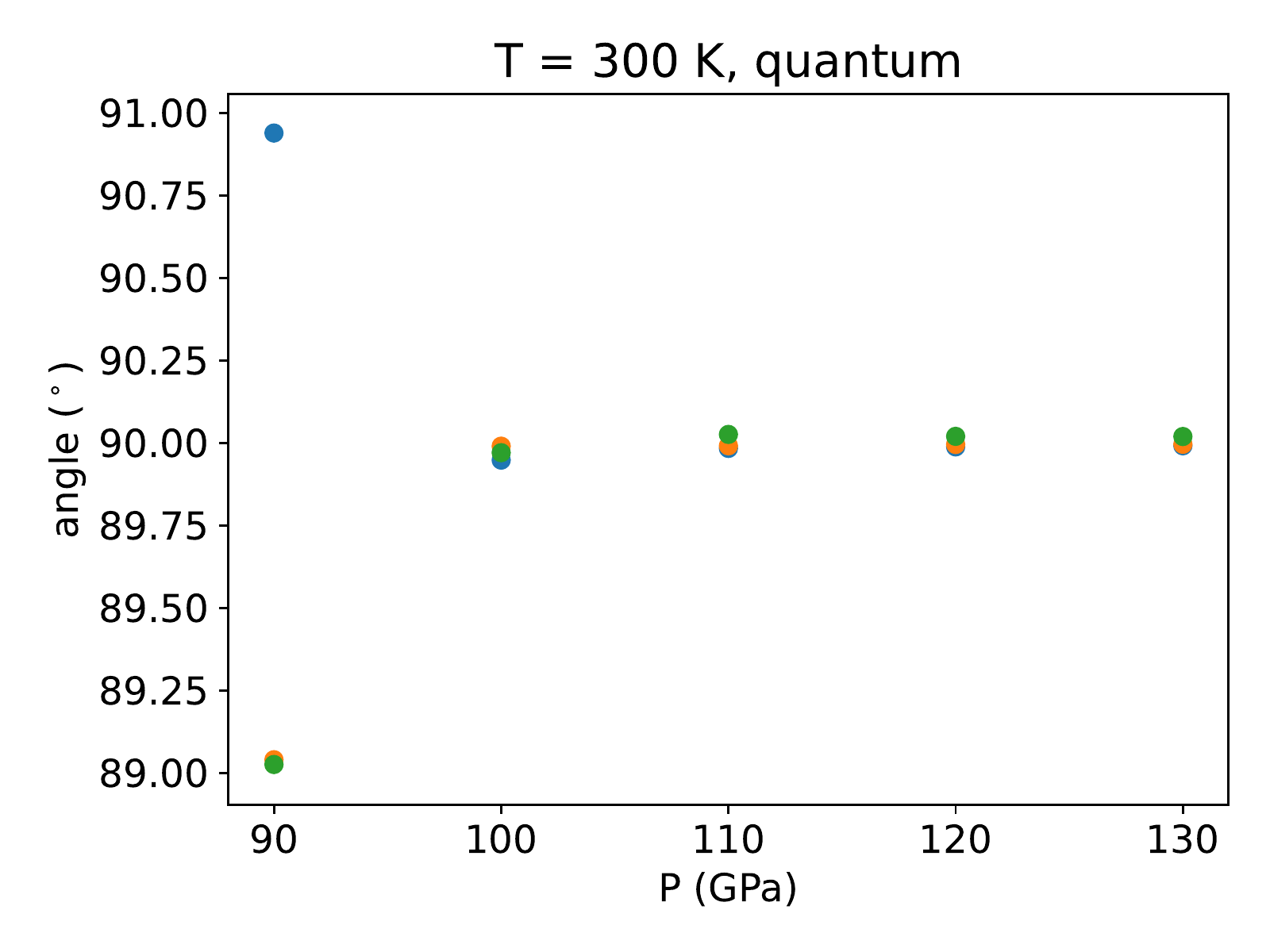}
        \caption{\label{fig:quantum_angles}}
    \end{subfigure}
    \begin{subfigure}{0.49\linewidth}
        \includegraphics[width=\textwidth]{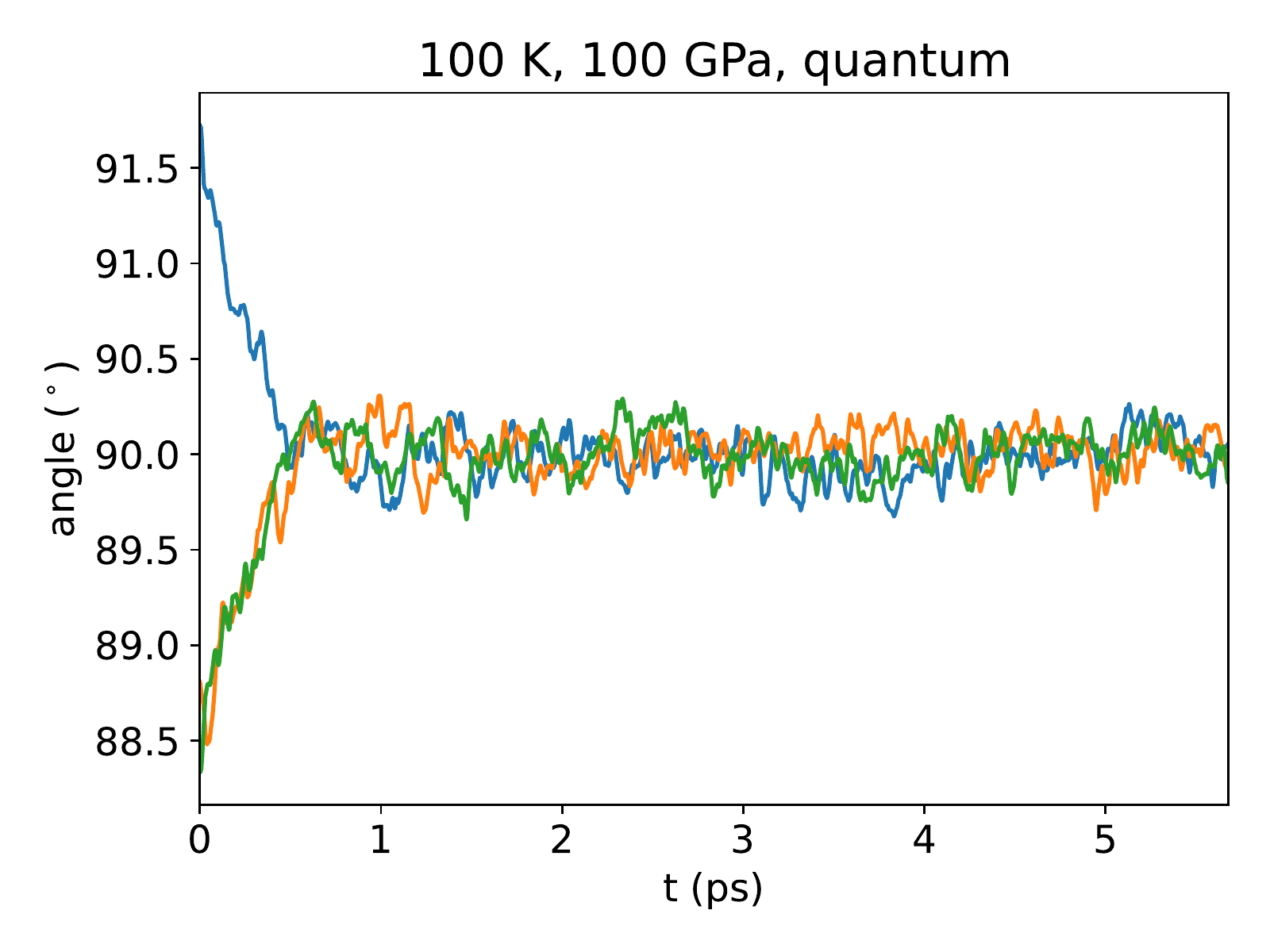}
        \caption{\label{fig:quantum_trace}}
    \end{subfigure}

    \begin{subfigure}{0.49\linewidth}
        \includegraphics[width=\textwidth]{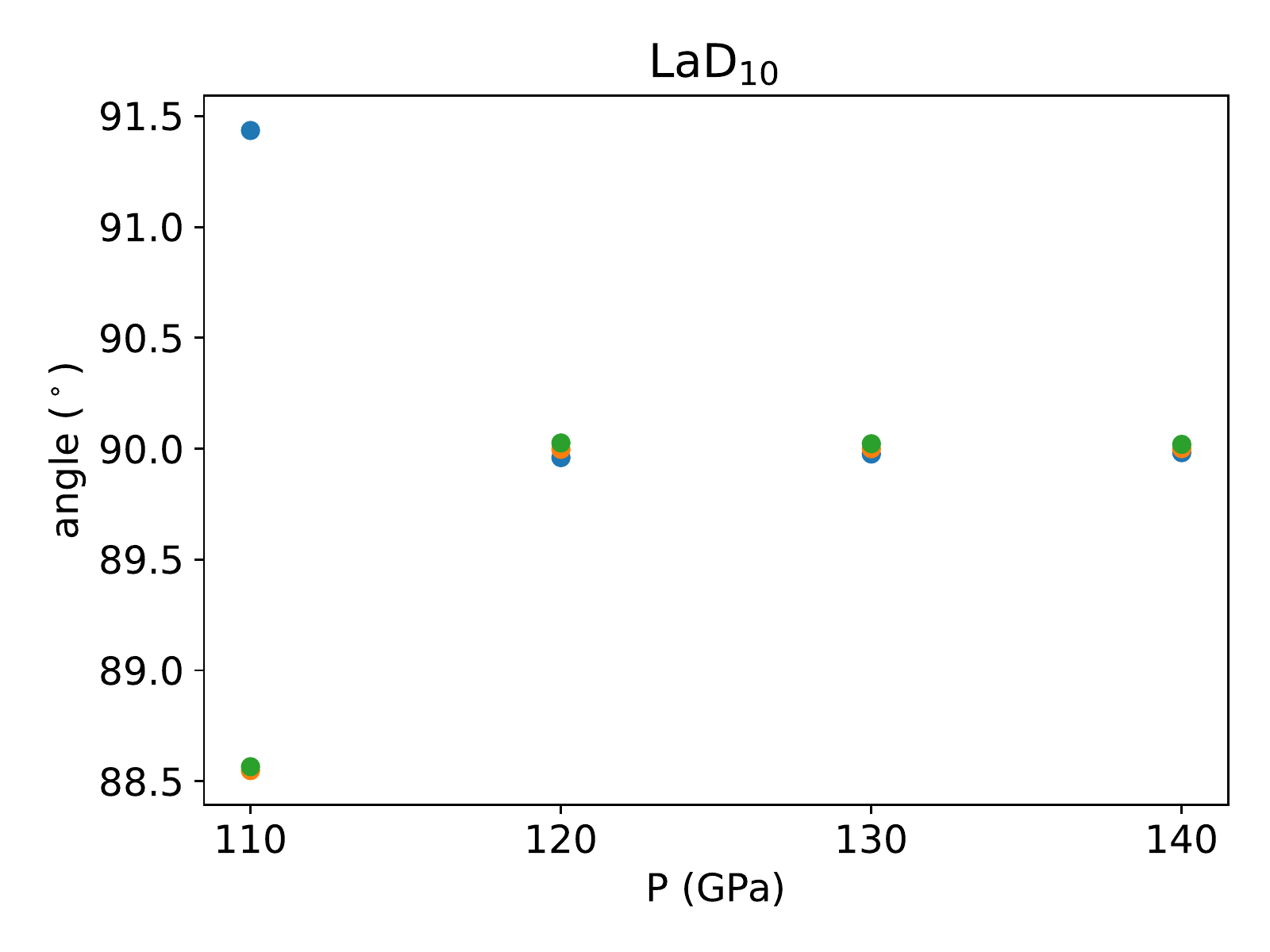}
        \caption{\label{fig:isotope}}
    \end{subfigure}
    \caption{\label{fig:quantum}
        (a) The mean angles at 300 K, with quantum effects.
        (b) Traces of the three angles in a simulation at 100 K and 100 GPa, with quantum effects.
        (c) The mean angles at 300 K for LaD$_{10}$.
    }

\end{figure}

We performed quantum simulations using PIMD with either 16 or 32 beads; convergence with the number of beads is shown in the supplementary material.
Shown in Figure \ref{fig:quantum_angles} is how the angles vary with pressure at 300 K.
As in the classical simulations, the cubic structure undergoes a rhombohedral distortion at low pressures, though in this case it must be below 100 GPa.
In other words, quantum effects stabilize the cubic structure down to even lower pressures.

A striking difference between the classical and path-integral simulations can be seen upon cooling.
Shown in Figure \ref{fig:quantum_trace} are the angles in a simulation at 100 K and 100 GPa, near the pressure where the distortion appears at 300 K.
The cubic structure still appears, despite initializing the simulation with a rhombohedral cell.
While cooling significantly destabilized the cubic cell classically, it appears to have little effect here.
Since 100 K is an order of magnitude smaller than the vibrational energy scales (c.f. the phonon band structures in \cite{LaH_SSCHA}), we believe that this picture may persist down to zero temperature.
That is, we believe it is the ZPM which stabilizes the cubic structure at lower pressures.

When substituting H with D, the pressure required to stabilize the cubic structure is increased.
Shown in Figure \ref{fig:isotope} are how the cell angles vary with pressure for LaD$_{10}$.
The cell becomes cubic just below 120 GPa, up from below 100 GPa in LaH$_{10}$.
This is roughly consistent with what we expect, since the substitution of the heavier isotope suppresses the ZPM, and the result lies between the quantum and classical limits for H.
This is also consistent with the SSCHA calculations, where an imaginary mode appears around 120 GPa.

\begin{figure}
    \begin{subfigure}{0.49\linewidth}
        \includegraphics[width=\textwidth]{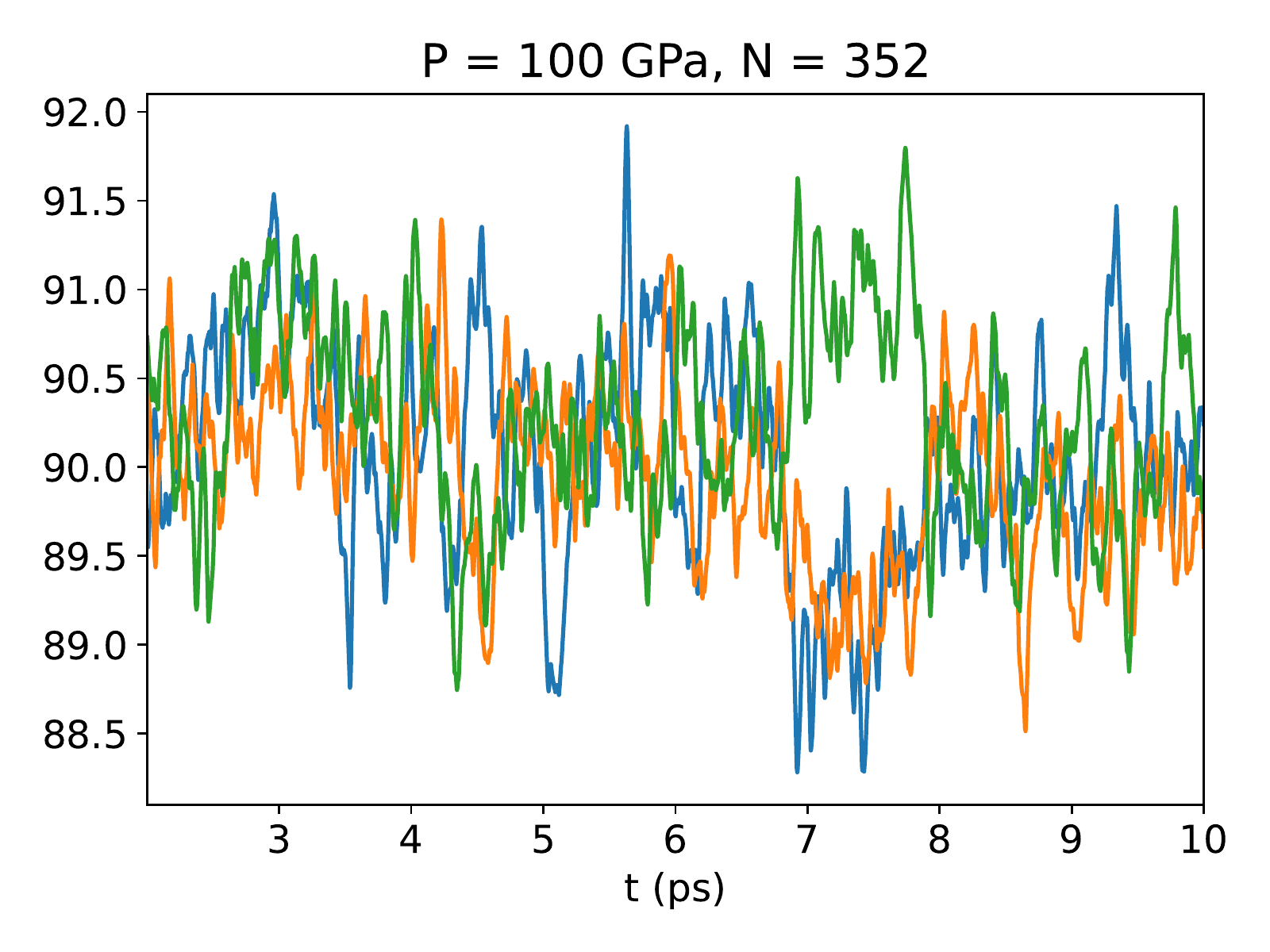}
        \caption{\label{fig:small_trace}}
    \end{subfigure}
    \begin{subfigure}{0.49\linewidth}
        \includegraphics[width=\textwidth]{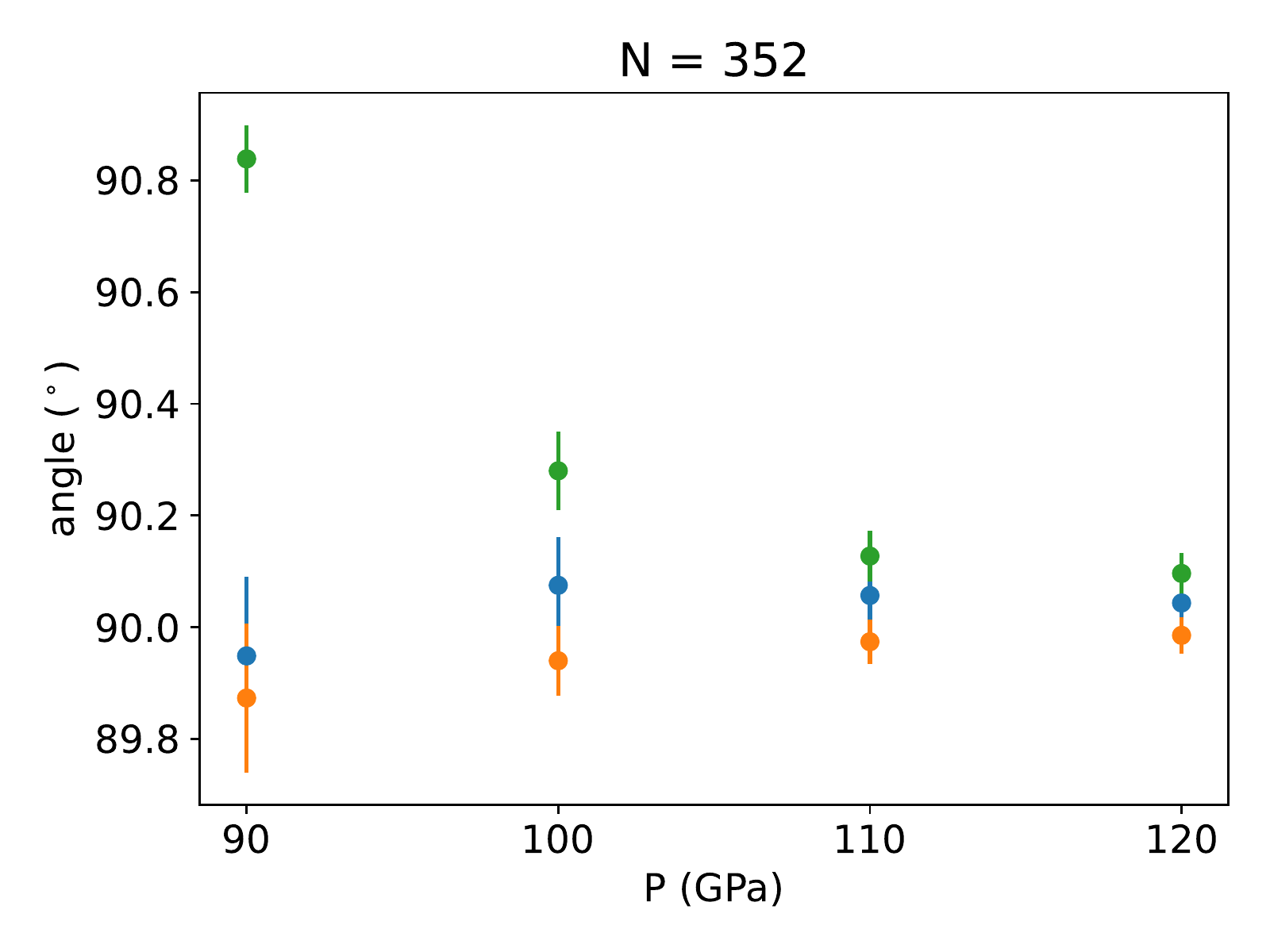}
        \caption{\label{fig:small_angles}}
    \end{subfigure}
    \caption{\label{fig:small}
        (a) Traces of the three angles in a simulation of a smaller system, $N = 352$.
        (b) Mean angles at various pressures in the smaller system.
    }
\end{figure}

All of the above results were obtained in simulations with $N = 2816$ atoms.
For smaller systems, the large cell fluctuations make it difficult to resolve the structure.
Shown in Figure \ref{fig:small} are path-integral simulations performed with $N = 352$.
The large fluctuations are apparent in the vertical scale in Figure \ref{fig:small_trace} and the error bars in Figure \ref{fig:small_angles}.
Below 100 GPa the simulation shuttles between the distorted and cubic structure, obscuring the nature of the distortion.
This behavior disappears at $N = 2816$.
In \cite{AIMD_2022} $N = 44$ was used, so we suspect that these fluctuations obscured the resolution of the low pressure structure.

\begin{figure}
    \begin{subfigure}{0.49\linewidth}
        \includegraphics[width=\textwidth]{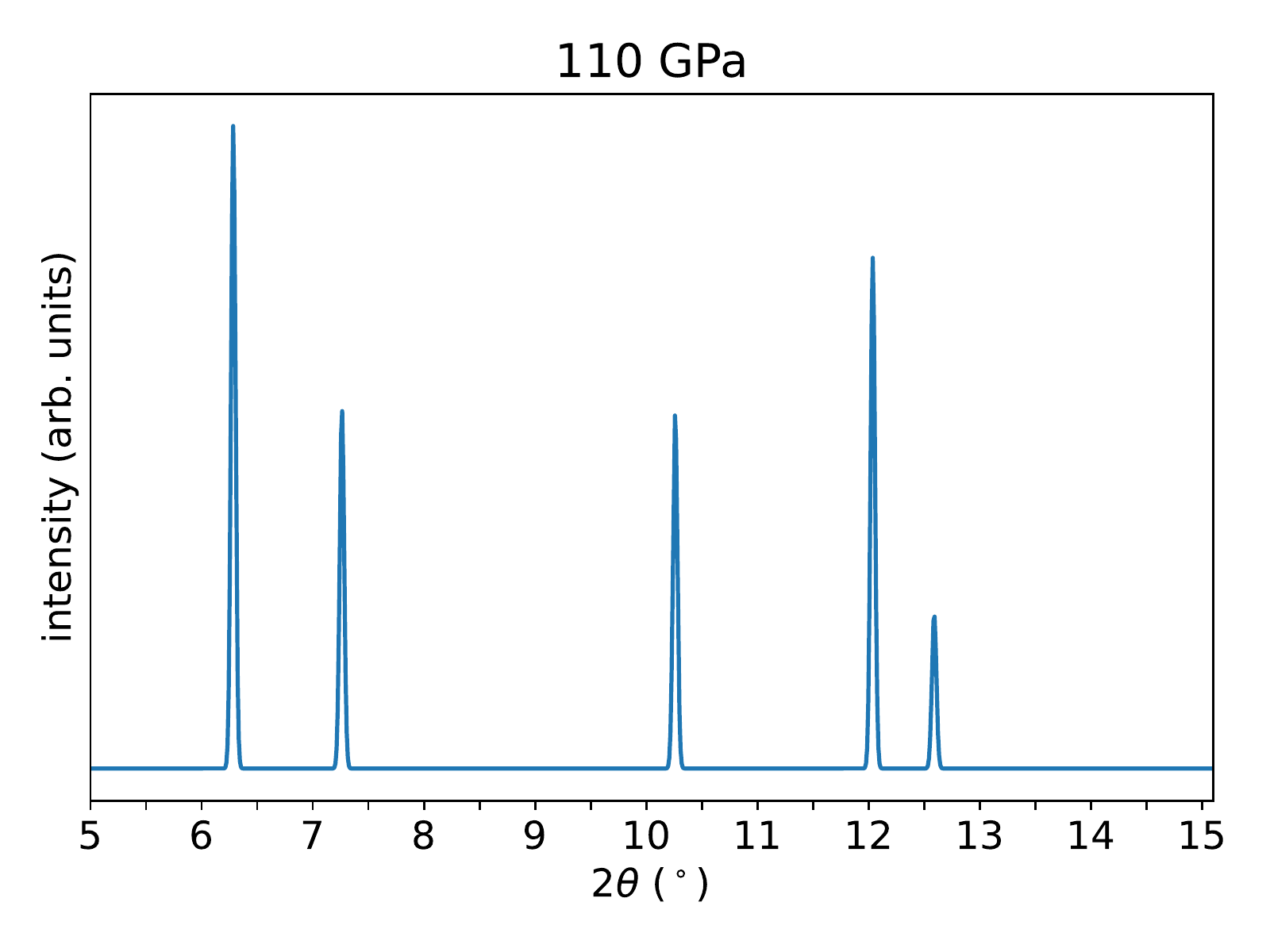}
        \caption{\label{fig:xrd_110}}
    \end{subfigure}
    \begin{subfigure}{0.49\linewidth}
        \includegraphics[width=\textwidth]{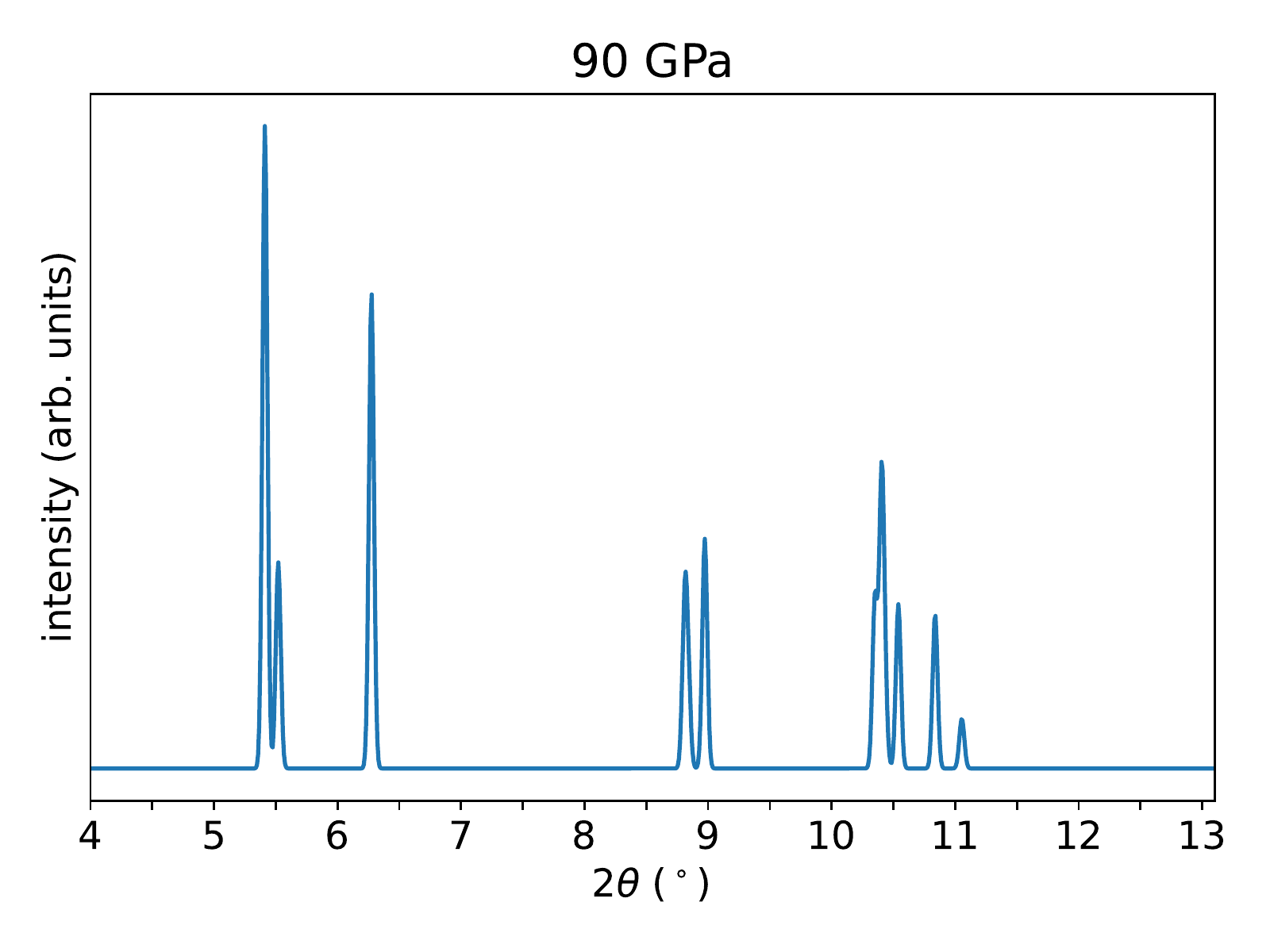}
        \caption{\label{fig:xrd_90}}
    \end{subfigure}
    \caption{\label{fig:xrd}
        Simulated diffraction patterns for (a) the cubic structure and (b) the distorted structure.
        For comparison with the patterns of \cite{LaH_structural_instability}, we used $\lambda = 0.3344$ \AA \ in (a) and $\lambda = 0.2952$ \AA \ in (b).
    }
\end{figure}

Shown in Figure \ref{fig:xrd} are diffraction patterns simulated from our path-integral simulations.
For comparison with the x-ray diffraction (XRD) results of \cite{LaH_structural_instability}, we show patterns based only on the La-La static structure factor, since scattering from the hydrogen sublattice has not been experimentally detected.
The behavior of the peaks as the pressure is lowered into the distorted structure is roughly consistent with what is observed in experiments.
The reflections in Figure \ref{fig:xrd_90} can be reproduced by taking a conventional fcc cell and distorting it according to the angles obtained at 90 GPa.

\begin{figure}
    \begin{subfigure}{0.49\linewidth}
        \includegraphics[width=\textwidth]{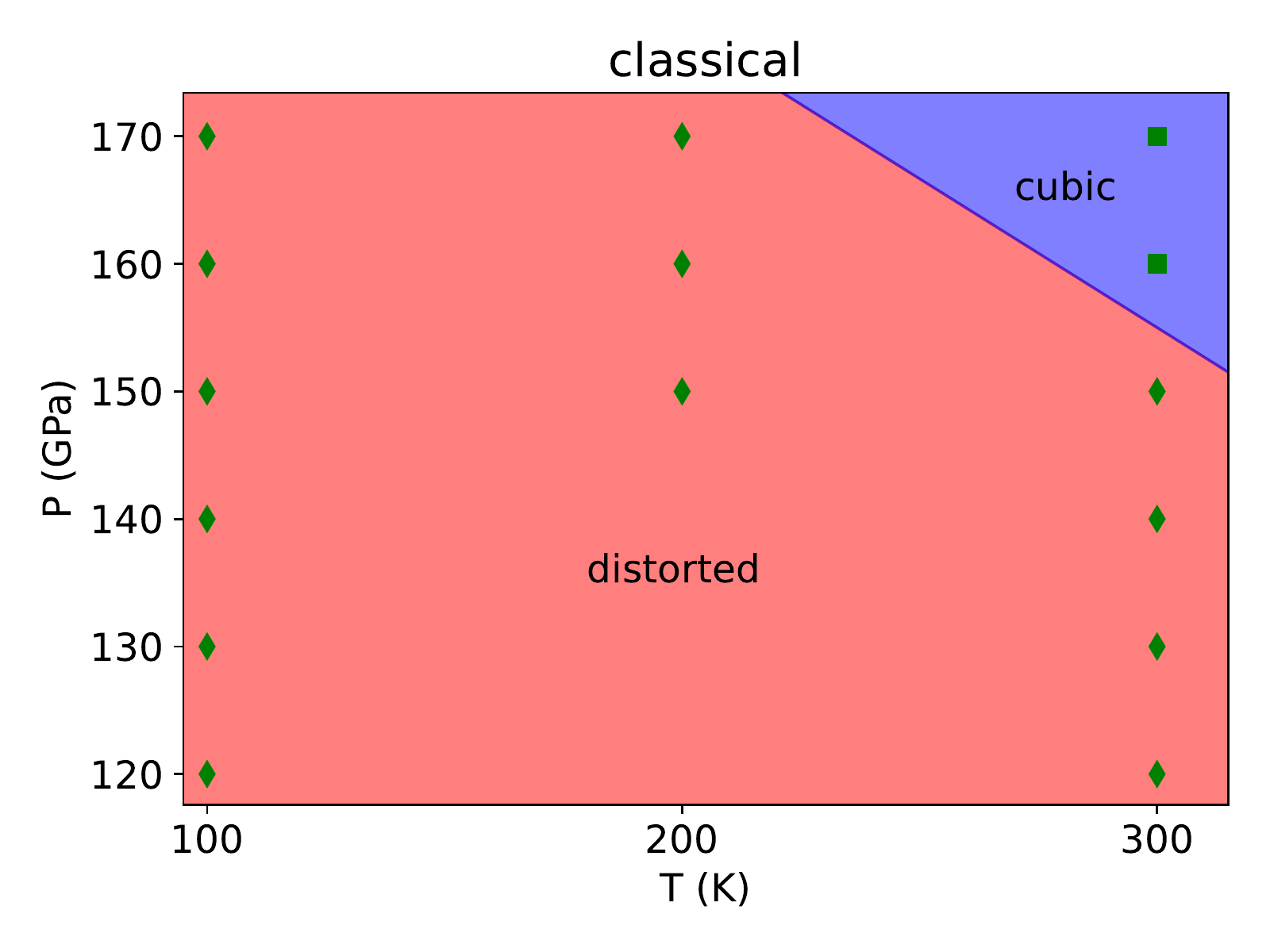}
        \caption{\label{fig:classical_diagram}}
    \end{subfigure}
    \begin{subfigure}{0.49\linewidth}
        \includegraphics[width=\textwidth]{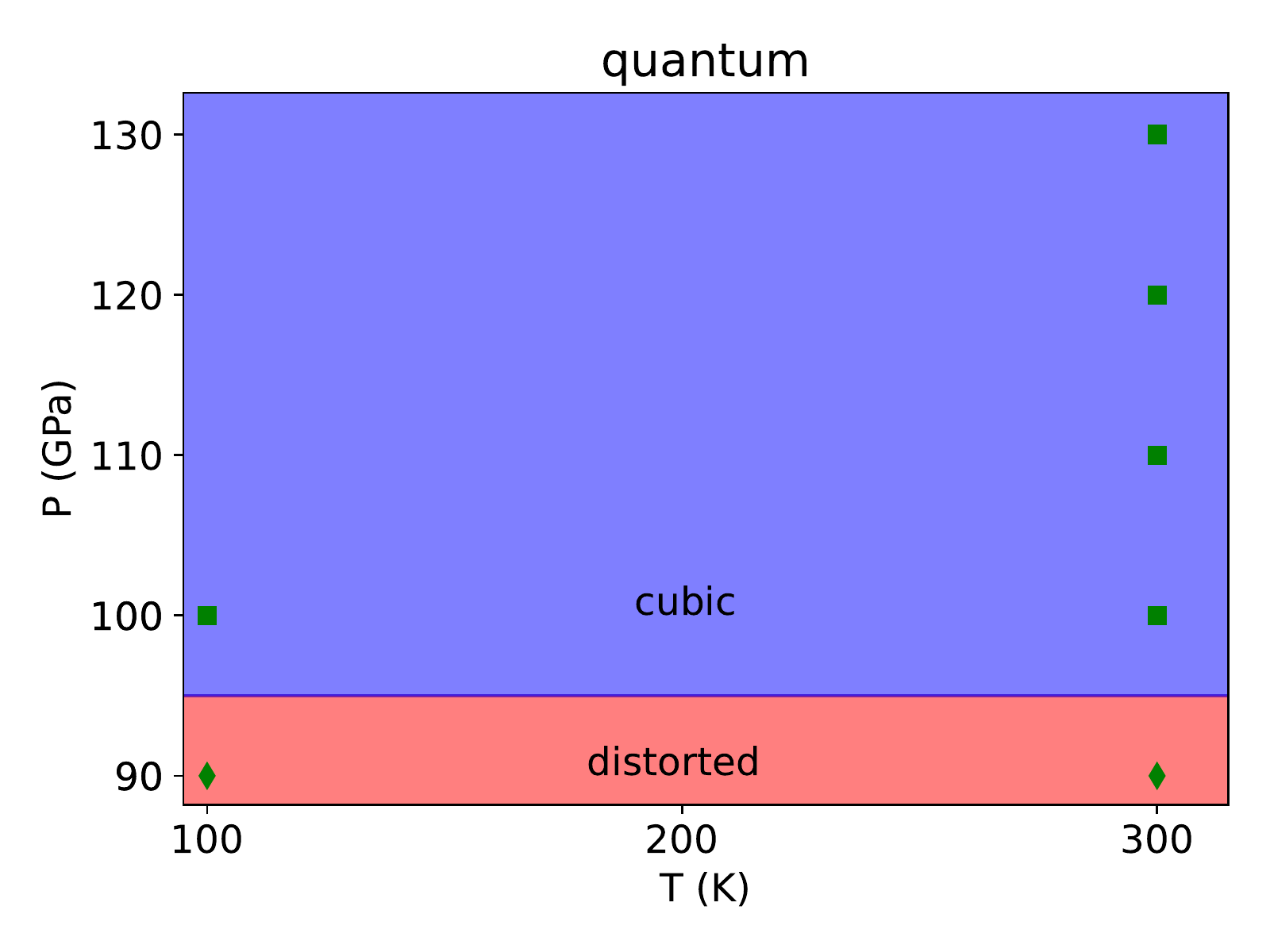}
        \caption{\label{fig:quantum_diagram}}
    \end{subfigure}
    \caption{\label{fig:phase_diagrams}
        (a) Classical and (b) quantum phase diagrams.
        The points indicate where our simulations were performed.
    }
\end{figure}

Shown in Figure \ref{fig:phase_diagrams} are phase diagrams summarizing our results.
The phase boundaries are drawn to be linear, though there are not enough points to determine the actual shape.
They are meant only to illustrate that the cubic structure can be stabilized with increasing temperature and pressure.
They also illustrate the significant difference between the classical and quantum simulations, where the quantum phase boundary has a weak dependence on temperature.

\section{\label{sec:conclusion}Conclusion}

Using PIMD, we showed that fcc-LaH$_{10}$ experiences a rhombohedral distortion at low pressures.
Even without quantum effects, the cubic structure can be stabilized down to nearly 150 GPa at 300 K.
Upon cooling, the difference between our classical and quantum simulations suggests that ZPM is significant in extending the range of stability.
This may account for the relative difficulty reported in experiments in synthesizing the isotopic analogue LaD$_{10}$, as the substitution of deuterium suppresses the role of the ZPM.
This differs from the case of the superconducting sulfur hydride, in which \emph{both} H$_3$S and D$_3$S undergo a cubic-to-rhombohedral distortion upon decompression at nearly the same pressure \cite{sulfur_hydride}.

The pressure at which we observed the distortion is noticeably lower than that reported by experiments.
We believe this is due to the underlying density functional PBE, which is overestimating the stability of the cubic structure.
We performed some simulations with an LDA-based model, and the distortion never appears.
It is known that DFT simulations of dense hydrogen can vary significantly with the choice of density functional \cite{xc_benchmarks}, and the same appears to be true here.
For example, we find significantly different pressure estimates between LDA and PBE.

In \cite{LaH_structural_instability} the authors suggest that the distortion is not rhombohedral but instead monoclinic.
Both distortions produce similar XRD patterns, and it is not obvious to us that the rhombohedral distortion can be ruled out.
It is not surprising that our simulations favor the more symmetric rhombohedral structure, given that our model cannot account for the various other defects that appear in experiments.
Nevertheless, we have shown that, upon decompression, the destabilization of the cubic structure does not strictly require factors like variable hydrogen content or external anisotropy.

We have found that, where applicable, our MLP-based simulations are consistent with previous SSCHA and AIMD calculations.
We believe that, constructed carefully, MLPs are viable in performing path-integral simulations for future studies in designing hydrogen-based superconductors.
Recent studies explored candidate structures based on fcc-LaH$_{10}$ with the goal of finding superconductors at lower pressures \cite{lower_pressure_H} or higher temperatures \cite{doping}.
Since these calculations worked within the harmonic approximation, it may be worthwhile to revisit some of these candidates with path-integral simulations.
It is possible that the inclusion of anharmonicity and quantum effects, particularly ZPM, will lower the pressure requirements more.

\section{\label{sec:acknowledgements}Acknowledgements}

We would like to thank Hongwei Niu, Yubo Yang, and Scott Jensen for their insights on the usage of MLPs.
We would also like to thank Russell Hemley for his input from the experiments and related feedback.
This work was supported by the U.S. Department of Energy, Office of Science, Office of Basic Energy Sciences, Computation Materials Sciences program under Award Number DE-SC-0020177.
This work made use of the Illinois Campus Cluster, a computing resource that is operated by the Illinois Campus Cluster Program (ICCP) in conjunction with the National Center for Supercomputing Applications (NCSA) and that is supported by funds from the University of Illinois at Urbana-Champaign.
This work utilized resources supported by the National Science Foundation’s Major Research Instrumentation program, grant \#1725729, as well as the University of Illinois at Urbana-Champaign.

\bibliography{bibliography}

\begin{thebibliography}{45}%
\makeatletter
\providecommand \@ifxundefined [1]{%
 \@ifx{#1\undefined}
}%
\providecommand \@ifnum [1]{%
 \ifnum #1\expandafter \@firstoftwo
 \else \expandafter \@secondoftwo
 \fi
}%
\providecommand \@ifx [1]{%
 \ifx #1\expandafter \@firstoftwo
 \else \expandafter \@secondoftwo
 \fi
}%
\providecommand \natexlab [1]{#1}%
\providecommand \enquote  [1]{``#1''}%
\providecommand \bibnamefont  [1]{#1}%
\providecommand \bibfnamefont [1]{#1}%
\providecommand \citenamefont [1]{#1}%
\providecommand \href@noop [0]{\@secondoftwo}%
\providecommand \href [0]{\begingroup \@sanitize@url \@href}%
\providecommand \@href[1]{\@@startlink{#1}\@@href}%
\providecommand \@@href[1]{\endgroup#1\@@endlink}%
\providecommand \@sanitize@url [0]{\catcode `\\12\catcode `\$12\catcode
  `\&12\catcode `\#12\catcode `\^12\catcode `\_12\catcode `\%12\relax}%
\providecommand \@@startlink[1]{}%
\providecommand \@@endlink[0]{}%
\providecommand \url  [0]{\begingroup\@sanitize@url \@url }%
\providecommand \@url [1]{\endgroup\@href {#1}{\urlprefix }}%
\providecommand \urlprefix  [0]{URL }%
\providecommand \Eprint [0]{\href }%
\providecommand \doibase [0]{https://doi.org/}%
\providecommand \selectlanguage [0]{\@gobble}%
\providecommand \bibinfo  [0]{\@secondoftwo}%
\providecommand \bibfield  [0]{\@secondoftwo}%
\providecommand \translation [1]{[#1]}%
\providecommand \BibitemOpen [0]{}%
\providecommand \bibitemStop [0]{}%
\providecommand \bibitemNoStop [0]{.\EOS\space}%
\providecommand \EOS [0]{\spacefactor3000\relax}%
\providecommand \BibitemShut  [1]{\csname bibitem#1\endcsname}%
\let\auto@bib@innerbib\@empty
\bibitem [{\citenamefont {Ashcroft}(1968)}]{ashcroft_metallic_H}%
  \BibitemOpen
  \bibfield  {author} {\bibinfo {author} {\bibfnamefont {N.~W.}\ \bibnamefont
  {Ashcroft}},\ }\bibfield  {title} {\bibinfo {title} {Metallic hydrogen: A
  high-temperature superconductor?},\ }\href
  {https://doi.org/10.1103/PhysRevLett.21.1748} {\bibfield  {journal} {\bibinfo
   {journal} {Phys. Rev. Lett.}\ }\textbf {\bibinfo {volume} {21}},\ \bibinfo
  {pages} {1748} (\bibinfo {year} {1968})}\BibitemShut {NoStop}%
\bibitem [{\citenamefont {Ashcroft}(2004)}]{ashcroft_alloy}%
  \BibitemOpen
  \bibfield  {author} {\bibinfo {author} {\bibfnamefont {N.~W.}\ \bibnamefont
  {Ashcroft}},\ }\bibfield  {title} {\bibinfo {title} {Hydrogen dominant
  metallic alloys: High temperature superconductors?},\ }\href
  {https://doi.org/10.1103/PhysRevLett.92.187002} {\bibfield  {journal}
  {\bibinfo  {journal} {Phys. Rev. Lett.}\ }\textbf {\bibinfo {volume} {92}},\
  \bibinfo {pages} {187002} (\bibinfo {year} {2004})}\BibitemShut {NoStop}%
\bibitem [{\citenamefont {Peng}\ \emph {et~al.}(2017)\citenamefont {Peng},
  \citenamefont {Sun}, \citenamefont {Pickard}, \citenamefont {Needs},
  \citenamefont {Wu},\ and\ \citenamefont {Ma}}]{structure_search_PRL}%
  \BibitemOpen
  \bibfield  {author} {\bibinfo {author} {\bibfnamefont {F.}~\bibnamefont
  {Peng}}, \bibinfo {author} {\bibfnamefont {Y.}~\bibnamefont {Sun}}, \bibinfo
  {author} {\bibfnamefont {C.~J.}\ \bibnamefont {Pickard}}, \bibinfo {author}
  {\bibfnamefont {R.~J.}\ \bibnamefont {Needs}}, \bibinfo {author}
  {\bibfnamefont {Q.}~\bibnamefont {Wu}},\ and\ \bibinfo {author}
  {\bibfnamefont {Y.}~\bibnamefont {Ma}},\ }\bibfield  {title} {\bibinfo
  {title} {Hydrogen clathrate structures in rare earth hydrides at high
  pressures: Possible route to room-temperature superconductivity},\ }\href
  {https://doi.org/10.1103/PhysRevLett.119.107001} {\bibfield  {journal}
  {\bibinfo  {journal} {Phys. Rev. Lett.}\ }\textbf {\bibinfo {volume} {119}},\
  \bibinfo {pages} {107001} (\bibinfo {year} {2017})}\BibitemShut {NoStop}%
\bibitem [{\citenamefont {Liu}\ \emph {et~al.}(2017)\citenamefont {Liu},
  \citenamefont {Naumov}, \citenamefont {Hoffmann}, \citenamefont {Ashcroft},\
  and\ \citenamefont {Hemley}}]{structure_search_PNAS}%
  \BibitemOpen
  \bibfield  {author} {\bibinfo {author} {\bibfnamefont {H.}~\bibnamefont
  {Liu}}, \bibinfo {author} {\bibfnamefont {I.~I.}\ \bibnamefont {Naumov}},
  \bibinfo {author} {\bibfnamefont {R.}~\bibnamefont {Hoffmann}}, \bibinfo
  {author} {\bibfnamefont {N.~W.}\ \bibnamefont {Ashcroft}},\ and\ \bibinfo
  {author} {\bibfnamefont {R.~J.}\ \bibnamefont {Hemley}},\ }\bibfield  {title}
  {\bibinfo {title} {Potential high-tc superconducting lanthanum and yttrium
  hydrides at high pressure},\ }\href {https://doi.org/10.1073/pnas.1704505114}
  {\bibfield  {journal} {\bibinfo  {journal} {Proceedings of the National
  Academy of Sciences}\ }\textbf {\bibinfo {volume} {114}},\ \bibinfo {pages}
  {6990} (\bibinfo {year} {2017})},\ \Eprint
  {https://arxiv.org/abs/https://www.pnas.org/content/114/27/6990.full.pdf}
  {https://www.pnas.org/content/114/27/6990.full.pdf} \BibitemShut {NoStop}%
\bibitem [{\citenamefont {Flores-Livas}\ \emph {et~al.}(2020)\citenamefont
  {Flores-Livas}, \citenamefont {Boeri}, \citenamefont {Sanna}, \citenamefont
  {Profeta}, \citenamefont {Arita},\ and\ \citenamefont
  {Eremets}}]{review_high_pressure}%
  \BibitemOpen
  \bibfield  {author} {\bibinfo {author} {\bibfnamefont {J.~A.}\ \bibnamefont
  {Flores-Livas}}, \bibinfo {author} {\bibfnamefont {L.}~\bibnamefont {Boeri}},
  \bibinfo {author} {\bibfnamefont {A.}~\bibnamefont {Sanna}}, \bibinfo
  {author} {\bibfnamefont {G.}~\bibnamefont {Profeta}}, \bibinfo {author}
  {\bibfnamefont {R.}~\bibnamefont {Arita}},\ and\ \bibinfo {author}
  {\bibfnamefont {M.}~\bibnamefont {Eremets}},\ }\bibfield  {title} {\bibinfo
  {title} {A perspective on conventional high-temperature superconductors at
  high pressure: Methods and materials},\ }\href
  {https://doi.org/https://doi.org/10.1016/j.physrep.2020.02.003} {\bibfield
  {journal} {\bibinfo  {journal} {Physics Reports}\ }\textbf {\bibinfo {volume}
  {856}},\ \bibinfo {pages} {1} (\bibinfo {year} {2020})}\BibitemShut {NoStop}%
\bibitem [{\citenamefont {Somayazulu}\ \emph {et~al.}(2019)\citenamefont
  {Somayazulu}, \citenamefont {Ahart}, \citenamefont {Mishra}, \citenamefont
  {Geballe}, \citenamefont {Baldini}, \citenamefont {Meng}, \citenamefont
  {Struzhkin},\ and\ \citenamefont {Hemley}}]{LaH_PRL}%
  \BibitemOpen
  \bibfield  {author} {\bibinfo {author} {\bibfnamefont {M.}~\bibnamefont
  {Somayazulu}}, \bibinfo {author} {\bibfnamefont {M.}~\bibnamefont {Ahart}},
  \bibinfo {author} {\bibfnamefont {A.~K.}\ \bibnamefont {Mishra}}, \bibinfo
  {author} {\bibfnamefont {Z.~M.}\ \bibnamefont {Geballe}}, \bibinfo {author}
  {\bibfnamefont {M.}~\bibnamefont {Baldini}}, \bibinfo {author} {\bibfnamefont
  {Y.}~\bibnamefont {Meng}}, \bibinfo {author} {\bibfnamefont {V.~V.}\
  \bibnamefont {Struzhkin}},\ and\ \bibinfo {author} {\bibfnamefont {R.~J.}\
  \bibnamefont {Hemley}},\ }\bibfield  {title} {\bibinfo {title} {Evidence for
  superconductivity above 260 k in lanthanum superhydride at megabar
  pressures},\ }\href {https://doi.org/10.1103/PhysRevLett.122.027001}
  {\bibfield  {journal} {\bibinfo  {journal} {Phys. Rev. Lett.}\ }\textbf
  {\bibinfo {volume} {122}},\ \bibinfo {pages} {027001} (\bibinfo {year}
  {2019})}\BibitemShut {NoStop}%
\bibitem [{\citenamefont {Drozdov}\ \emph {et~al.}(2019)\citenamefont
  {Drozdov}, \citenamefont {Kong}, \citenamefont {Minkov}, \citenamefont
  {Besedin}, \citenamefont {Kuzovnikov}, \citenamefont {Mozaffari},
  \citenamefont {Balicas}, \citenamefont {Balakirev}, \citenamefont {Graf},
  \citenamefont {Prakapenka}, \citenamefont {Greenberg}, \citenamefont
  {Knyazev}, \citenamefont {Tkacz},\ and\ \citenamefont
  {Eremets}}]{LaH_Nature}%
  \BibitemOpen
  \bibfield  {author} {\bibinfo {author} {\bibfnamefont {A.~P.}\ \bibnamefont
  {Drozdov}}, \bibinfo {author} {\bibfnamefont {P.~P.}\ \bibnamefont {Kong}},
  \bibinfo {author} {\bibfnamefont {V.~S.}\ \bibnamefont {Minkov}}, \bibinfo
  {author} {\bibfnamefont {S.~P.}\ \bibnamefont {Besedin}}, \bibinfo {author}
  {\bibfnamefont {M.~A.}\ \bibnamefont {Kuzovnikov}}, \bibinfo {author}
  {\bibfnamefont {S.}~\bibnamefont {Mozaffari}}, \bibinfo {author}
  {\bibfnamefont {L.}~\bibnamefont {Balicas}}, \bibinfo {author} {\bibfnamefont
  {F.~F.}\ \bibnamefont {Balakirev}}, \bibinfo {author} {\bibfnamefont {D.~E.}\
  \bibnamefont {Graf}}, \bibinfo {author} {\bibfnamefont {V.~B.}\ \bibnamefont
  {Prakapenka}}, \bibinfo {author} {\bibfnamefont {E.}~\bibnamefont
  {Greenberg}}, \bibinfo {author} {\bibfnamefont {D.~A.}\ \bibnamefont
  {Knyazev}}, \bibinfo {author} {\bibfnamefont {M.}~\bibnamefont {Tkacz}},\
  and\ \bibinfo {author} {\bibfnamefont {M.~I.}\ \bibnamefont {Eremets}},\
  }\bibfield  {title} {\bibinfo {title} {Superconductivity at 250 k in
  lanthanum hydride under high pressures},\ }\href
  {https://doi.org/10.1038/s41586-019-1201-8} {\bibfield  {journal} {\bibinfo
  {journal} {Nature}\ }\textbf {\bibinfo {volume} {569}},\ \bibinfo {pages}
  {528} (\bibinfo {year} {2019})}\BibitemShut {NoStop}%
\bibitem [{\citenamefont {Liu}\ \emph {et~al.}(2018)\citenamefont {Liu},
  \citenamefont {Naumov}, \citenamefont {Geballe}, \citenamefont {Somayazulu},
  \citenamefont {Tse},\ and\ \citenamefont {Hemley}}]{LaH_AIMD}%
  \BibitemOpen
  \bibfield  {author} {\bibinfo {author} {\bibfnamefont {H.}~\bibnamefont
  {Liu}}, \bibinfo {author} {\bibfnamefont {I.~I.}\ \bibnamefont {Naumov}},
  \bibinfo {author} {\bibfnamefont {Z.~M.}\ \bibnamefont {Geballe}}, \bibinfo
  {author} {\bibfnamefont {M.}~\bibnamefont {Somayazulu}}, \bibinfo {author}
  {\bibfnamefont {J.~S.}\ \bibnamefont {Tse}},\ and\ \bibinfo {author}
  {\bibfnamefont {R.~J.}\ \bibnamefont {Hemley}},\ }\bibfield  {title}
  {\bibinfo {title} {Dynamics and superconductivity in compressed lanthanum
  superhydride},\ }\href {https://doi.org/10.1103/PhysRevB.98.100102}
  {\bibfield  {journal} {\bibinfo  {journal} {Phys. Rev. B}\ }\textbf {\bibinfo
  {volume} {98}},\ \bibinfo {pages} {100102} (\bibinfo {year}
  {2018})}\BibitemShut {NoStop}%
\bibitem [{\citenamefont {Errea}\ \emph {et~al.}(2020)\citenamefont {Errea},
  \citenamefont {Belli}, \citenamefont {Monacelli}, \citenamefont {Sanna},
  \citenamefont {Koretsune}, \citenamefont {Tadano}, \citenamefont {Bianco},
  \citenamefont {Calandra}, \citenamefont {Arita}, \citenamefont {Mauri},\ and\
  \citenamefont {Flores-Livas}}]{LaH_SSCHA}%
  \BibitemOpen
  \bibfield  {author} {\bibinfo {author} {\bibfnamefont {I.}~\bibnamefont
  {Errea}}, \bibinfo {author} {\bibfnamefont {F.}~\bibnamefont {Belli}},
  \bibinfo {author} {\bibfnamefont {L.}~\bibnamefont {Monacelli}}, \bibinfo
  {author} {\bibfnamefont {A.}~\bibnamefont {Sanna}}, \bibinfo {author}
  {\bibfnamefont {T.}~\bibnamefont {Koretsune}}, \bibinfo {author}
  {\bibfnamefont {T.}~\bibnamefont {Tadano}}, \bibinfo {author} {\bibfnamefont
  {R.}~\bibnamefont {Bianco}}, \bibinfo {author} {\bibfnamefont
  {M.}~\bibnamefont {Calandra}}, \bibinfo {author} {\bibfnamefont
  {R.}~\bibnamefont {Arita}}, \bibinfo {author} {\bibfnamefont
  {F.}~\bibnamefont {Mauri}},\ and\ \bibinfo {author} {\bibfnamefont {J.~A.}\
  \bibnamefont {Flores-Livas}},\ }\bibfield  {title} {\bibinfo {title} {Quantum
  crystal structure in the 250-kelvin superconducting lanthanum hydride},\
  }\href {https://doi.org/10.1038/s41586-020-1955-z} {\bibfield  {journal}
  {\bibinfo  {journal} {Nature}\ }\textbf {\bibinfo {volume} {578}},\ \bibinfo
  {pages} {66} (\bibinfo {year} {2020})}\BibitemShut {NoStop}%
\bibitem [{\citenamefont {Geballe}\ \emph {et~al.}(2018)\citenamefont
  {Geballe}, \citenamefont {Liu}, \citenamefont {Mishra}, \citenamefont
  {Ahart}, \citenamefont {Somayazulu}, \citenamefont {Meng}, \citenamefont
  {Baldini},\ and\ \citenamefont {Hemley}}]{LaH_synthesis}%
  \BibitemOpen
  \bibfield  {author} {\bibinfo {author} {\bibfnamefont {Z.~M.}\ \bibnamefont
  {Geballe}}, \bibinfo {author} {\bibfnamefont {H.}~\bibnamefont {Liu}},
  \bibinfo {author} {\bibfnamefont {A.~K.}\ \bibnamefont {Mishra}}, \bibinfo
  {author} {\bibfnamefont {M.}~\bibnamefont {Ahart}}, \bibinfo {author}
  {\bibfnamefont {M.}~\bibnamefont {Somayazulu}}, \bibinfo {author}
  {\bibfnamefont {Y.}~\bibnamefont {Meng}}, \bibinfo {author} {\bibfnamefont
  {M.}~\bibnamefont {Baldini}},\ and\ \bibinfo {author} {\bibfnamefont {R.~J.}\
  \bibnamefont {Hemley}},\ }\bibfield  {title} {\bibinfo {title} {Synthesis and
  stability of lanthanum superhydrides},\ }\href@noop {} {\bibfield  {journal}
  {\bibinfo  {journal} {Angew. Chem. Int. Ed.}\ }\textbf {\bibinfo {volume}
  {57}},\ \bibinfo {pages} {688} (\bibinfo {year} {2018})}\BibitemShut
  {NoStop}%
\bibitem [{\citenamefont {Sun}\ \emph {et~al.}(2021)\citenamefont {Sun},
  \citenamefont {Minkov}, \citenamefont {Mozaffari}, \citenamefont {Sun},
  \citenamefont {Ma}, \citenamefont {Chariton}, \citenamefont {Prakapenka},
  \citenamefont {Eremets}, \citenamefont {Balicas},\ and\ \citenamefont
  {Balakirev}}]{LaH_structural_instability}%
  \BibitemOpen
  \bibfield  {author} {\bibinfo {author} {\bibfnamefont {D.}~\bibnamefont
  {Sun}}, \bibinfo {author} {\bibfnamefont {V.~S.}\ \bibnamefont {Minkov}},
  \bibinfo {author} {\bibfnamefont {S.}~\bibnamefont {Mozaffari}}, \bibinfo
  {author} {\bibfnamefont {Y.}~\bibnamefont {Sun}}, \bibinfo {author}
  {\bibfnamefont {Y.}~\bibnamefont {Ma}}, \bibinfo {author} {\bibfnamefont
  {S.}~\bibnamefont {Chariton}}, \bibinfo {author} {\bibfnamefont {V.~B.}\
  \bibnamefont {Prakapenka}}, \bibinfo {author} {\bibfnamefont {M.~I.}\
  \bibnamefont {Eremets}}, \bibinfo {author} {\bibfnamefont {L.}~\bibnamefont
  {Balicas}},\ and\ \bibinfo {author} {\bibfnamefont {F.~F.}\ \bibnamefont
  {Balakirev}},\ }\bibfield  {title} {\bibinfo {title} {High-temperature
  superconductivity on the verge of a structural instability in lanthanum
  superhydride},\ }\bibfield  {journal} {\bibinfo  {journal} {Nature
  Communications}\ }\textbf {\bibinfo {volume} {12}},\ \href
  {https://doi.org/10.1038/s41467-021-26706-w} {10.1038/s41467-021-26706-w}
  (\bibinfo {year} {2021})\BibitemShut {NoStop}%
\bibitem [{\citenamefont {Watanabe}\ \emph {et~al.}(2022)\citenamefont
  {Watanabe}, \citenamefont {Nomoto},\ and\ \citenamefont {Arita}}]{AIMD_2022}%
  \BibitemOpen
  \bibfield  {author} {\bibinfo {author} {\bibfnamefont {Y.}~\bibnamefont
  {Watanabe}}, \bibinfo {author} {\bibfnamefont {T.}~\bibnamefont {Nomoto}},\
  and\ \bibinfo {author} {\bibfnamefont {R.}~\bibnamefont {Arita}},\ }\bibfield
   {title} {\bibinfo {title} {Quantum and temperature effects on the crystal
  structure of superhydride ${\mathrm{lah}}_{10}$: A path integral molecular
  dynamics study},\ }\href {https://doi.org/10.1103/PhysRevB.105.174111}
  {\bibfield  {journal} {\bibinfo  {journal} {Phys. Rev. B}\ }\textbf {\bibinfo
  {volume} {105}},\ \bibinfo {pages} {174111} (\bibinfo {year}
  {2022})}\BibitemShut {NoStop}%
\bibitem [{\citenamefont {Morresi}\ \emph {et~al.}(2021)\citenamefont
  {Morresi}, \citenamefont {Paulatto}, \citenamefont {Vuilleumier},\ and\
  \citenamefont {Casula}}]{metallic_H_anharmonicity}%
  \BibitemOpen
  \bibfield  {author} {\bibinfo {author} {\bibfnamefont {T.}~\bibnamefont
  {Morresi}}, \bibinfo {author} {\bibfnamefont {L.}~\bibnamefont {Paulatto}},
  \bibinfo {author} {\bibfnamefont {R.}~\bibnamefont {Vuilleumier}},\ and\
  \bibinfo {author} {\bibfnamefont {M.}~\bibnamefont {Casula}},\ }\bibfield
  {title} {\bibinfo {title} {Probing anharmonic phonons by quantum correlators:
  A path integral approach},\ }\href {https://doi.org/10.1063/5.0050450}
  {\bibfield  {journal} {\bibinfo  {journal} {The Journal of Chemical Physics}\
  }\textbf {\bibinfo {volume} {154}},\ \bibinfo {pages} {224108} (\bibinfo
  {year} {2021})},\ \Eprint
  {https://arxiv.org/abs/https://doi.org/10.1063/5.0050450}
  {https://doi.org/10.1063/5.0050450} \BibitemShut {NoStop}%
\bibitem [{\citenamefont {Westermayr}\ \emph {et~al.}(2021)\citenamefont
  {Westermayr}, \citenamefont {Gastegger}, \citenamefont {Schütt},\ and\
  \citenamefont {Maurer}}]{MLP_review_JCP}%
  \BibitemOpen
  \bibfield  {author} {\bibinfo {author} {\bibfnamefont {J.}~\bibnamefont
  {Westermayr}}, \bibinfo {author} {\bibfnamefont {M.}~\bibnamefont
  {Gastegger}}, \bibinfo {author} {\bibfnamefont {K.~T.}\ \bibnamefont
  {Schütt}},\ and\ \bibinfo {author} {\bibfnamefont {R.~J.}\ \bibnamefont
  {Maurer}},\ }\bibfield  {title} {\bibinfo {title} {Perspective on integrating
  machine learning into computational chemistry and materials science},\ }\href
  {https://doi.org/10.1063/5.0047760} {\bibfield  {journal} {\bibinfo
  {journal} {The Journal of Chemical Physics}\ }\textbf {\bibinfo {volume}
  {154}},\ \bibinfo {pages} {230903} (\bibinfo {year} {2021})},\ \Eprint
  {https://arxiv.org/abs/https://doi.org/10.1063/5.0047760}
  {https://doi.org/10.1063/5.0047760} \BibitemShut {NoStop}%
\bibitem [{\citenamefont {Behler}(2015)}]{NNP_review}%
  \BibitemOpen
  \bibfield  {author} {\bibinfo {author} {\bibfnamefont {J.}~\bibnamefont
  {Behler}},\ }\bibfield  {title} {\bibinfo {title} {Constructing
  high-dimensional neural network potentials: A tutorial review},\ }\href@noop
  {} {\bibfield  {journal} {\bibinfo  {journal} {International Journal of
  Quantum Chemistry}\ }\textbf {\bibinfo {volume} {115}},\ \bibinfo {pages}
  {1032} (\bibinfo {year} {2015})}\BibitemShut {NoStop}%
\bibitem [{\citenamefont {Zuo}\ \emph {et~al.}(2020)\citenamefont {Zuo},
  \citenamefont {Chen}, \citenamefont {Li}, \citenamefont {Deng}, \citenamefont
  {Chen}, \citenamefont {Behler}, \citenamefont {Csányi}, \citenamefont
  {Shapeev}, \citenamefont {Thompson}, \citenamefont {Wood},\ and\
  \citenamefont {Ong}}]{MLP_benchmarks}%
  \BibitemOpen
  \bibfield  {author} {\bibinfo {author} {\bibfnamefont {Y.}~\bibnamefont
  {Zuo}}, \bibinfo {author} {\bibfnamefont {C.}~\bibnamefont {Chen}}, \bibinfo
  {author} {\bibfnamefont {X.}~\bibnamefont {Li}}, \bibinfo {author}
  {\bibfnamefont {Z.}~\bibnamefont {Deng}}, \bibinfo {author} {\bibfnamefont
  {Y.}~\bibnamefont {Chen}}, \bibinfo {author} {\bibfnamefont {J.}~\bibnamefont
  {Behler}}, \bibinfo {author} {\bibfnamefont {G.}~\bibnamefont {Csányi}},
  \bibinfo {author} {\bibfnamefont {A.~V.}\ \bibnamefont {Shapeev}}, \bibinfo
  {author} {\bibfnamefont {A.~P.}\ \bibnamefont {Thompson}}, \bibinfo {author}
  {\bibfnamefont {M.~A.}\ \bibnamefont {Wood}},\ and\ \bibinfo {author}
  {\bibfnamefont {S.~P.}\ \bibnamefont {Ong}},\ }\bibfield  {title} {\bibinfo
  {title} {Performance and cost assessment of machine learning interatomic
  potentials},\ }\href {https://doi.org/10.1021/acs.jpca.9b08723} {\bibfield
  {journal} {\bibinfo  {journal} {The Journal of Physical Chemistry A}\
  }\textbf {\bibinfo {volume} {124}},\ \bibinfo {pages} {731} (\bibinfo {year}
  {2020})},\ \bibinfo {note} {pMID: 31916773},\ \Eprint
  {https://arxiv.org/abs/https://doi.org/10.1021/acs.jpca.9b08723}
  {https://doi.org/10.1021/acs.jpca.9b08723} \BibitemShut {NoStop}%
\bibitem [{\citenamefont {Musil}\ \emph {et~al.}(2021)\citenamefont {Musil},
  \citenamefont {Grisafi}, \citenamefont {Bartók}, \citenamefont {Ortner},
  \citenamefont {Csányi},\ and\ \citenamefont {Ceriotti}}]{representations}%
  \BibitemOpen
  \bibfield  {author} {\bibinfo {author} {\bibfnamefont {F.}~\bibnamefont
  {Musil}}, \bibinfo {author} {\bibfnamefont {A.}~\bibnamefont {Grisafi}},
  \bibinfo {author} {\bibfnamefont {A.~P.}\ \bibnamefont {Bartók}}, \bibinfo
  {author} {\bibfnamefont {C.}~\bibnamefont {Ortner}}, \bibinfo {author}
  {\bibfnamefont {G.}~\bibnamefont {Csányi}},\ and\ \bibinfo {author}
  {\bibfnamefont {M.}~\bibnamefont {Ceriotti}},\ }\bibfield  {title} {\bibinfo
  {title} {Physics-inspired structural representations for molecules and
  materials},\ }\href {https://doi.org/10.1021/acs.chemrev.1c00021} {\bibfield
  {journal} {\bibinfo  {journal} {Chemical Reviews}\ }\textbf {\bibinfo
  {volume} {121}},\ \bibinfo {pages} {9759} (\bibinfo {year} {2021})},\
  \bibinfo {note} {pMID: 34310133},\ \Eprint
  {https://arxiv.org/abs/https://doi.org/10.1021/acs.chemrev.1c00021}
  {https://doi.org/10.1021/acs.chemrev.1c00021} \BibitemShut {NoStop}%
\bibitem [{\citenamefont {Giannozzi}\ \emph {et~al.}(2009)\citenamefont
  {Giannozzi}, \citenamefont {Baroni}, \citenamefont {Bonini}, \citenamefont
  {Calandra}, \citenamefont {Car}, \citenamefont {Cavazzoni}, \citenamefont
  {Ceresoli}, \citenamefont {Chiarotti}, \citenamefont {Cococcioni},
  \citenamefont {Dabo}, \citenamefont {Corso}, \citenamefont {de~Gironcoli},
  \citenamefont {Fabris}, \citenamefont {Fratesi}, \citenamefont {Gebauer},
  \citenamefont {Gerstmann}, \citenamefont {Gougoussis}, \citenamefont
  {Kokalj}, \citenamefont {Lazzeri}, \citenamefont {Martin-Samos},
  \citenamefont {Marzari}, \citenamefont {Mauri}, \citenamefont {Mazzarello},
  \citenamefont {Paolini}, \citenamefont {Pasquarello}, \citenamefont
  {Paulatto}, \citenamefont {Sbraccia}, \citenamefont {Scandolo}, \citenamefont
  {Sclauzero}, \citenamefont {Seitsonen}, \citenamefont {Smogunov},
  \citenamefont {Umari},\ and\ \citenamefont {Wentzcovitch}}]{QE_2009}%
  \BibitemOpen
  \bibfield  {author} {\bibinfo {author} {\bibfnamefont {P.}~\bibnamefont
  {Giannozzi}}, \bibinfo {author} {\bibfnamefont {S.}~\bibnamefont {Baroni}},
  \bibinfo {author} {\bibfnamefont {N.}~\bibnamefont {Bonini}}, \bibinfo
  {author} {\bibfnamefont {M.}~\bibnamefont {Calandra}}, \bibinfo {author}
  {\bibfnamefont {R.}~\bibnamefont {Car}}, \bibinfo {author} {\bibfnamefont
  {C.}~\bibnamefont {Cavazzoni}}, \bibinfo {author} {\bibfnamefont
  {D.}~\bibnamefont {Ceresoli}}, \bibinfo {author} {\bibfnamefont {G.~L.}\
  \bibnamefont {Chiarotti}}, \bibinfo {author} {\bibfnamefont {M.}~\bibnamefont
  {Cococcioni}}, \bibinfo {author} {\bibfnamefont {I.}~\bibnamefont {Dabo}},
  \bibinfo {author} {\bibfnamefont {A.~D.}\ \bibnamefont {Corso}}, \bibinfo
  {author} {\bibfnamefont {S.}~\bibnamefont {de~Gironcoli}}, \bibinfo {author}
  {\bibfnamefont {S.}~\bibnamefont {Fabris}}, \bibinfo {author} {\bibfnamefont
  {G.}~\bibnamefont {Fratesi}}, \bibinfo {author} {\bibfnamefont
  {R.}~\bibnamefont {Gebauer}}, \bibinfo {author} {\bibfnamefont
  {U.}~\bibnamefont {Gerstmann}}, \bibinfo {author} {\bibfnamefont
  {C.}~\bibnamefont {Gougoussis}}, \bibinfo {author} {\bibfnamefont
  {A.}~\bibnamefont {Kokalj}}, \bibinfo {author} {\bibfnamefont
  {M.}~\bibnamefont {Lazzeri}}, \bibinfo {author} {\bibfnamefont
  {L.}~\bibnamefont {Martin-Samos}}, \bibinfo {author} {\bibfnamefont
  {N.}~\bibnamefont {Marzari}}, \bibinfo {author} {\bibfnamefont
  {F.}~\bibnamefont {Mauri}}, \bibinfo {author} {\bibfnamefont
  {R.}~\bibnamefont {Mazzarello}}, \bibinfo {author} {\bibfnamefont
  {S.}~\bibnamefont {Paolini}}, \bibinfo {author} {\bibfnamefont
  {A.}~\bibnamefont {Pasquarello}}, \bibinfo {author} {\bibfnamefont
  {L.}~\bibnamefont {Paulatto}}, \bibinfo {author} {\bibfnamefont
  {C.}~\bibnamefont {Sbraccia}}, \bibinfo {author} {\bibfnamefont
  {S.}~\bibnamefont {Scandolo}}, \bibinfo {author} {\bibfnamefont
  {G.}~\bibnamefont {Sclauzero}}, \bibinfo {author} {\bibfnamefont {A.~P.}\
  \bibnamefont {Seitsonen}}, \bibinfo {author} {\bibfnamefont {A.}~\bibnamefont
  {Smogunov}}, \bibinfo {author} {\bibfnamefont {P.}~\bibnamefont {Umari}},\
  and\ \bibinfo {author} {\bibfnamefont {R.~M.}\ \bibnamefont {Wentzcovitch}},\
  }\bibfield  {title} {\bibinfo {title} {{QUANTUM} {ESPRESSO}: a modular and
  open-source software project for quantum simulations of materials},\ }\href
  {https://doi.org/10.1088/0953-8984/21/39/395502} {\bibfield  {journal}
  {\bibinfo  {journal} {Journal of Physics: Condensed Matter}\ }\textbf
  {\bibinfo {volume} {21}},\ \bibinfo {pages} {395502} (\bibinfo {year}
  {2009})}\BibitemShut {NoStop}%
\bibitem [{\citenamefont {Giannozzi}\ \emph {et~al.}(2017)\citenamefont
  {Giannozzi}, \citenamefont {Andreussi}, \citenamefont {Brumme}, \citenamefont
  {Bunau}, \citenamefont {Nardelli}, \citenamefont {Calandra}, \citenamefont
  {Car}, \citenamefont {Cavazzoni}, \citenamefont {Ceresoli}, \citenamefont
  {Cococcioni}, \citenamefont {Colonna}, \citenamefont {Carnimeo},
  \citenamefont {Corso}, \citenamefont {de~Gironcoli}, \citenamefont {Delugas},
  \citenamefont {DiStasio}, \citenamefont {Ferretti}, \citenamefont {Floris},
  \citenamefont {Fratesi}, \citenamefont {Fugallo}, \citenamefont {Gebauer},
  \citenamefont {Gerstmann}, \citenamefont {Giustino}, \citenamefont {Gorni},
  \citenamefont {Jia}, \citenamefont {Kawamura}, \citenamefont {Ko},
  \citenamefont {Kokalj}, \citenamefont {Kü{\c{c}}ükbenli}, \citenamefont
  {Lazzeri}, \citenamefont {Marsili}, \citenamefont {Marzari}, \citenamefont
  {Mauri}, \citenamefont {Nguyen}, \citenamefont {Nguyen}, \citenamefont {de-la
  Roza}, \citenamefont {Paulatto}, \citenamefont {Ponc{\'{e}}}, \citenamefont
  {Rocca}, \citenamefont {Sabatini}, \citenamefont {Santra}, \citenamefont
  {Schlipf}, \citenamefont {Seitsonen}, \citenamefont {Smogunov}, \citenamefont
  {Timrov}, \citenamefont {Thonhauser}, \citenamefont {Umari}, \citenamefont
  {Vast}, \citenamefont {Wu},\ and\ \citenamefont {Baroni}}]{QE_2017}%
  \BibitemOpen
  \bibfield  {author} {\bibinfo {author} {\bibfnamefont {P.}~\bibnamefont
  {Giannozzi}}, \bibinfo {author} {\bibfnamefont {O.}~\bibnamefont
  {Andreussi}}, \bibinfo {author} {\bibfnamefont {T.}~\bibnamefont {Brumme}},
  \bibinfo {author} {\bibfnamefont {O.}~\bibnamefont {Bunau}}, \bibinfo
  {author} {\bibfnamefont {M.~B.}\ \bibnamefont {Nardelli}}, \bibinfo {author}
  {\bibfnamefont {M.}~\bibnamefont {Calandra}}, \bibinfo {author}
  {\bibfnamefont {R.}~\bibnamefont {Car}}, \bibinfo {author} {\bibfnamefont
  {C.}~\bibnamefont {Cavazzoni}}, \bibinfo {author} {\bibfnamefont
  {D.}~\bibnamefont {Ceresoli}}, \bibinfo {author} {\bibfnamefont
  {M.}~\bibnamefont {Cococcioni}}, \bibinfo {author} {\bibfnamefont
  {N.}~\bibnamefont {Colonna}}, \bibinfo {author} {\bibfnamefont
  {I.}~\bibnamefont {Carnimeo}}, \bibinfo {author} {\bibfnamefont {A.~D.}\
  \bibnamefont {Corso}}, \bibinfo {author} {\bibfnamefont {S.}~\bibnamefont
  {de~Gironcoli}}, \bibinfo {author} {\bibfnamefont {P.}~\bibnamefont
  {Delugas}}, \bibinfo {author} {\bibfnamefont {R.~A.}\ \bibnamefont
  {DiStasio}}, \bibinfo {author} {\bibfnamefont {A.}~\bibnamefont {Ferretti}},
  \bibinfo {author} {\bibfnamefont {A.}~\bibnamefont {Floris}}, \bibinfo
  {author} {\bibfnamefont {G.}~\bibnamefont {Fratesi}}, \bibinfo {author}
  {\bibfnamefont {G.}~\bibnamefont {Fugallo}}, \bibinfo {author} {\bibfnamefont
  {R.}~\bibnamefont {Gebauer}}, \bibinfo {author} {\bibfnamefont
  {U.}~\bibnamefont {Gerstmann}}, \bibinfo {author} {\bibfnamefont
  {F.}~\bibnamefont {Giustino}}, \bibinfo {author} {\bibfnamefont
  {T.}~\bibnamefont {Gorni}}, \bibinfo {author} {\bibfnamefont
  {J.}~\bibnamefont {Jia}}, \bibinfo {author} {\bibfnamefont {M.}~\bibnamefont
  {Kawamura}}, \bibinfo {author} {\bibfnamefont {H.-Y.}\ \bibnamefont {Ko}},
  \bibinfo {author} {\bibfnamefont {A.}~\bibnamefont {Kokalj}}, \bibinfo
  {author} {\bibfnamefont {E.}~\bibnamefont {Kü{\c{c}}ükbenli}}, \bibinfo
  {author} {\bibfnamefont {M.}~\bibnamefont {Lazzeri}}, \bibinfo {author}
  {\bibfnamefont {M.}~\bibnamefont {Marsili}}, \bibinfo {author} {\bibfnamefont
  {N.}~\bibnamefont {Marzari}}, \bibinfo {author} {\bibfnamefont
  {F.}~\bibnamefont {Mauri}}, \bibinfo {author} {\bibfnamefont {N.~L.}\
  \bibnamefont {Nguyen}}, \bibinfo {author} {\bibfnamefont {H.-V.}\
  \bibnamefont {Nguyen}}, \bibinfo {author} {\bibfnamefont {A.~O.}\
  \bibnamefont {de-la Roza}}, \bibinfo {author} {\bibfnamefont
  {L.}~\bibnamefont {Paulatto}}, \bibinfo {author} {\bibfnamefont
  {S.}~\bibnamefont {Ponc{\'{e}}}}, \bibinfo {author} {\bibfnamefont
  {D.}~\bibnamefont {Rocca}}, \bibinfo {author} {\bibfnamefont
  {R.}~\bibnamefont {Sabatini}}, \bibinfo {author} {\bibfnamefont
  {B.}~\bibnamefont {Santra}}, \bibinfo {author} {\bibfnamefont
  {M.}~\bibnamefont {Schlipf}}, \bibinfo {author} {\bibfnamefont {A.~P.}\
  \bibnamefont {Seitsonen}}, \bibinfo {author} {\bibfnamefont {A.}~\bibnamefont
  {Smogunov}}, \bibinfo {author} {\bibfnamefont {I.}~\bibnamefont {Timrov}},
  \bibinfo {author} {\bibfnamefont {T.}~\bibnamefont {Thonhauser}}, \bibinfo
  {author} {\bibfnamefont {P.}~\bibnamefont {Umari}}, \bibinfo {author}
  {\bibfnamefont {N.}~\bibnamefont {Vast}}, \bibinfo {author} {\bibfnamefont
  {X.}~\bibnamefont {Wu}},\ and\ \bibinfo {author} {\bibfnamefont
  {S.}~\bibnamefont {Baroni}},\ }\bibfield  {title} {\bibinfo {title} {Advanced
  capabilities for materials modelling with quantum {ESPRESSO}},\ }\href
  {https://doi.org/10.1088/1361-648x/aa8f79} {\bibfield  {journal} {\bibinfo
  {journal} {Journal of Physics: Condensed Matter}\ }\textbf {\bibinfo {volume}
  {29}},\ \bibinfo {pages} {465901} (\bibinfo {year} {2017})}\BibitemShut
  {NoStop}%
\bibitem [{\citenamefont {{Dal Corso}}(2014)}]{pslibrary}%
  \BibitemOpen
  \bibfield  {author} {\bibinfo {author} {\bibfnamefont {A.}~\bibnamefont {{Dal
  Corso}}},\ }\bibfield  {title} {\bibinfo {title} {Pseudopotentials periodic
  table: From h to pu},\ }\href
  {https://doi.org/https://doi.org/10.1016/j.commatsci.2014.07.043} {\bibfield
  {journal} {\bibinfo  {journal} {Computational Materials Science}\ }\textbf
  {\bibinfo {volume} {95}},\ \bibinfo {pages} {337} (\bibinfo {year}
  {2014})}\BibitemShut {NoStop}%
\bibitem [{\citenamefont {Perdew}\ \emph {et~al.}(1996)\citenamefont {Perdew},
  \citenamefont {Burke},\ and\ \citenamefont {Ernzerhof}}]{PBE}%
  \BibitemOpen
  \bibfield  {author} {\bibinfo {author} {\bibfnamefont {J.~P.}\ \bibnamefont
  {Perdew}}, \bibinfo {author} {\bibfnamefont {K.}~\bibnamefont {Burke}},\ and\
  \bibinfo {author} {\bibfnamefont {M.}~\bibnamefont {Ernzerhof}},\ }\bibfield
  {title} {\bibinfo {title} {Generalized gradient approximation made simple},\
  }\href {https://doi.org/10.1103/PhysRevLett.77.3865} {\bibfield  {journal}
  {\bibinfo  {journal} {Phys. Rev. Lett.}\ }\textbf {\bibinfo {volume} {77}},\
  \bibinfo {pages} {3865} (\bibinfo {year} {1996})}\BibitemShut {NoStop}%
\bibitem [{\citenamefont {Perdew}\ and\ \citenamefont {Zunger}(1981)}]{PZ}%
  \BibitemOpen
  \bibfield  {author} {\bibinfo {author} {\bibfnamefont {J.~P.}\ \bibnamefont
  {Perdew}}\ and\ \bibinfo {author} {\bibfnamefont {A.}~\bibnamefont
  {Zunger}},\ }\bibfield  {title} {\bibinfo {title} {Self-interaction
  correction to density-functional approximations for many-electron systems},\
  }\href {https://doi.org/10.1103/PhysRevB.23.5048} {\bibfield  {journal}
  {\bibinfo  {journal} {Phys. Rev. B}\ }\textbf {\bibinfo {volume} {23}},\
  \bibinfo {pages} {5048} (\bibinfo {year} {1981})}\BibitemShut {NoStop}%
\bibitem [{\citenamefont {Zhang}\ \emph
  {et~al.}(2018{\natexlab{a}})\citenamefont {Zhang}, \citenamefont {Han},
  \citenamefont {Wang}, \citenamefont {Car},\ and\ \citenamefont {E}}]{DP}%
  \BibitemOpen
  \bibfield  {author} {\bibinfo {author} {\bibfnamefont {L.}~\bibnamefont
  {Zhang}}, \bibinfo {author} {\bibfnamefont {J.}~\bibnamefont {Han}}, \bibinfo
  {author} {\bibfnamefont {H.}~\bibnamefont {Wang}}, \bibinfo {author}
  {\bibfnamefont {R.}~\bibnamefont {Car}},\ and\ \bibinfo {author}
  {\bibfnamefont {W.}~\bibnamefont {E}},\ }\bibfield  {title} {\bibinfo {title}
  {Deep potential molecular dynamics: A scalable model with the accuracy of
  quantum mechanics},\ }\href {https://doi.org/10.1103/PhysRevLett.120.143001}
  {\bibfield  {journal} {\bibinfo  {journal} {Phys. Rev. Lett.}\ }\textbf
  {\bibinfo {volume} {120}},\ \bibinfo {pages} {143001} (\bibinfo {year}
  {2018}{\natexlab{a}})}\BibitemShut {NoStop}%
\bibitem [{\citenamefont {Zhang}\ \emph
  {et~al.}(2018{\natexlab{b}})\citenamefont {Zhang}, \citenamefont {Han},
  \citenamefont {Wang}, \citenamefont {Saidi}, \citenamefont {Car},\ and\
  \citenamefont {E}}]{DP_SE}%
  \BibitemOpen
  \bibfield  {author} {\bibinfo {author} {\bibfnamefont {L.}~\bibnamefont
  {Zhang}}, \bibinfo {author} {\bibfnamefont {J.}~\bibnamefont {Han}}, \bibinfo
  {author} {\bibfnamefont {H.}~\bibnamefont {Wang}}, \bibinfo {author}
  {\bibfnamefont {W.}~\bibnamefont {Saidi}}, \bibinfo {author} {\bibfnamefont
  {R.}~\bibnamefont {Car}},\ and\ \bibinfo {author} {\bibfnamefont
  {W.}~\bibnamefont {E}},\ }\bibfield  {title} {\bibinfo {title} {End-to-end
  symmetry preserving inter-atomic potential energy model for finite and
  extended systems},\ }in\ \href
  {https://proceedings.neurips.cc/paper/2018/file/e2ad76f2326fbc6b56a45a56c59fafdb-Paper.pdf}
  {\emph {\bibinfo {booktitle} {Advances in Neural Information Processing
  Systems}}},\ Vol.~\bibinfo {volume} {31},\ \bibinfo {editor} {edited by\
  \bibinfo {editor} {\bibfnamefont {S.}~\bibnamefont {Bengio}}, \bibinfo
  {editor} {\bibfnamefont {H.}~\bibnamefont {Wallach}}, \bibinfo {editor}
  {\bibfnamefont {H.}~\bibnamefont {Larochelle}}, \bibinfo {editor}
  {\bibfnamefont {K.}~\bibnamefont {Grauman}}, \bibinfo {editor} {\bibfnamefont
  {N.}~\bibnamefont {Cesa-Bianchi}},\ and\ \bibinfo {editor} {\bibfnamefont
  {R.}~\bibnamefont {Garnett}}}\ (\bibinfo  {publisher} {Curran Associates,
  Inc.},\ \bibinfo {year} {2018})\BibitemShut {NoStop}%
\bibitem [{\citenamefont {Behler}\ and\ \citenamefont {Parrinello}(2007)}]{BP}%
  \BibitemOpen
  \bibfield  {author} {\bibinfo {author} {\bibfnamefont {J.}~\bibnamefont
  {Behler}}\ and\ \bibinfo {author} {\bibfnamefont {M.}~\bibnamefont
  {Parrinello}},\ }\bibfield  {title} {\bibinfo {title} {Generalized
  neural-network representation of high-dimensional potential-energy
  surfaces},\ }\href {https://doi.org/10.1103/PhysRevLett.98.146401} {\bibfield
   {journal} {\bibinfo  {journal} {Phys. Rev. Lett.}\ }\textbf {\bibinfo
  {volume} {98}},\ \bibinfo {pages} {146401} (\bibinfo {year}
  {2007})}\BibitemShut {NoStop}%
\bibitem [{\citenamefont {Bart\'ok}\ \emph {et~al.}(2013)\citenamefont
  {Bart\'ok}, \citenamefont {Kondor},\ and\ \citenamefont {Cs\'anyi}}]{SOAP}%
  \BibitemOpen
  \bibfield  {author} {\bibinfo {author} {\bibfnamefont {A.~P.}\ \bibnamefont
  {Bart\'ok}}, \bibinfo {author} {\bibfnamefont {R.}~\bibnamefont {Kondor}},\
  and\ \bibinfo {author} {\bibfnamefont {G.}~\bibnamefont {Cs\'anyi}},\
  }\bibfield  {title} {\bibinfo {title} {On representing chemical
  environments},\ }\href {https://doi.org/10.1103/PhysRevB.87.184115}
  {\bibfield  {journal} {\bibinfo  {journal} {Phys. Rev. B}\ }\textbf {\bibinfo
  {volume} {87}},\ \bibinfo {pages} {184115} (\bibinfo {year}
  {2013})}\BibitemShut {NoStop}%
\bibitem [{\citenamefont {Tang}\ \emph {et~al.}(2021)\citenamefont {Tang},
  \citenamefont {Ho},\ and\ \citenamefont {Wang}}]{AlCe}%
  \BibitemOpen
  \bibfield  {author} {\bibinfo {author} {\bibfnamefont {L.}~\bibnamefont
  {Tang}}, \bibinfo {author} {\bibfnamefont {K.~M.}\ \bibnamefont {Ho}},\ and\
  \bibinfo {author} {\bibfnamefont {C.~Z.}\ \bibnamefont {Wang}},\ }\bibfield
  {title} {\bibinfo {title} {Molecular dynamics simulation of metallic al–ce
  liquids using a neural network machine learning interatomic potential},\
  }\href {https://doi.org/10.1063/5.0066061} {\bibfield  {journal} {\bibinfo
  {journal} {The Journal of Chemical Physics}\ }\textbf {\bibinfo {volume}
  {155}},\ \bibinfo {pages} {194503} (\bibinfo {year} {2021})},\ \Eprint
  {https://arxiv.org/abs/https://doi.org/10.1063/5.0066061}
  {https://doi.org/10.1063/5.0066061} \BibitemShut {NoStop}%
\bibitem [{\citenamefont {Yang}\ \emph {et~al.}(2021)\citenamefont {Yang},
  \citenamefont {Karmakar},\ and\ \citenamefont {Parrinello}}]{phosphorus}%
  \BibitemOpen
  \bibfield  {author} {\bibinfo {author} {\bibfnamefont {M.}~\bibnamefont
  {Yang}}, \bibinfo {author} {\bibfnamefont {T.}~\bibnamefont {Karmakar}},\
  and\ \bibinfo {author} {\bibfnamefont {M.}~\bibnamefont {Parrinello}},\
  }\bibfield  {title} {\bibinfo {title} {Liquid-liquid critical point in
  phosphorus},\ }\href {https://doi.org/10.1103/PhysRevLett.127.080603}
  {\bibfield  {journal} {\bibinfo  {journal} {Phys. Rev. Lett.}\ }\textbf
  {\bibinfo {volume} {127}},\ \bibinfo {pages} {080603} (\bibinfo {year}
  {2021})}\BibitemShut {NoStop}%
\bibitem [{\citenamefont {Zhang}\ \emph {et~al.}(2021)\citenamefont {Zhang},
  \citenamefont {Wang}, \citenamefont {Car},\ and\ \citenamefont
  {E}}]{DP_water}%
  \BibitemOpen
  \bibfield  {author} {\bibinfo {author} {\bibfnamefont {L.}~\bibnamefont
  {Zhang}}, \bibinfo {author} {\bibfnamefont {H.}~\bibnamefont {Wang}},
  \bibinfo {author} {\bibfnamefont {R.}~\bibnamefont {Car}},\ and\ \bibinfo
  {author} {\bibfnamefont {W.}~\bibnamefont {E}},\ }\bibfield  {title}
  {\bibinfo {title} {Phase diagram of a deep potential water model},\ }\href
  {https://doi.org/10.1103/PhysRevLett.126.236001} {\bibfield  {journal}
  {\bibinfo  {journal} {Phys. Rev. Lett.}\ }\textbf {\bibinfo {volume} {126}},\
  \bibinfo {pages} {236001} (\bibinfo {year} {2021})}\BibitemShut {NoStop}%
\bibitem [{\citenamefont {Myung}\ \emph {et~al.}(2022)\citenamefont {Myung},
  \citenamefont {Hirshberg},\ and\ \citenamefont {Parrinello}}]{supersolid_D}%
  \BibitemOpen
  \bibfield  {author} {\bibinfo {author} {\bibfnamefont {C.~W.}\ \bibnamefont
  {Myung}}, \bibinfo {author} {\bibfnamefont {B.}~\bibnamefont {Hirshberg}},\
  and\ \bibinfo {author} {\bibfnamefont {M.}~\bibnamefont {Parrinello}},\
  }\bibfield  {title} {\bibinfo {title} {Prediction of a supersolid phase in
  high-pressure deuterium},\ }\href
  {https://doi.org/10.1103/PhysRevLett.128.045301} {\bibfield  {journal}
  {\bibinfo  {journal} {Phys. Rev. Lett.}\ }\textbf {\bibinfo {volume} {128}},\
  \bibinfo {pages} {045301} (\bibinfo {year} {2022})}\BibitemShut {NoStop}%
\bibitem [{\citenamefont {He}\ \emph {et~al.}(2022)\citenamefont {He},
  \citenamefont {Wu}, \citenamefont {Zhang}, \citenamefont {Wang},
  \citenamefont {Fu}, \citenamefont {Liu},\ and\ \citenamefont {Zhong}}]{STO}%
  \BibitemOpen
  \bibfield  {author} {\bibinfo {author} {\bibfnamefont {R.}~\bibnamefont
  {He}}, \bibinfo {author} {\bibfnamefont {H.}~\bibnamefont {Wu}}, \bibinfo
  {author} {\bibfnamefont {L.}~\bibnamefont {Zhang}}, \bibinfo {author}
  {\bibfnamefont {X.}~\bibnamefont {Wang}}, \bibinfo {author} {\bibfnamefont
  {F.}~\bibnamefont {Fu}}, \bibinfo {author} {\bibfnamefont {S.}~\bibnamefont
  {Liu}},\ and\ \bibinfo {author} {\bibfnamefont {Z.}~\bibnamefont {Zhong}},\
  }\bibfield  {title} {\bibinfo {title} {Structural phase transitions in
  $\mathrm{SrTi}{\mathrm{o}}_{3}$ from deep potential molecular dynamics},\
  }\href {https://doi.org/10.1103/PhysRevB.105.064104} {\bibfield  {journal}
  {\bibinfo  {journal} {Phys. Rev. B}\ }\textbf {\bibinfo {volume} {105}},\
  \bibinfo {pages} {064104} (\bibinfo {year} {2022})}\BibitemShut {NoStop}%
\bibitem [{\citenamefont {Sun}(2015)}]{libcint}%
  \BibitemOpen
  \bibfield  {author} {\bibinfo {author} {\bibfnamefont {Q.}~\bibnamefont
  {Sun}},\ }\bibfield  {title} {\bibinfo {title} {Libcint: An efficient general
  integral library for gaussian basis functions},\ }\href@noop {} {\bibfield
  {journal} {\bibinfo  {journal} {J. Comp. Chem.}\ }\textbf {\bibinfo {volume}
  {36}},\ \bibinfo {pages} {1664} (\bibinfo {year} {2015})}\BibitemShut
  {NoStop}%
\bibitem [{\citenamefont {Sun}\ \emph {et~al.}(2018)\citenamefont {Sun},
  \citenamefont {Berkelbach}, \citenamefont {Blunt}, \citenamefont {Booth},
  \citenamefont {Guo}, \citenamefont {Li}, \citenamefont {Liu}, \citenamefont
  {McClain}, \citenamefont {Sharma}, \citenamefont {Wouters},\ and\
  \citenamefont {Chan}}]{pyscf_2018}%
  \BibitemOpen
  \bibfield  {author} {\bibinfo {author} {\bibfnamefont {Q.}~\bibnamefont
  {Sun}}, \bibinfo {author} {\bibfnamefont {T.~C.}\ \bibnamefont {Berkelbach}},
  \bibinfo {author} {\bibfnamefont {N.~S.}\ \bibnamefont {Blunt}}, \bibinfo
  {author} {\bibfnamefont {G.~H.}\ \bibnamefont {Booth}}, \bibinfo {author}
  {\bibfnamefont {S.}~\bibnamefont {Guo}}, \bibinfo {author} {\bibfnamefont
  {Z.}~\bibnamefont {Li}}, \bibinfo {author} {\bibfnamefont {J.}~\bibnamefont
  {Liu}}, \bibinfo {author} {\bibfnamefont {J.}~\bibnamefont {McClain}},
  \bibinfo {author} {\bibfnamefont {S.}~\bibnamefont {Sharma}}, \bibinfo
  {author} {\bibfnamefont {S.}~\bibnamefont {Wouters}},\ and\ \bibinfo {author}
  {\bibfnamefont {G.~K.-L.}\ \bibnamefont {Chan}},\ }\bibfield  {title}
  {\bibinfo {title} {Pyscf: the python-based simulations of chemistry
  framework},\ }\href@noop {} {\bibfield  {journal} {\bibinfo  {journal} {WIREs
  Comput. Mol. Sci.}\ }\textbf {\bibinfo {volume} {8}},\ \bibinfo {pages}
  {e1340} (\bibinfo {year} {2018})}\BibitemShut {NoStop}%
\bibitem [{\citenamefont {Sun}\ \emph {et~al.}(2020)\citenamefont {Sun},
  \citenamefont {Zhang}, \citenamefont {Banerjee}, \citenamefont {Bao},
  \citenamefont {Barbry}, \citenamefont {Blunt}, \citenamefont {Bogdanov},
  \citenamefont {Booth}, \citenamefont {Chen}, \citenamefont {Cui},
  \citenamefont {Eriksen}, \citenamefont {Gao}, \citenamefont {Guo},
  \citenamefont {Hermann}, \citenamefont {Hermes}, \citenamefont {Koh},
  \citenamefont {Koval}, \citenamefont {Lehtola}, \citenamefont {Li},
  \citenamefont {Liu}, \citenamefont {Mardirossian}, \citenamefont {McClain},
  \citenamefont {Motta}, \citenamefont {Mussard}, \citenamefont {Pham},
  \citenamefont {Pulkin}, \citenamefont {Purwanto}, \citenamefont {Robinson},
  \citenamefont {Ronca}, \citenamefont {Sayfutyarova}, \citenamefont
  {Scheurer}, \citenamefont {Schurkus}, \citenamefont {Smith}, \citenamefont
  {Sun}, \citenamefont {Sun}, \citenamefont {Upadhyay}, \citenamefont {Wagner},
  \citenamefont {Wang}, \citenamefont {White}, \citenamefont {Whitfield},
  \citenamefont {Williamson}, \citenamefont {Wouters}, \citenamefont {Yang},
  \citenamefont {Yu}, \citenamefont {Zhu}, \citenamefont {Berkelbach},
  \citenamefont {Sharma}, \citenamefont {Sokolov},\ and\ \citenamefont
  {Chan}}]{pyscf_2020}%
  \BibitemOpen
  \bibfield  {author} {\bibinfo {author} {\bibfnamefont {Q.}~\bibnamefont
  {Sun}}, \bibinfo {author} {\bibfnamefont {X.}~\bibnamefont {Zhang}}, \bibinfo
  {author} {\bibfnamefont {S.}~\bibnamefont {Banerjee}}, \bibinfo {author}
  {\bibfnamefont {P.}~\bibnamefont {Bao}}, \bibinfo {author} {\bibfnamefont
  {M.}~\bibnamefont {Barbry}}, \bibinfo {author} {\bibfnamefont {N.~S.}\
  \bibnamefont {Blunt}}, \bibinfo {author} {\bibfnamefont {N.~A.}\ \bibnamefont
  {Bogdanov}}, \bibinfo {author} {\bibfnamefont {G.~H.}\ \bibnamefont {Booth}},
  \bibinfo {author} {\bibfnamefont {J.}~\bibnamefont {Chen}}, \bibinfo {author}
  {\bibfnamefont {Z.-H.}\ \bibnamefont {Cui}}, \bibinfo {author} {\bibfnamefont
  {J.~J.}\ \bibnamefont {Eriksen}}, \bibinfo {author} {\bibfnamefont
  {Y.}~\bibnamefont {Gao}}, \bibinfo {author} {\bibfnamefont {S.}~\bibnamefont
  {Guo}}, \bibinfo {author} {\bibfnamefont {J.}~\bibnamefont {Hermann}},
  \bibinfo {author} {\bibfnamefont {M.~R.}\ \bibnamefont {Hermes}}, \bibinfo
  {author} {\bibfnamefont {K.}~\bibnamefont {Koh}}, \bibinfo {author}
  {\bibfnamefont {P.}~\bibnamefont {Koval}}, \bibinfo {author} {\bibfnamefont
  {S.}~\bibnamefont {Lehtola}}, \bibinfo {author} {\bibfnamefont
  {Z.}~\bibnamefont {Li}}, \bibinfo {author} {\bibfnamefont {J.}~\bibnamefont
  {Liu}}, \bibinfo {author} {\bibfnamefont {N.}~\bibnamefont {Mardirossian}},
  \bibinfo {author} {\bibfnamefont {J.~D.}\ \bibnamefont {McClain}}, \bibinfo
  {author} {\bibfnamefont {M.}~\bibnamefont {Motta}}, \bibinfo {author}
  {\bibfnamefont {B.}~\bibnamefont {Mussard}}, \bibinfo {author} {\bibfnamefont
  {H.~Q.}\ \bibnamefont {Pham}}, \bibinfo {author} {\bibfnamefont
  {A.}~\bibnamefont {Pulkin}}, \bibinfo {author} {\bibfnamefont
  {W.}~\bibnamefont {Purwanto}}, \bibinfo {author} {\bibfnamefont {P.~J.}\
  \bibnamefont {Robinson}}, \bibinfo {author} {\bibfnamefont {E.}~\bibnamefont
  {Ronca}}, \bibinfo {author} {\bibfnamefont {E.~R.}\ \bibnamefont
  {Sayfutyarova}}, \bibinfo {author} {\bibfnamefont {M.}~\bibnamefont
  {Scheurer}}, \bibinfo {author} {\bibfnamefont {H.~F.}\ \bibnamefont
  {Schurkus}}, \bibinfo {author} {\bibfnamefont {J.~E.~T.}\ \bibnamefont
  {Smith}}, \bibinfo {author} {\bibfnamefont {C.}~\bibnamefont {Sun}}, \bibinfo
  {author} {\bibfnamefont {S.-N.}\ \bibnamefont {Sun}}, \bibinfo {author}
  {\bibfnamefont {S.}~\bibnamefont {Upadhyay}}, \bibinfo {author}
  {\bibfnamefont {L.~K.}\ \bibnamefont {Wagner}}, \bibinfo {author}
  {\bibfnamefont {X.}~\bibnamefont {Wang}}, \bibinfo {author} {\bibfnamefont
  {A.}~\bibnamefont {White}}, \bibinfo {author} {\bibfnamefont {J.~D.}\
  \bibnamefont {Whitfield}}, \bibinfo {author} {\bibfnamefont {M.~J.}\
  \bibnamefont {Williamson}}, \bibinfo {author} {\bibfnamefont
  {S.}~\bibnamefont {Wouters}}, \bibinfo {author} {\bibfnamefont
  {J.}~\bibnamefont {Yang}}, \bibinfo {author} {\bibfnamefont {J.~M.}\
  \bibnamefont {Yu}}, \bibinfo {author} {\bibfnamefont {T.}~\bibnamefont
  {Zhu}}, \bibinfo {author} {\bibfnamefont {T.~C.}\ \bibnamefont {Berkelbach}},
  \bibinfo {author} {\bibfnamefont {S.}~\bibnamefont {Sharma}}, \bibinfo
  {author} {\bibfnamefont {A.~Y.}\ \bibnamefont {Sokolov}},\ and\ \bibinfo
  {author} {\bibfnamefont {G.~K.-L.}\ \bibnamefont {Chan}},\ }\bibfield
  {title} {\bibinfo {title} {Recent developments in the pyscf program
  package},\ }\href {https://doi.org/10.1063/5.0006074} {\bibfield  {journal}
  {\bibinfo  {journal} {The Journal of Chemical Physics}\ }\textbf {\bibinfo
  {volume} {153}},\ \bibinfo {pages} {024109} (\bibinfo {year} {2020})},\
  \Eprint {https://arxiv.org/abs/https://doi.org/10.1063/5.0006074}
  {https://doi.org/10.1063/5.0006074} \BibitemShut {NoStop}%
\bibitem [{\citenamefont {Zhang}\ \emph {et~al.}(2019)\citenamefont {Zhang},
  \citenamefont {Lin}, \citenamefont {Wang}, \citenamefont {Car},\ and\
  \citenamefont {E}}]{DPGEN}%
  \BibitemOpen
  \bibfield  {author} {\bibinfo {author} {\bibfnamefont {L.}~\bibnamefont
  {Zhang}}, \bibinfo {author} {\bibfnamefont {D.-Y.}\ \bibnamefont {Lin}},
  \bibinfo {author} {\bibfnamefont {H.}~\bibnamefont {Wang}}, \bibinfo {author}
  {\bibfnamefont {R.}~\bibnamefont {Car}},\ and\ \bibinfo {author}
  {\bibfnamefont {W.}~\bibnamefont {E}},\ }\bibfield  {title} {\bibinfo {title}
  {Active learning of uniformly accurate interatomic potentials for materials
  simulation},\ }\href {https://doi.org/10.1103/PhysRevMaterials.3.023804}
  {\bibfield  {journal} {\bibinfo  {journal} {Phys. Rev. Materials}\ }\textbf
  {\bibinfo {volume} {3}},\ \bibinfo {pages} {023804} (\bibinfo {year}
  {2019})}\BibitemShut {NoStop}%
\bibitem [{\citenamefont {Cheng}\ \emph {et~al.}(2020)\citenamefont {Cheng},
  \citenamefont {Griffiths}, \citenamefont {Wengert}, \citenamefont {Kunkel},
  \citenamefont {Stenczel}, \citenamefont {Zhu}, \citenamefont {Deringer},
  \citenamefont {Bernstein}, \citenamefont {Margraf}, \citenamefont {Reuter},\
  and\ \citenamefont {Csanyi}}]{ASAP}%
  \BibitemOpen
  \bibfield  {author} {\bibinfo {author} {\bibfnamefont {B.}~\bibnamefont
  {Cheng}}, \bibinfo {author} {\bibfnamefont {R.-R.}\ \bibnamefont
  {Griffiths}}, \bibinfo {author} {\bibfnamefont {S.}~\bibnamefont {Wengert}},
  \bibinfo {author} {\bibfnamefont {C.}~\bibnamefont {Kunkel}}, \bibinfo
  {author} {\bibfnamefont {T.}~\bibnamefont {Stenczel}}, \bibinfo {author}
  {\bibfnamefont {B.}~\bibnamefont {Zhu}}, \bibinfo {author} {\bibfnamefont
  {V.~L.}\ \bibnamefont {Deringer}}, \bibinfo {author} {\bibfnamefont
  {N.}~\bibnamefont {Bernstein}}, \bibinfo {author} {\bibfnamefont {J.~T.}\
  \bibnamefont {Margraf}}, \bibinfo {author} {\bibfnamefont {K.}~\bibnamefont
  {Reuter}},\ and\ \bibinfo {author} {\bibfnamefont {G.}~\bibnamefont
  {Csanyi}},\ }\bibfield  {title} {\bibinfo {title} {Mapping materials and
  molecules},\ }\href {https://doi.org/10.1021/acs.accounts.0c00403} {\bibfield
   {journal} {\bibinfo  {journal} {Accounts of Chemical Research}\ }\textbf
  {\bibinfo {volume} {53}},\ \bibinfo {pages} {1981} (\bibinfo {year}
  {2020})},\ \bibinfo {note} {pMID: 32794697},\ \Eprint
  {https://arxiv.org/abs/https://doi.org/10.1021/acs.accounts.0c00403}
  {https://doi.org/10.1021/acs.accounts.0c00403} \BibitemShut {NoStop}%
\bibitem [{\citenamefont {McMahon}\ and\ \citenamefont
  {Ceperley}(2011)}]{cs_iv_structure}%
  \BibitemOpen
  \bibfield  {author} {\bibinfo {author} {\bibfnamefont {J.~M.}\ \bibnamefont
  {McMahon}}\ and\ \bibinfo {author} {\bibfnamefont {D.~M.}\ \bibnamefont
  {Ceperley}},\ }\bibfield  {title} {\bibinfo {title} {Ground-state structures
  of atomic metallic hydrogen},\ }\href
  {https://doi.org/10.1103/PhysRevLett.106.165302} {\bibfield  {journal}
  {\bibinfo  {journal} {Phys. Rev. Lett.}\ }\textbf {\bibinfo {volume} {106}},\
  \bibinfo {pages} {165302} (\bibinfo {year} {2011})}\BibitemShut {NoStop}%
\bibitem [{\citenamefont {Chen}\ \emph {et~al.}(2020)\citenamefont {Chen},
  \citenamefont {Semenok}, \citenamefont {Troyan}, \citenamefont {Ivanova},
  \citenamefont {Huang}, \citenamefont {Oganov},\ and\ \citenamefont
  {Cui}}]{La_EOS}%
  \BibitemOpen
  \bibfield  {author} {\bibinfo {author} {\bibfnamefont {W.}~\bibnamefont
  {Chen}}, \bibinfo {author} {\bibfnamefont {D.~V.}\ \bibnamefont {Semenok}},
  \bibinfo {author} {\bibfnamefont {I.~A.}\ \bibnamefont {Troyan}}, \bibinfo
  {author} {\bibfnamefont {A.~G.}\ \bibnamefont {Ivanova}}, \bibinfo {author}
  {\bibfnamefont {X.}~\bibnamefont {Huang}}, \bibinfo {author} {\bibfnamefont
  {A.~R.}\ \bibnamefont {Oganov}},\ and\ \bibinfo {author} {\bibfnamefont
  {T.}~\bibnamefont {Cui}},\ }\bibfield  {title} {\bibinfo {title}
  {Superconductivity and equation of state of lanthanum at megabar pressures},\
  }\href {https://doi.org/10.1103/PhysRevB.102.134510} {\bibfield  {journal}
  {\bibinfo  {journal} {Phys. Rev. B}\ }\textbf {\bibinfo {volume} {102}},\
  \bibinfo {pages} {134510} (\bibinfo {year} {2020})}\BibitemShut {NoStop}%
\bibitem [{\citenamefont {Thompson}\ \emph {et~al.}(2022)\citenamefont
  {Thompson}, \citenamefont {Aktulga}, \citenamefont {Berger}, \citenamefont
  {Bolintineanu}, \citenamefont {Brown}, \citenamefont {Crozier}, \citenamefont
  {in~'t Veld}, \citenamefont {Kohlmeyer}, \citenamefont {Moore}, \citenamefont
  {Nguyen}, \citenamefont {Shan}, \citenamefont {Stevens}, \citenamefont
  {Tranchida}, \citenamefont {Trott},\ and\ \citenamefont {Plimpton}}]{LAMMPS}%
  \BibitemOpen
  \bibfield  {author} {\bibinfo {author} {\bibfnamefont {A.~P.}\ \bibnamefont
  {Thompson}}, \bibinfo {author} {\bibfnamefont {H.~M.}\ \bibnamefont
  {Aktulga}}, \bibinfo {author} {\bibfnamefont {R.}~\bibnamefont {Berger}},
  \bibinfo {author} {\bibfnamefont {D.~S.}\ \bibnamefont {Bolintineanu}},
  \bibinfo {author} {\bibfnamefont {W.~M.}\ \bibnamefont {Brown}}, \bibinfo
  {author} {\bibfnamefont {P.~S.}\ \bibnamefont {Crozier}}, \bibinfo {author}
  {\bibfnamefont {P.~J.}\ \bibnamefont {in~'t Veld}}, \bibinfo {author}
  {\bibfnamefont {A.}~\bibnamefont {Kohlmeyer}}, \bibinfo {author}
  {\bibfnamefont {S.~G.}\ \bibnamefont {Moore}}, \bibinfo {author}
  {\bibfnamefont {T.~D.}\ \bibnamefont {Nguyen}}, \bibinfo {author}
  {\bibfnamefont {R.}~\bibnamefont {Shan}}, \bibinfo {author} {\bibfnamefont
  {M.~J.}\ \bibnamefont {Stevens}}, \bibinfo {author} {\bibfnamefont
  {J.}~\bibnamefont {Tranchida}}, \bibinfo {author} {\bibfnamefont
  {C.}~\bibnamefont {Trott}},\ and\ \bibinfo {author} {\bibfnamefont {S.~J.}\
  \bibnamefont {Plimpton}},\ }\bibfield  {title} {\bibinfo {title} {{LAMMPS} -
  a flexible simulation tool for particle-based materials modeling at the
  atomic, meso, and continuum scales},\ }\href
  {https://doi.org/10.1016/j.cpc.2021.108171} {\bibfield  {journal} {\bibinfo
  {journal} {Comp. Phys. Comm.}\ }\textbf {\bibinfo {volume} {271}},\ \bibinfo
  {pages} {108171} (\bibinfo {year} {2022})}\BibitemShut {NoStop}%
\bibitem [{\citenamefont {Kapil}\ \emph {et~al.}(2019)\citenamefont {Kapil},
  \citenamefont {Rossi}, \citenamefont {Marsalek}, \citenamefont {Petraglia},
  \citenamefont {Litman}, \citenamefont {Spura}, \citenamefont {Cheng},
  \citenamefont {Cuzzocrea}, \citenamefont {Meißner}, \citenamefont {Wilkins},
  \citenamefont {Helfrecht}, \citenamefont {Juda}, \citenamefont {Bienvenue},
  \citenamefont {Fang}, \citenamefont {Kessler}, \citenamefont {Poltavsky},
  \citenamefont {Vandenbrande}, \citenamefont {Wieme}, \citenamefont
  {Corminboeuf}, \citenamefont {Kühne}, \citenamefont {Manolopoulos},
  \citenamefont {Markland}, \citenamefont {Richardson}, \citenamefont
  {Tkatchenko}, \citenamefont {Tribello}, \citenamefont {{Van Speybroeck}},\
  and\ \citenamefont {Ceriotti}}]{i-pi}%
  \BibitemOpen
  \bibfield  {author} {\bibinfo {author} {\bibfnamefont {V.}~\bibnamefont
  {Kapil}}, \bibinfo {author} {\bibfnamefont {M.}~\bibnamefont {Rossi}},
  \bibinfo {author} {\bibfnamefont {O.}~\bibnamefont {Marsalek}}, \bibinfo
  {author} {\bibfnamefont {R.}~\bibnamefont {Petraglia}}, \bibinfo {author}
  {\bibfnamefont {Y.}~\bibnamefont {Litman}}, \bibinfo {author} {\bibfnamefont
  {T.}~\bibnamefont {Spura}}, \bibinfo {author} {\bibfnamefont
  {B.}~\bibnamefont {Cheng}}, \bibinfo {author} {\bibfnamefont
  {A.}~\bibnamefont {Cuzzocrea}}, \bibinfo {author} {\bibfnamefont {R.~H.}\
  \bibnamefont {Meißner}}, \bibinfo {author} {\bibfnamefont {D.~M.}\
  \bibnamefont {Wilkins}}, \bibinfo {author} {\bibfnamefont {B.~A.}\
  \bibnamefont {Helfrecht}}, \bibinfo {author} {\bibfnamefont {P.}~\bibnamefont
  {Juda}}, \bibinfo {author} {\bibfnamefont {S.~P.}\ \bibnamefont {Bienvenue}},
  \bibinfo {author} {\bibfnamefont {W.}~\bibnamefont {Fang}}, \bibinfo {author}
  {\bibfnamefont {J.}~\bibnamefont {Kessler}}, \bibinfo {author} {\bibfnamefont
  {I.}~\bibnamefont {Poltavsky}}, \bibinfo {author} {\bibfnamefont
  {S.}~\bibnamefont {Vandenbrande}}, \bibinfo {author} {\bibfnamefont
  {J.}~\bibnamefont {Wieme}}, \bibinfo {author} {\bibfnamefont
  {C.}~\bibnamefont {Corminboeuf}}, \bibinfo {author} {\bibfnamefont {T.~D.}\
  \bibnamefont {Kühne}}, \bibinfo {author} {\bibfnamefont {D.~E.}\
  \bibnamefont {Manolopoulos}}, \bibinfo {author} {\bibfnamefont {T.~E.}\
  \bibnamefont {Markland}}, \bibinfo {author} {\bibfnamefont {J.~O.}\
  \bibnamefont {Richardson}}, \bibinfo {author} {\bibfnamefont
  {A.}~\bibnamefont {Tkatchenko}}, \bibinfo {author} {\bibfnamefont {G.~A.}\
  \bibnamefont {Tribello}}, \bibinfo {author} {\bibfnamefont {V.}~\bibnamefont
  {{Van Speybroeck}}},\ and\ \bibinfo {author} {\bibfnamefont {M.}~\bibnamefont
  {Ceriotti}},\ }\bibfield  {title} {\bibinfo {title} {i-pi 2.0: A universal
  force engine for advanced molecular simulations},\ }\href
  {https://doi.org/https://doi.org/10.1016/j.cpc.2018.09.020} {\bibfield
  {journal} {\bibinfo  {journal} {Computer Physics Communications}\ }\textbf
  {\bibinfo {volume} {236}},\ \bibinfo {pages} {214} (\bibinfo {year}
  {2019})}\BibitemShut {NoStop}%
\bibitem [{\citenamefont {Martyna}\ \emph {et~al.}(1999)\citenamefont
  {Martyna}, \citenamefont {Hughes},\ and\ \citenamefont
  {Tuckerman}}]{barostat}%
  \BibitemOpen
  \bibfield  {author} {\bibinfo {author} {\bibfnamefont {G.~J.}\ \bibnamefont
  {Martyna}}, \bibinfo {author} {\bibfnamefont {A.}~\bibnamefont {Hughes}},\
  and\ \bibinfo {author} {\bibfnamefont {M.~E.}\ \bibnamefont {Tuckerman}},\
  }\bibfield  {title} {\bibinfo {title} {Molecular dynamics algorithms for path
  integrals at constant pressure},\ }\href {https://doi.org/10.1063/1.478193}
  {\bibfield  {journal} {\bibinfo  {journal} {The Journal of Chemical Physics}\
  }\textbf {\bibinfo {volume} {110}},\ \bibinfo {pages} {3275} (\bibinfo {year}
  {1999})},\ \Eprint {https://arxiv.org/abs/https://doi.org/10.1063/1.478193}
  {https://doi.org/10.1063/1.478193} \BibitemShut {NoStop}%
\bibitem [{\citenamefont {Minkov}\ \emph {et~al.}(2020)\citenamefont {Minkov},
  \citenamefont {Prakapenka}, \citenamefont {Greenberg},\ and\ \citenamefont
  {Eremets}}]{sulfur_hydride}%
  \BibitemOpen
  \bibfield  {author} {\bibinfo {author} {\bibfnamefont {V.~S.}\ \bibnamefont
  {Minkov}}, \bibinfo {author} {\bibfnamefont {V.~B.}\ \bibnamefont
  {Prakapenka}}, \bibinfo {author} {\bibfnamefont {E.}~\bibnamefont
  {Greenberg}},\ and\ \bibinfo {author} {\bibfnamefont {M.~I.}\ \bibnamefont
  {Eremets}},\ }\bibfield  {title} {\bibinfo {title} {A boosted critical
  temperature of 166 k in superconducting d$_3$s synthesized from elemental
  sulfur and hydrogen},\ }\href@noop {} {\bibfield  {journal} {\bibinfo
  {journal} {Angew. Chem. Int. Ed.}\ }\textbf {\bibinfo {volume} {59}},\
  \bibinfo {pages} {18970} (\bibinfo {year} {2020})}\BibitemShut {NoStop}%
\bibitem [{\citenamefont {Clay}\ \emph {et~al.}(2014)\citenamefont {Clay},
  \citenamefont {Mcminis}, \citenamefont {McMahon}, \citenamefont {Pierleoni},
  \citenamefont {Ceperley},\ and\ \citenamefont {Morales}}]{xc_benchmarks}%
  \BibitemOpen
  \bibfield  {author} {\bibinfo {author} {\bibfnamefont {R.~C.}\ \bibnamefont
  {Clay}}, \bibinfo {author} {\bibfnamefont {J.}~\bibnamefont {Mcminis}},
  \bibinfo {author} {\bibfnamefont {J.~M.}\ \bibnamefont {McMahon}}, \bibinfo
  {author} {\bibfnamefont {C.}~\bibnamefont {Pierleoni}}, \bibinfo {author}
  {\bibfnamefont {D.~M.}\ \bibnamefont {Ceperley}},\ and\ \bibinfo {author}
  {\bibfnamefont {M.~A.}\ \bibnamefont {Morales}},\ }\bibfield  {title}
  {\bibinfo {title} {Benchmarking exchange-correlation functionals for hydrogen
  at high pressures using quantum monte carlo},\ }\href
  {https://doi.org/10.1103/PhysRevB.89.184106} {\bibfield  {journal} {\bibinfo
  {journal} {Phys. Rev. B}\ }\textbf {\bibinfo {volume} {89}},\ \bibinfo
  {pages} {184106} (\bibinfo {year} {2014})}\BibitemShut {NoStop}%
\bibitem [{\citenamefont {Zhang}\ \emph {et~al.}(2022)\citenamefont {Zhang},
  \citenamefont {Cui}, \citenamefont {Hutcheon}, \citenamefont {Shipley},
  \citenamefont {Song}, \citenamefont {Du}, \citenamefont {Kresin},
  \citenamefont {Duan}, \citenamefont {Pickard},\ and\ \citenamefont
  {Yao}}]{lower_pressure_H}%
  \BibitemOpen
  \bibfield  {author} {\bibinfo {author} {\bibfnamefont {Z.}~\bibnamefont
  {Zhang}}, \bibinfo {author} {\bibfnamefont {T.}~\bibnamefont {Cui}}, \bibinfo
  {author} {\bibfnamefont {M.~J.}\ \bibnamefont {Hutcheon}}, \bibinfo {author}
  {\bibfnamefont {A.~M.}\ \bibnamefont {Shipley}}, \bibinfo {author}
  {\bibfnamefont {H.}~\bibnamefont {Song}}, \bibinfo {author} {\bibfnamefont
  {M.}~\bibnamefont {Du}}, \bibinfo {author} {\bibfnamefont {V.~Z.}\
  \bibnamefont {Kresin}}, \bibinfo {author} {\bibfnamefont {D.}~\bibnamefont
  {Duan}}, \bibinfo {author} {\bibfnamefont {C.~J.}\ \bibnamefont {Pickard}},\
  and\ \bibinfo {author} {\bibfnamefont {Y.}~\bibnamefont {Yao}},\ }\bibfield
  {title} {\bibinfo {title} {Design principles for high-temperature
  superconductors with a hydrogen-based alloy backbone at moderate pressure},\
  }\href {https://doi.org/10.1103/PhysRevLett.128.047001} {\bibfield  {journal}
  {\bibinfo  {journal} {Phys. Rev. Lett.}\ }\textbf {\bibinfo {volume} {128}},\
  \bibinfo {pages} {047001} (\bibinfo {year} {2022})}\BibitemShut {NoStop}%
\bibitem [{\citenamefont {Wang}\ \emph {et~al.}(2022)\citenamefont {Wang},
  \citenamefont {Flores-Livas}, \citenamefont {Nomoto}, \citenamefont {Ma},
  \citenamefont {Koretsune},\ and\ \citenamefont {Arita}}]{doping}%
  \BibitemOpen
  \bibfield  {author} {\bibinfo {author} {\bibfnamefont {T.}~\bibnamefont
  {Wang}}, \bibinfo {author} {\bibfnamefont {J.~A.}\ \bibnamefont
  {Flores-Livas}}, \bibinfo {author} {\bibfnamefont {T.}~\bibnamefont
  {Nomoto}}, \bibinfo {author} {\bibfnamefont {Y.}~\bibnamefont {Ma}}, \bibinfo
  {author} {\bibfnamefont {T.}~\bibnamefont {Koretsune}},\ and\ \bibinfo
  {author} {\bibfnamefont {R.}~\bibnamefont {Arita}},\ }\bibfield  {title}
  {\bibinfo {title} {Optimal alloying in hydrides: Reaching room-temperature
  superconductivity in ${\mathrm{lah}}_{10}$},\ }\href
  {https://doi.org/10.1103/PhysRevB.105.174516} {\bibfield  {journal} {\bibinfo
   {journal} {Phys. Rev. B}\ }\textbf {\bibinfo {volume} {105}},\ \bibinfo
  {pages} {174516} (\bibinfo {year} {2022})}\BibitemShut {NoStop}%
\end{thebibliography}%


\begin{thebibliography}{4}%
\makeatletter
\providecommand \@ifxundefined [1]{%
 \@ifx{#1\undefined}
}%
\providecommand \@ifnum [1]{%
 \ifnum #1\expandafter \@firstoftwo
 \else \expandafter \@secondoftwo
 \fi
}%
\providecommand \@ifx [1]{%
 \ifx #1\expandafter \@firstoftwo
 \else \expandafter \@secondoftwo
 \fi
}%
\providecommand \natexlab [1]{#1}%
\providecommand \enquote  [1]{``#1''}%
\providecommand \bibnamefont  [1]{#1}%
\providecommand \bibfnamefont [1]{#1}%
\providecommand \citenamefont [1]{#1}%
\providecommand \href@noop [0]{\@secondoftwo}%
\providecommand \href [0]{\begingroup \@sanitize@url \@href}%
\providecommand \@href[1]{\@@startlink{#1}\@@href}%
\providecommand \@@href[1]{\endgroup#1\@@endlink}%
\providecommand \@sanitize@url [0]{\catcode `\\12\catcode `\$12\catcode
  `\&12\catcode `\#12\catcode `\^12\catcode `\_12\catcode `\%12\relax}%
\providecommand \@@startlink[1]{}%
\providecommand \@@endlink[0]{}%
\providecommand \url  [0]{\begingroup\@sanitize@url \@url }%
\providecommand \@url [1]{\endgroup\@href {#1}{\urlprefix }}%
\providecommand \urlprefix  [0]{URL }%
\providecommand \Eprint [0]{\href }%
\providecommand \doibase [0]{https://doi.org/}%
\providecommand \selectlanguage [0]{\@gobble}%
\providecommand \bibinfo  [0]{\@secondoftwo}%
\providecommand \bibfield  [0]{\@secondoftwo}%
\providecommand \translation [1]{[#1]}%
\providecommand \BibitemOpen [0]{}%
\providecommand \bibitemStop [0]{}%
\providecommand \bibitemNoStop [0]{.\EOS\space}%
\providecommand \EOS [0]{\spacefactor3000\relax}%
\providecommand \BibitemShut  [1]{\csname bibitem#1\endcsname}%
\let\auto@bib@innerbib\@empty
\bibitem [{\citenamefont {Fan}\ \emph {et~al.}(2021)\citenamefont {Fan},
  \citenamefont {Zeng}, \citenamefont {Zhang}, \citenamefont {Wang},
  \citenamefont {Song}, \citenamefont {Dong}, \citenamefont {Chen},\ and\
  \citenamefont {Ala-Nissila}}]{NEP1}%
  \BibitemOpen
  \bibfield  {author} {\bibinfo {author} {\bibfnamefont {Z.}~\bibnamefont
  {Fan}}, \bibinfo {author} {\bibfnamefont {Z.}~\bibnamefont {Zeng}}, \bibinfo
  {author} {\bibfnamefont {C.}~\bibnamefont {Zhang}}, \bibinfo {author}
  {\bibfnamefont {Y.}~\bibnamefont {Wang}}, \bibinfo {author} {\bibfnamefont
  {K.}~\bibnamefont {Song}}, \bibinfo {author} {\bibfnamefont {H.}~\bibnamefont
  {Dong}}, \bibinfo {author} {\bibfnamefont {Y.}~\bibnamefont {Chen}},\ and\
  \bibinfo {author} {\bibfnamefont {T.}~\bibnamefont {Ala-Nissila}},\
  }\bibfield  {title} {\bibinfo {title} {Neuroevolution machine learning
  potentials: Combining high accuracy and low cost in atomistic simulations and
  application to heat transport},\ }\href
  {https://doi.org/10.1103/PhysRevB.104.104309} {\bibfield  {journal} {\bibinfo
   {journal} {Phys. Rev. B}\ }\textbf {\bibinfo {volume} {104}},\ \bibinfo
  {pages} {104309} (\bibinfo {year} {2021})}\BibitemShut {NoStop}%
\bibitem [{\citenamefont {Fan}(2022)}]{NEP2}%
  \BibitemOpen
  \bibfield  {author} {\bibinfo {author} {\bibfnamefont {Z.}~\bibnamefont
  {Fan}},\ }\bibfield  {title} {\bibinfo {title} {Improving the accuracy of the
  neuroevolution machine learning potential for multi-component systems},\
  }\href {https://doi.org/10.1088/1361-648x/ac462b} {\bibfield  {journal}
  {\bibinfo  {journal} {Journal of Physics: Condensed Matter}\ }\textbf
  {\bibinfo {volume} {34}},\ \bibinfo {pages} {125902} (\bibinfo {year}
  {2022})}\BibitemShut {NoStop}%
\bibitem [{\citenamefont {Liu}\ \emph {et~al.}(2018)\citenamefont {Liu},
  \citenamefont {Naumov}, \citenamefont {Geballe}, \citenamefont {Somayazulu},
  \citenamefont {Tse},\ and\ \citenamefont {Hemley}}]{LaH_AIMD}%
  \BibitemOpen
  \bibfield  {author} {\bibinfo {author} {\bibfnamefont {H.}~\bibnamefont
  {Liu}}, \bibinfo {author} {\bibfnamefont {I.~I.}\ \bibnamefont {Naumov}},
  \bibinfo {author} {\bibfnamefont {Z.~M.}\ \bibnamefont {Geballe}}, \bibinfo
  {author} {\bibfnamefont {M.}~\bibnamefont {Somayazulu}}, \bibinfo {author}
  {\bibfnamefont {J.~S.}\ \bibnamefont {Tse}},\ and\ \bibinfo {author}
  {\bibfnamefont {R.~J.}\ \bibnamefont {Hemley}},\ }\bibfield  {title}
  {\bibinfo {title} {Dynamics and superconductivity in compressed lanthanum
  superhydride},\ }\href {https://doi.org/10.1103/PhysRevB.98.100102}
  {\bibfield  {journal} {\bibinfo  {journal} {Phys. Rev. B}\ }\textbf {\bibinfo
  {volume} {98}},\ \bibinfo {pages} {100102} (\bibinfo {year}
  {2018})}\BibitemShut {NoStop}%
\bibitem [{\citenamefont {Errea}\ \emph {et~al.}(2020)\citenamefont {Errea},
  \citenamefont {Belli}, \citenamefont {Monacelli}, \citenamefont {Sanna},
  \citenamefont {Koretsune}, \citenamefont {Tadano}, \citenamefont {Bianco},
  \citenamefont {Calandra}, \citenamefont {Arita}, \citenamefont {Mauri},\ and\
  \citenamefont {Flores-Livas}}]{LaH_SSCHA}%
  \BibitemOpen
  \bibfield  {author} {\bibinfo {author} {\bibfnamefont {I.}~\bibnamefont
  {Errea}}, \bibinfo {author} {\bibfnamefont {F.}~\bibnamefont {Belli}},
  \bibinfo {author} {\bibfnamefont {L.}~\bibnamefont {Monacelli}}, \bibinfo
  {author} {\bibfnamefont {A.}~\bibnamefont {Sanna}}, \bibinfo {author}
  {\bibfnamefont {T.}~\bibnamefont {Koretsune}}, \bibinfo {author}
  {\bibfnamefont {T.}~\bibnamefont {Tadano}}, \bibinfo {author} {\bibfnamefont
  {R.}~\bibnamefont {Bianco}}, \bibinfo {author} {\bibfnamefont
  {M.}~\bibnamefont {Calandra}}, \bibinfo {author} {\bibfnamefont
  {R.}~\bibnamefont {Arita}}, \bibinfo {author} {\bibfnamefont
  {F.}~\bibnamefont {Mauri}},\ and\ \bibinfo {author} {\bibfnamefont {J.~A.}\
  \bibnamefont {Flores-Livas}},\ }\bibfield  {title} {\bibinfo {title} {Quantum
  crystal structure in the 250-kelvin superconducting lanthanum hydride},\
  }\href {https://doi.org/10.1038/s41586-020-1955-z} {\bibfield  {journal}
  {\bibinfo  {journal} {Nature}\ }\textbf {\bibinfo {volume} {578}},\ \bibinfo
  {pages} {66} (\bibinfo {year} {2020})}\BibitemShut {NoStop}%
\end{thebibliography}%

\end{document}


\title{Stability and distortion of fcc-LaH$_{10}$ with path-integral molecular dynamics (supplemental material)}

\author{Kevin K. Ly}
\email{kkly2@illinois.edu}
\author{David M. Ceperley}
\email{ceperley@illinois.edu}
\affiliation{Department of Physics, University of Illinois at Urbana-Champaign}

\date{\today}

\maketitle

\section{\label{sec:DFT}DFPT}

\begin{figure}
    \begin{subfigure}{0.32\linewidth}
        \includegraphics[width=\textwidth]{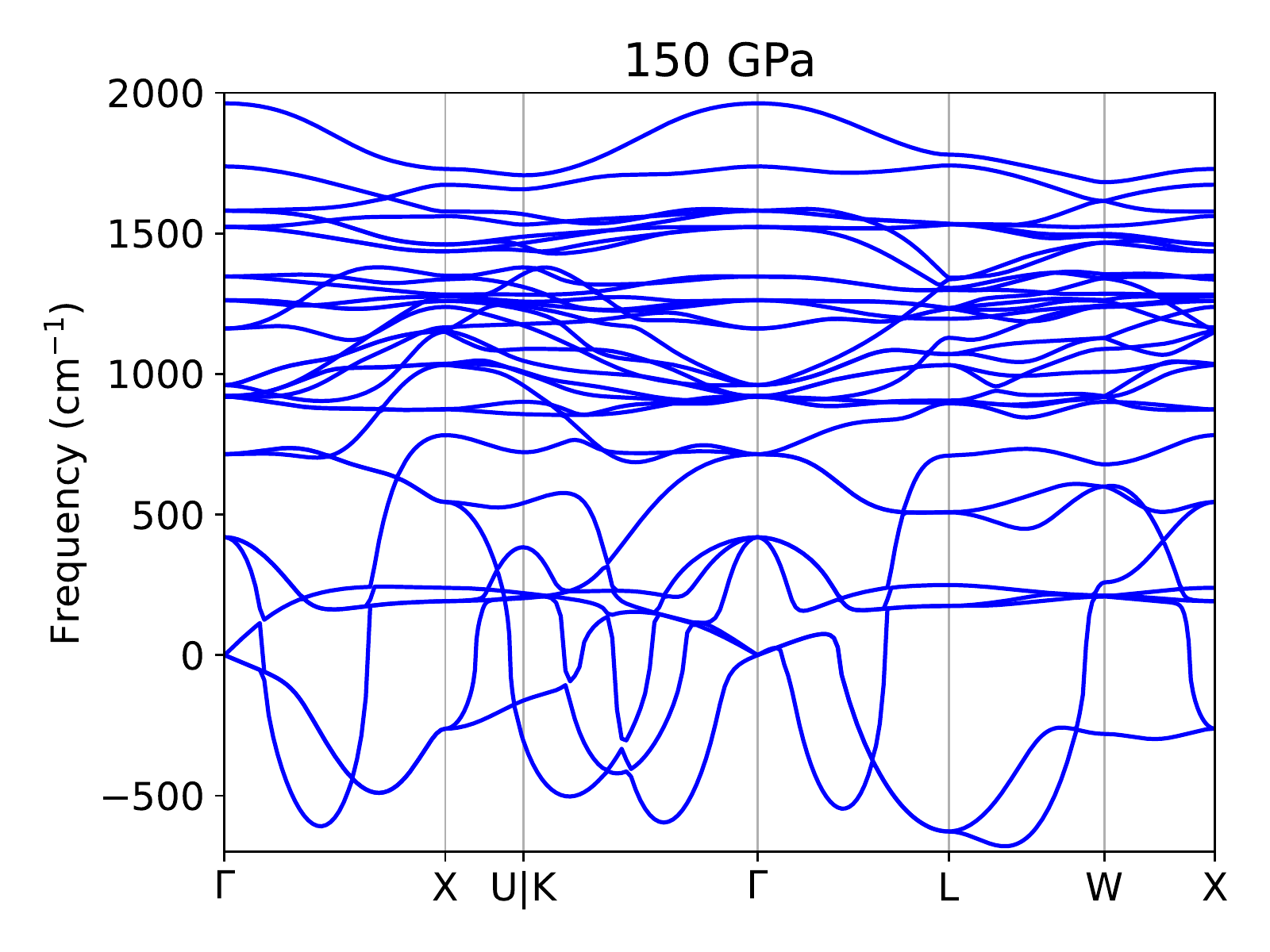}
        \caption{\label{fig:bands_150}}
    \end{subfigure}
    \begin{subfigure}{0.32\linewidth}
        \includegraphics[width=\textwidth]{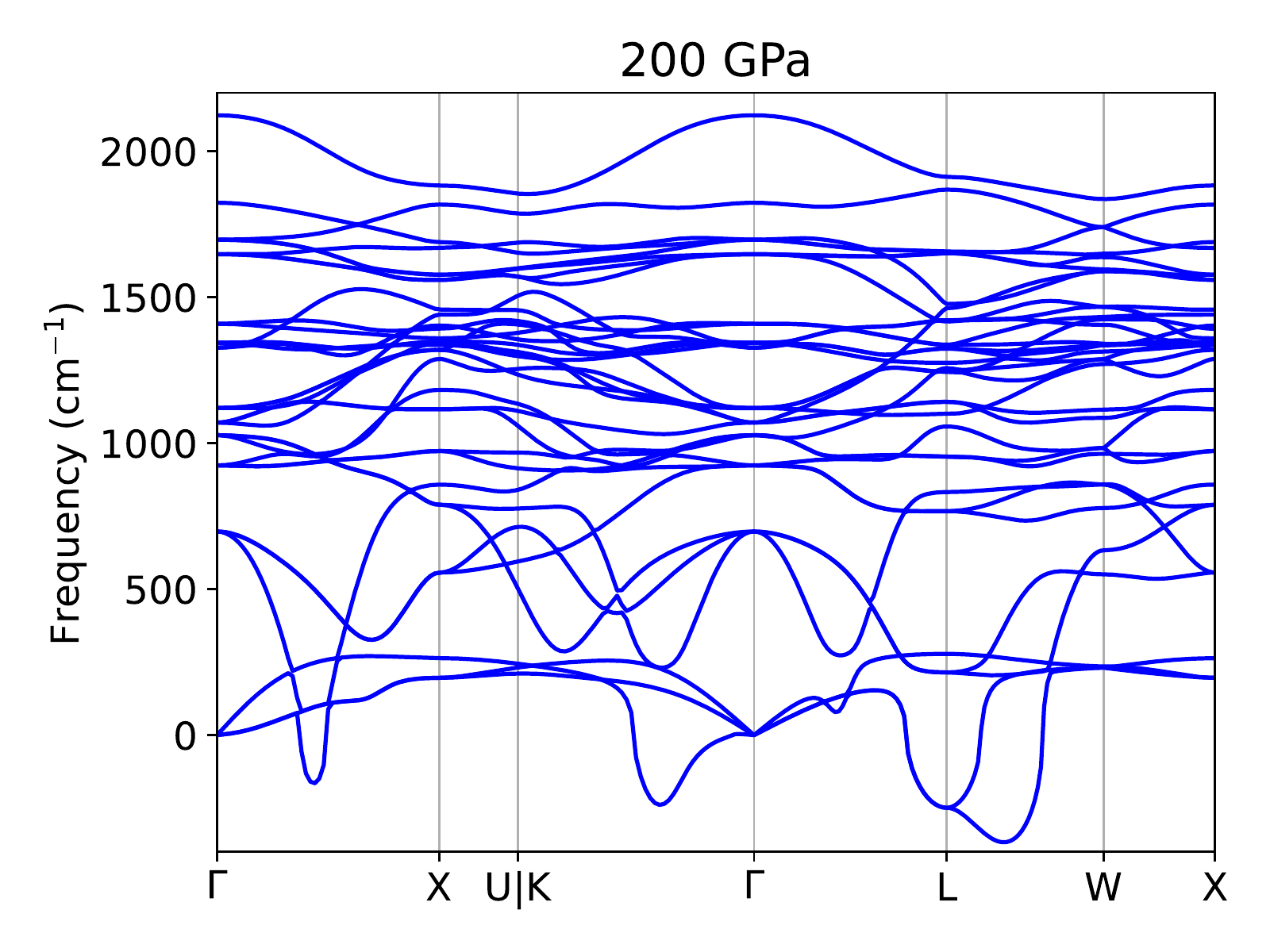}
        \caption{\label{fig:bands_200}}
    \end{subfigure}
    \begin{subfigure}{0.32\linewidth}
        \includegraphics[width=\textwidth]{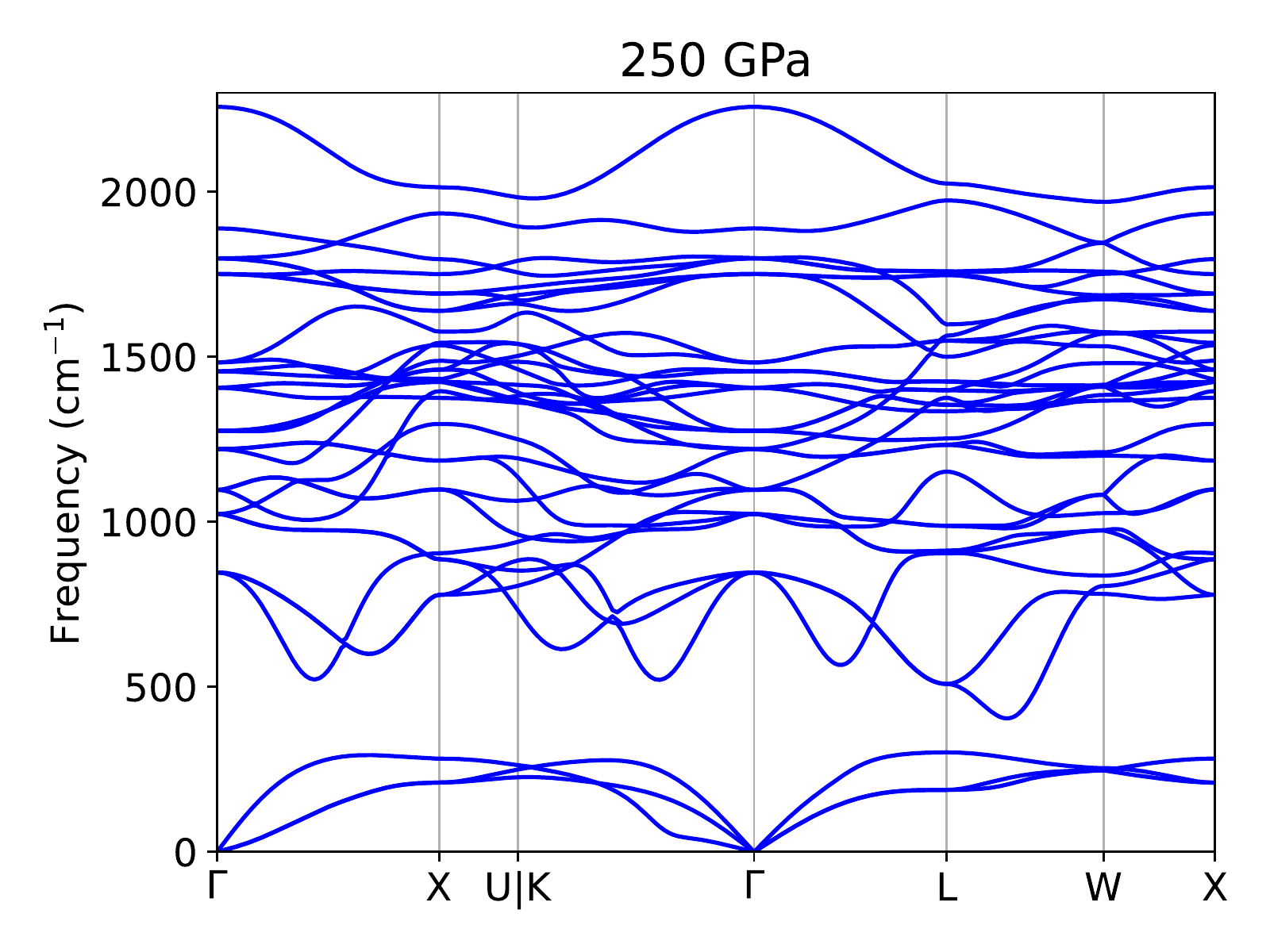}
        \caption{\label{fig:bands_250}}
    \end{subfigure}
    \caption{\label{fig:phonons}}
\end{figure}

Shown in Figure \ref{fig:phonons} are phonon band structures calculated with PBE DFT at three different pressures for fcc-LaH$_{10}$.
These are calculated within the harmonic approximation with DFPT, using a centered $4 \times 4 \times 4$ $\boldsymbol{q}$ grid.
Within this approximation the cubic structure is dynamically unstable until well above 200 GPa.

\section{\label{sec:pair}Short-ranged pair potentials}

\begin{figure}
    \begin{subfigure}{0.32\linewidth}
        \includegraphics[width=\textwidth]{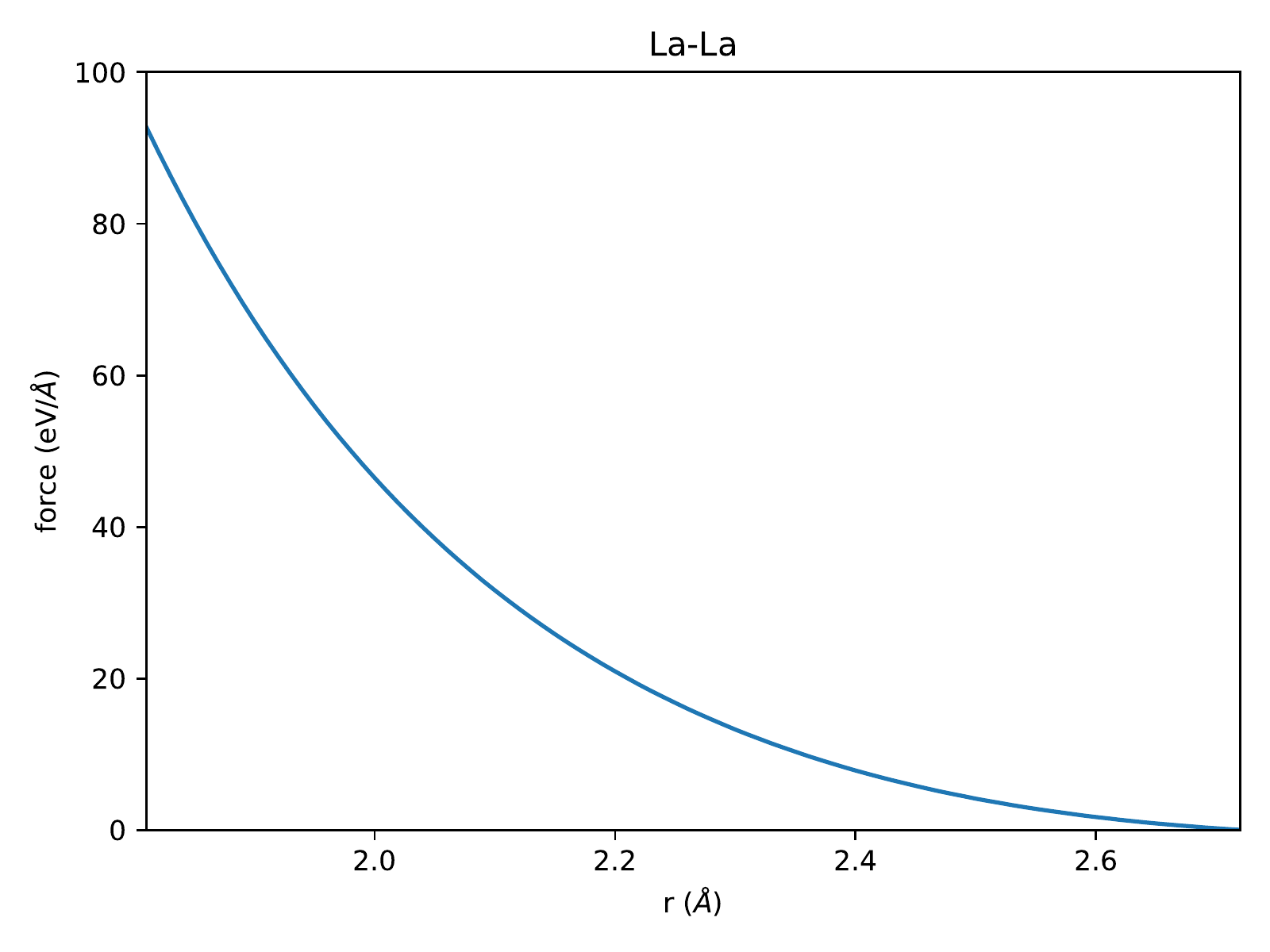}
        \caption{\label{fig:LL}}
    \end{subfigure}
    \begin{subfigure}{0.32\linewidth}
        \includegraphics[width=\textwidth]{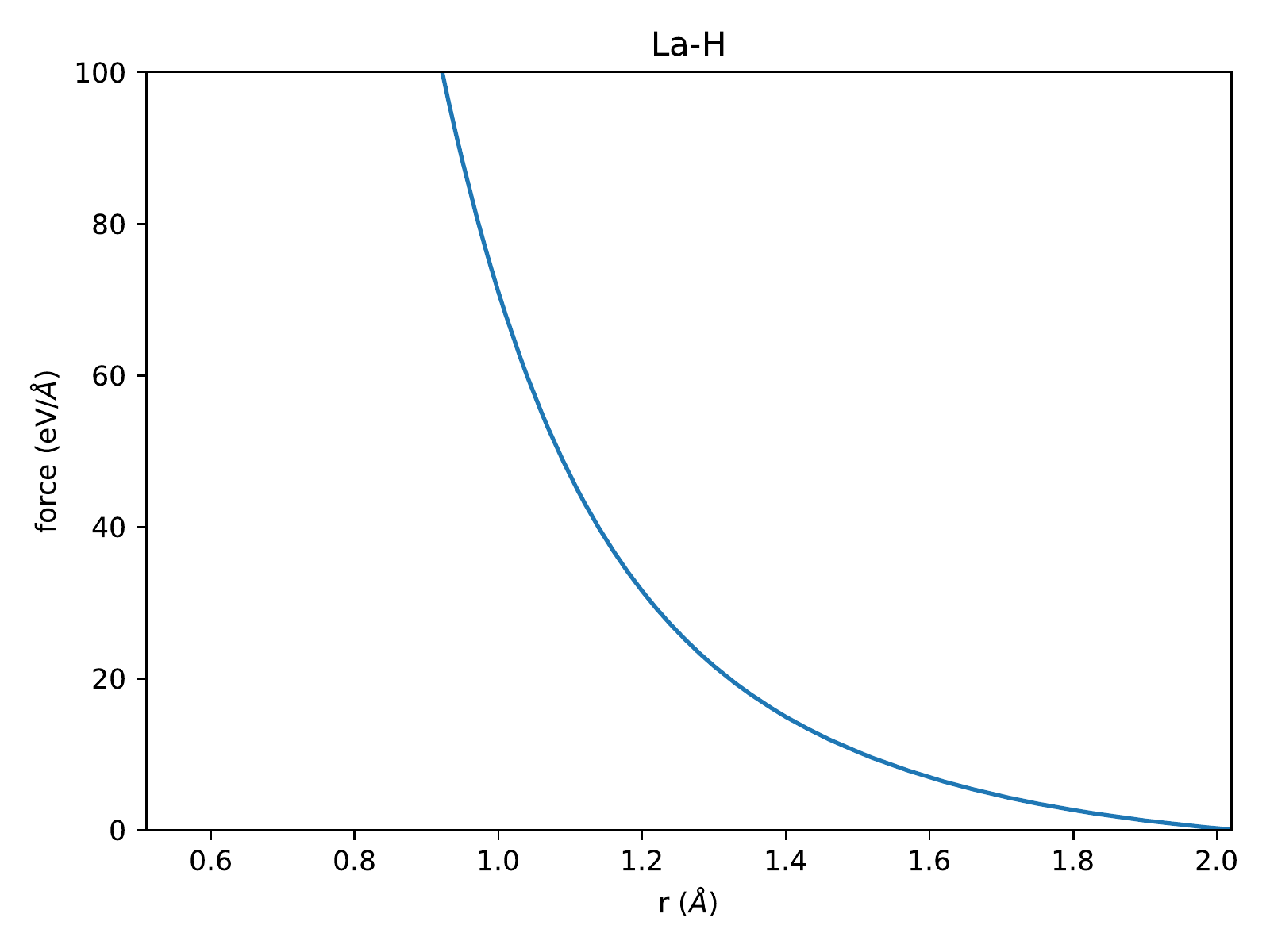}
        \caption{\label{fig:LH}}
    \end{subfigure}
    \begin{subfigure}{0.32\linewidth}
        \includegraphics[width=\textwidth]{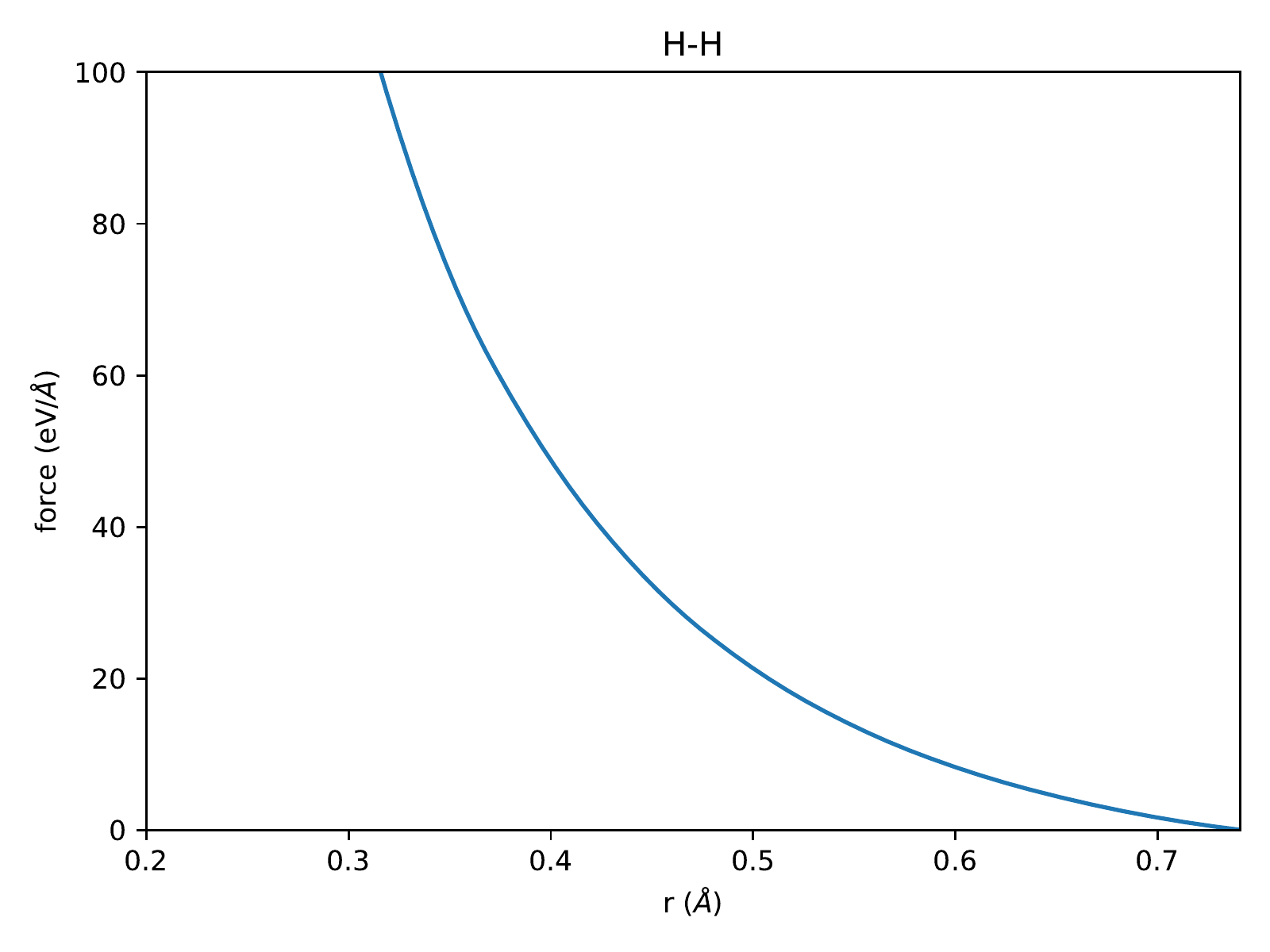}
        \caption{\label{fig:HH}}
    \end{subfigure}
    \caption{\label{fig:pairs}}
\end{figure}

To ensure repulsion at short distances, we train our neural network to learn the difference between the DFT energy / force and a short-ranged pair potential.
We only require that this potential approximate the repulsion between atoms as they get close, as the neural network will learn corrections to it.
To generate this potential we calculate the energy of the H-H, La-H, and La-La dimers with CCSD, varying the bond length.
The resulting forces are shown in Figure \ref{fig:pairs}.

\section{\label{sec:MLP}MLP}

\begin{figure}
    \begin{subfigure}{0.32\linewidth}
        \includegraphics[width=\textwidth]{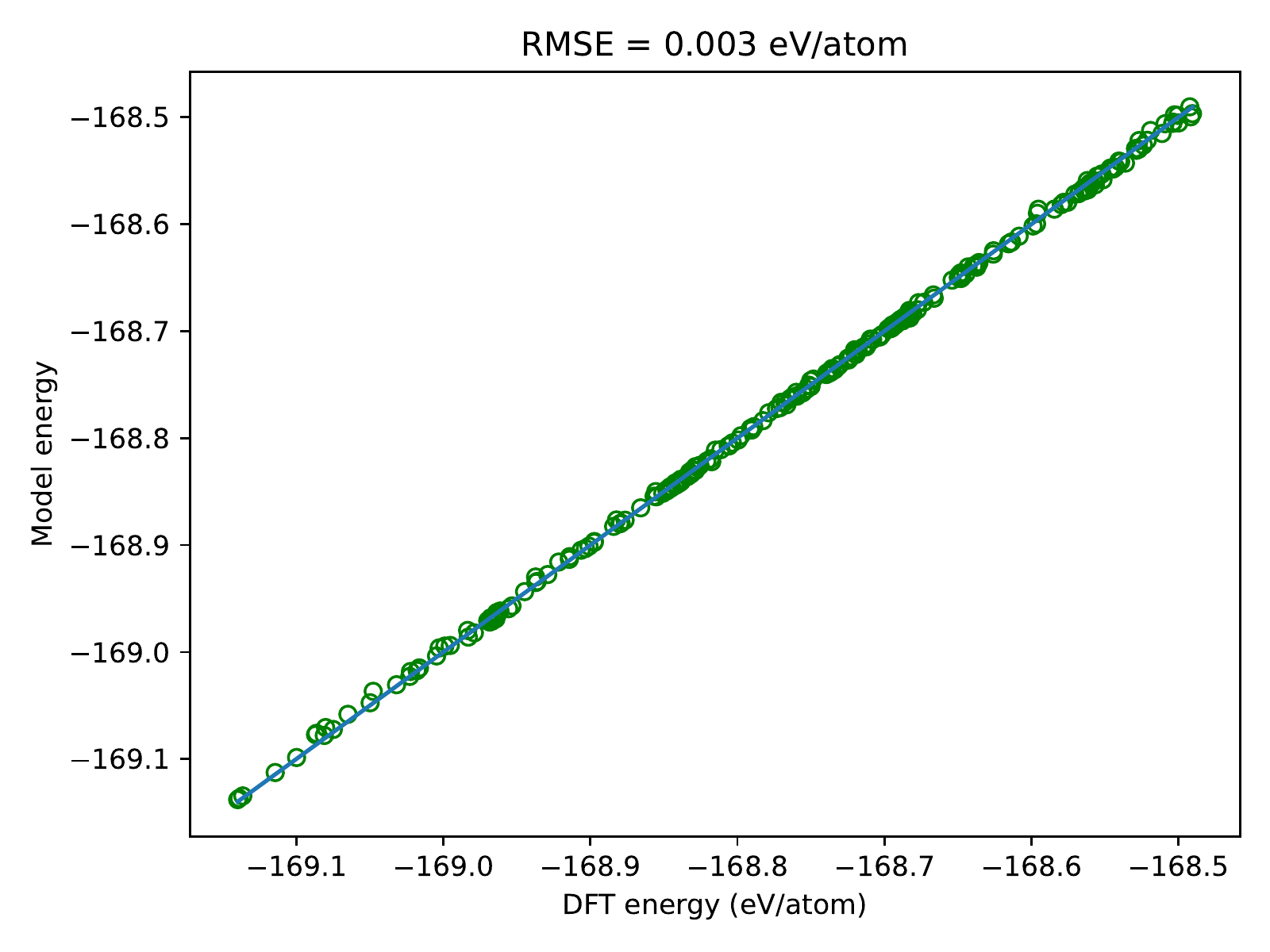}
        \caption{\label{fig:energy_fit}}
    \end{subfigure}
    \begin{subfigure}{0.32\linewidth}
        \includegraphics[width=\textwidth]{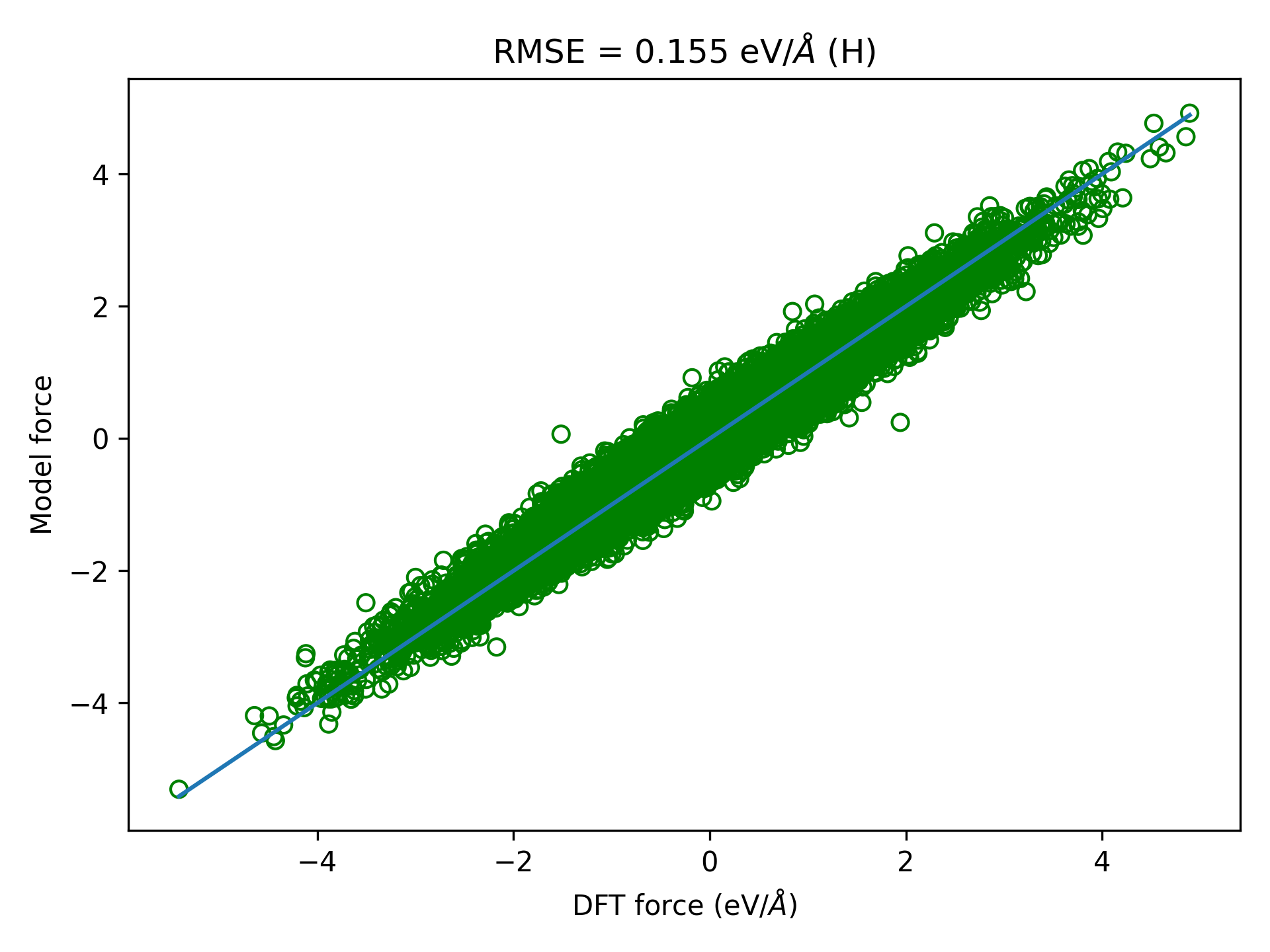}
        \caption{\label{fig:force_H_fit}}
    \end{subfigure}
    \begin{subfigure}{0.32\linewidth}
        \includegraphics[width=\textwidth]{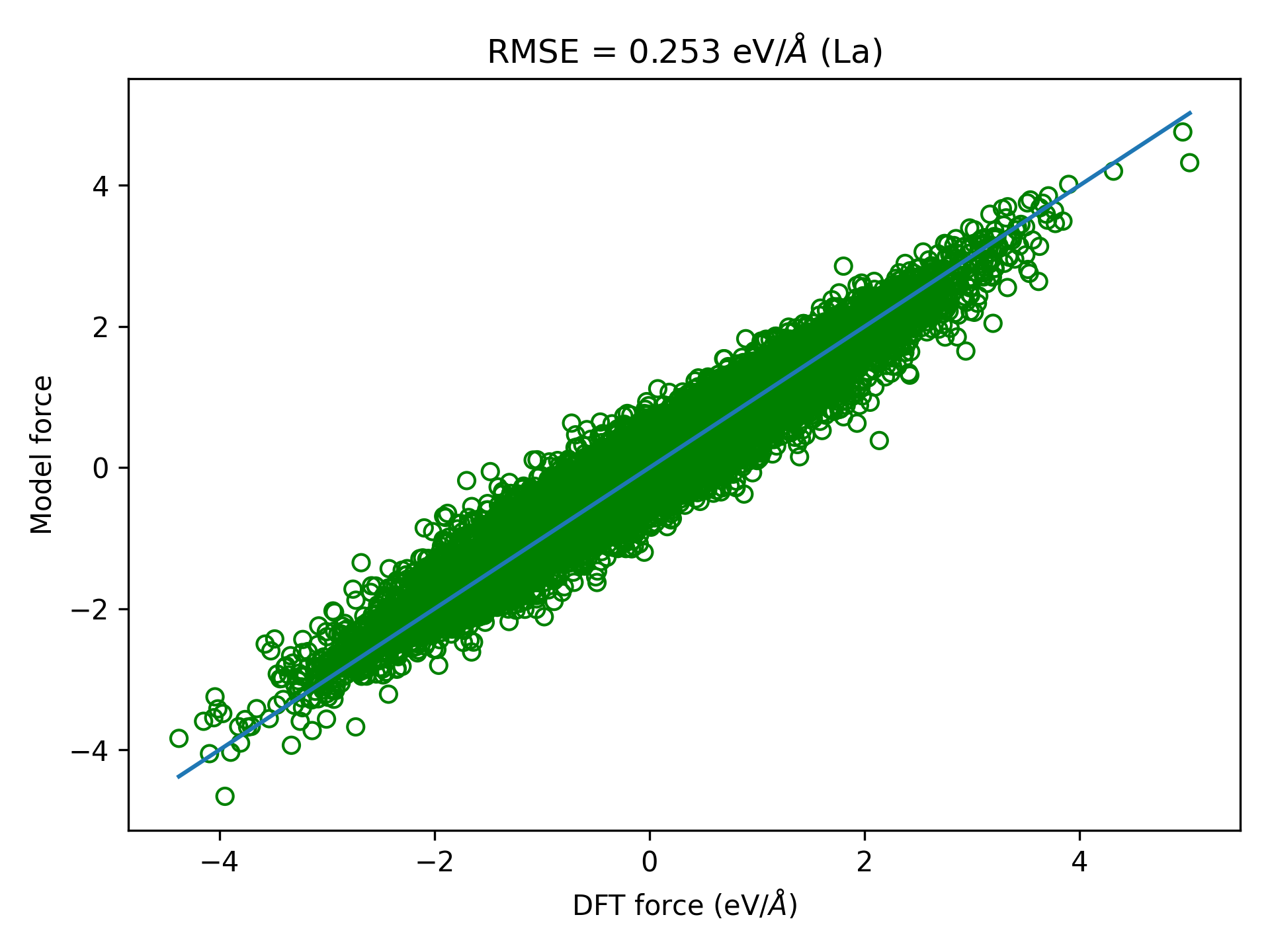}
        \caption{\label{fig:force_La_fit}}
    \end{subfigure}
    \caption{\label{fig:fits}}
\end{figure}

In Figure \ref{fig:fits} we show how our model fits energies and forces on a test set of structures which was not included in training.
Because of the difference in scales between the hydrogen and significantly heavier lanthanum atoms, we separate the force errors by atomic species.

We have also trained an NEP \cite{NEP1, NEP2} model as an independent test of our results.
Like DP, this is also a neural network potential.
One significant difference between NEP and DP is how atomic environments are represented.
Recall that in DP the feature space is determined by an embedding neural network.
In NEP, the features are two and three body descriptors calculated using a combination of Legendre and Chebyshev polynomials.
These descriptors are similar but not identical to Behler-Parrinello symmetry functions.

\begin{figure}
    \begin{subfigure}{0.49\linewidth}
        \includegraphics[width=\textwidth]{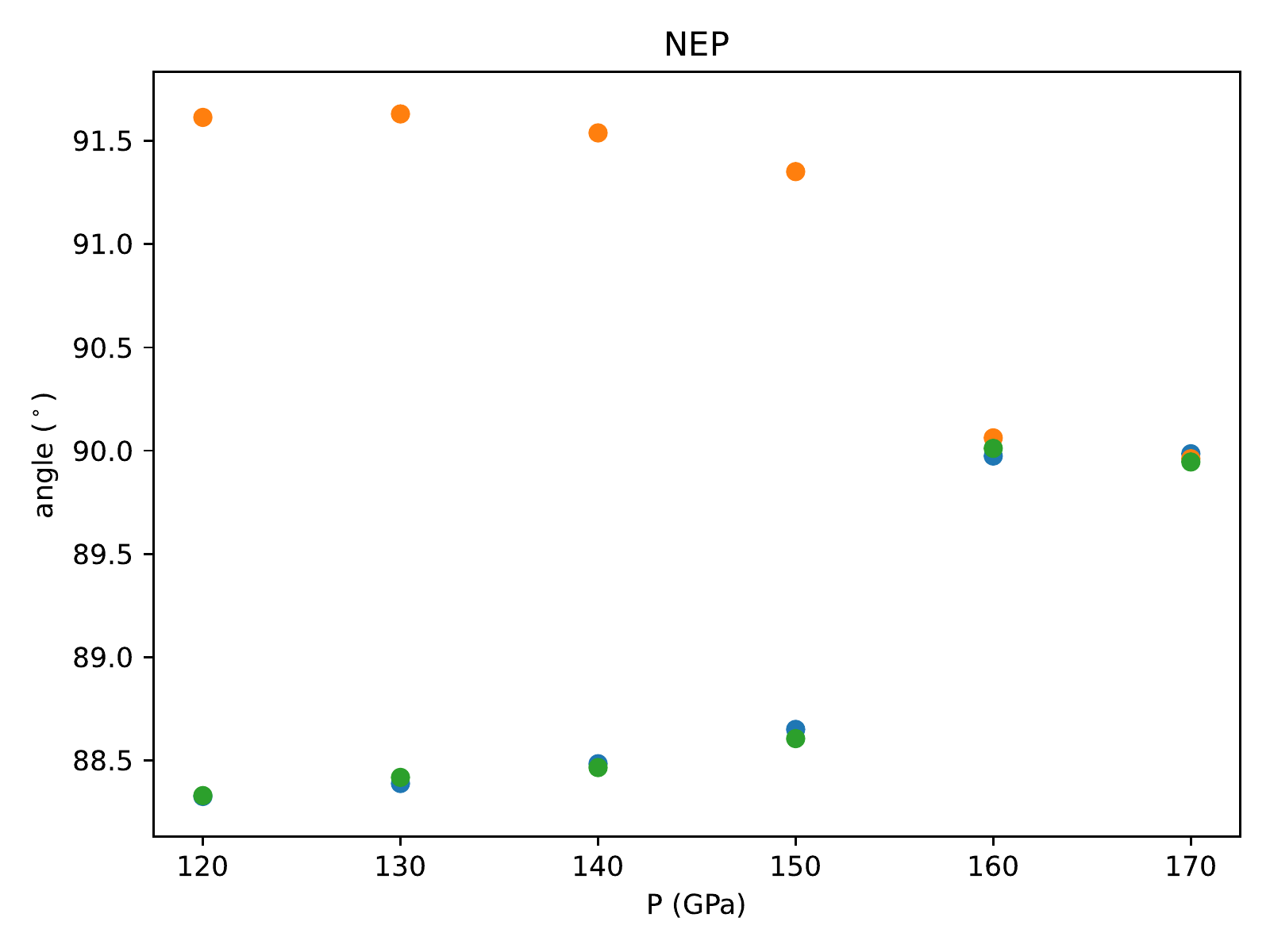}
        \caption{\label{fig:angles_nep}}
    \end{subfigure}
    \begin{subfigure}{0.49\linewidth}
        \includegraphics[width=\textwidth]{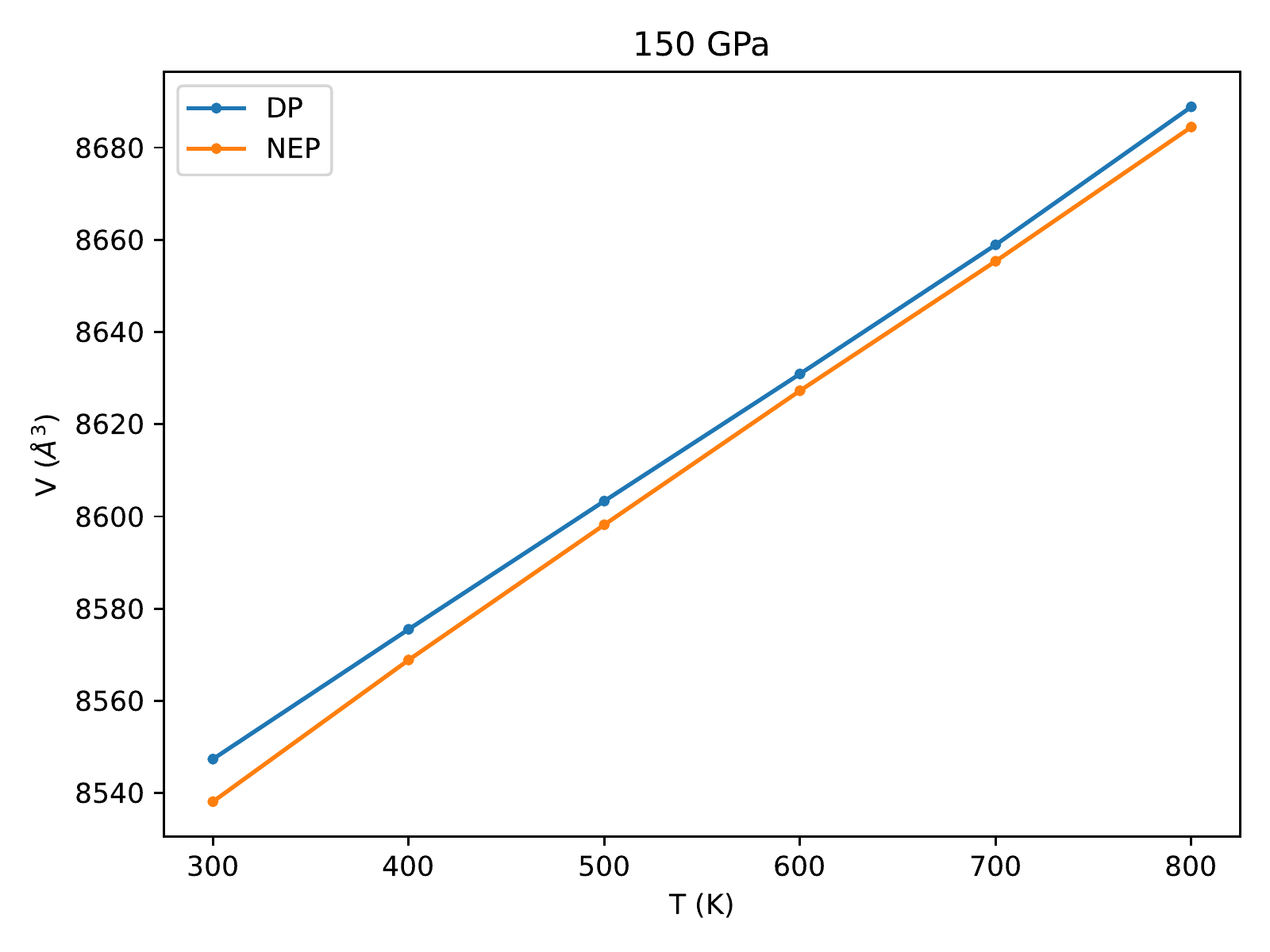}
        \caption{\label{fig:thermal_expansion}}
    \end{subfigure}
    \caption{\label{fig:nep}}
\end{figure}

NEP has been shown to have similar fitting performance as DP and we find that to be true in our case.
Our NEP model was trained on the same dataset that was used for our DP model, then employed in fixed pressure simulations to study the stability of fcc-LaH$_{10}$.
We also measured thermal expansion of fcc-LaH$_{10}$ with both models, to see how they capture anharmonicity.
Because NEP is not (yet) interfaced to a path-integral code, we show here only classical simulations.
In Figure \ref{fig:angles_nep} we show the cell angles at various pressures, where the rhombohedral cell transforms into the cubic structure above 150 GPa, just as in DP.
In Figure \ref{fig:thermal_expansion} we show the (supercell) volume at various temperatures and fixed pressure, where we find that the two models appear to be identical in their representation of anharmonicity.
We believe that the consistency in the results between these two distinct approaches means that the results are independent of the specific details of the models.
Instead, they are a consequence of the underlying dataset, a common factor between the models.
For this reason, we believe that our models are a good representation of the physics of the underlying DFT.

\section{\label{sec:diffusion}H diffusion}

\begin{figure}
    \begin{subfigure}{0.49\linewidth}
        \includegraphics[width=\textwidth]{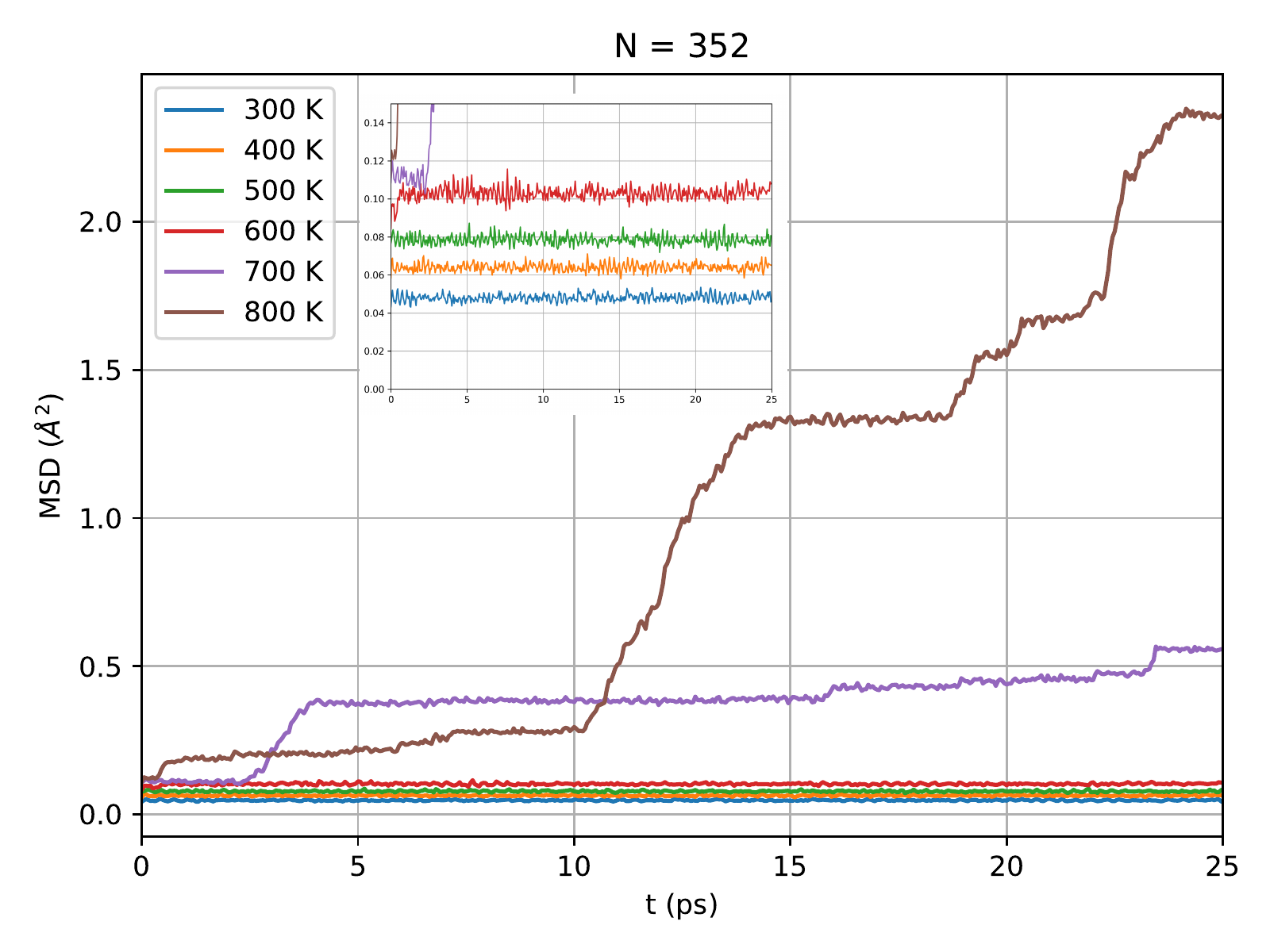}
        \caption{\label{fig:msd_352}}
    \end{subfigure}
    \begin{subfigure}{0.49\linewidth}
        \includegraphics[width=\textwidth]{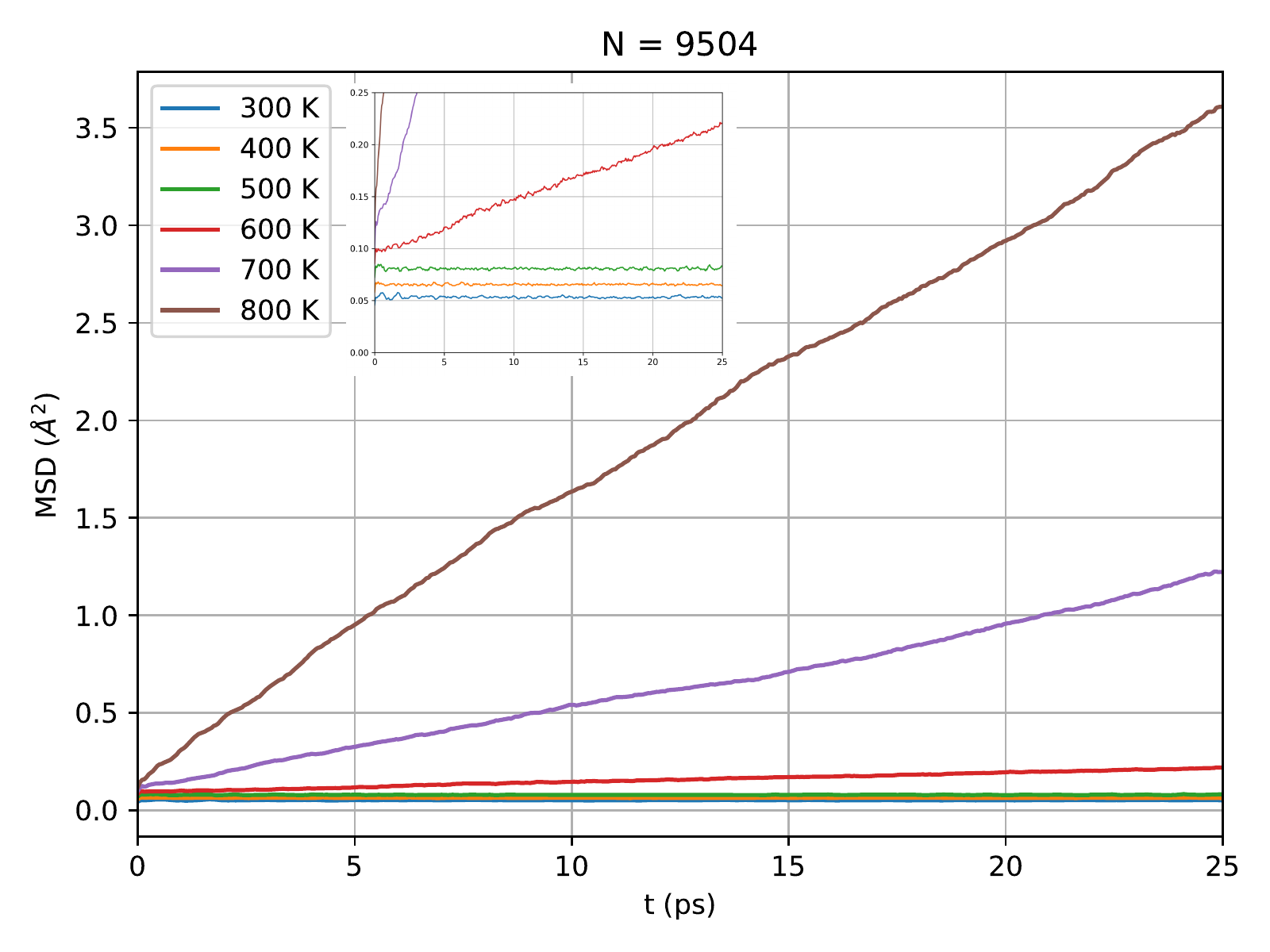}
        \caption{\label{fig:msd_9504}}
    \end{subfigure}
    \caption{\label{fig:msd}}
\end{figure}

Previous AIMD simulations found the onset of diffusion in the hydrogen sublattice at about 800 K \cite{LaH_AIMD}.
This is marked by a nonzero mean squared displacement (MSD) over time, which in this case displayed a staircase-like behavior.
The hydrogen sublattice is still ordered at these temperatures, and the diffusion occurs by protons hopping from site to site.
Our DP model reproduces this behavior.
In Figure \ref{fig:msd_352} we show the proton MSD over time in a simulation with a supercell of 352 atoms.
We find the onset of proton diffusion at around 700 K, slightly lower than was found in \cite{LaH_AIMD}.
However, we believe our result is consistent with the AIMD results, since the diffusion occurs in slow steps: notice that at 700 K the first jump does not occur until after 3 ps.
In other words, it is possible that if the AIMD simulations were run sufficiently long, such a jump would also be observed.
The frequency of proton hopping obviously increases with temperature, but we note that it is also a finite size effect.
In Figure \ref{fig:msd_9504} we show the MSD over time in a simulation with 9504 atoms.
In addition to smoothing out the staircase, the diffusion is increased (note the vertical axes).
Furthermore, the onset temperature is reduced to 600 K, and nonzero diffusion can be observed immediately after $t > 0$ ps.

\section{\label{sec:convergence}Path-integral convergence}

\begin{figure}
    \includegraphics[width=\textwidth]{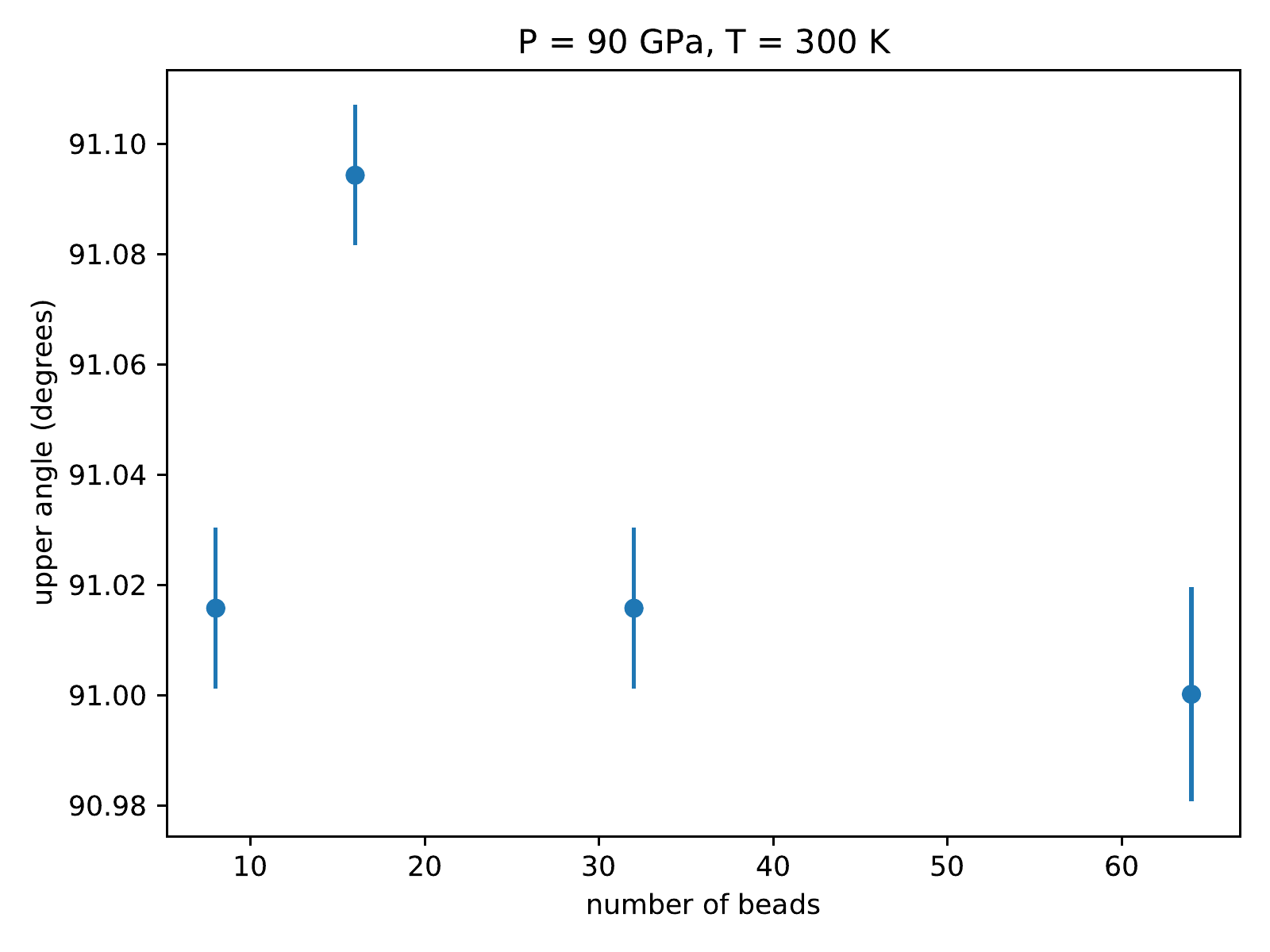}
    \caption{\label{fig:beads}}
\end{figure}

In path-integral simulations, the path is discretized into ``beads.''
A path with one bead yields the classical limit, while a fully quantum calculation would involve infinitely many beads.
In practice, one must settle for a finite number of beads such that the quantities of interest are sufficiently converged.
Since we found a distortion at 90 GPa in our path-integral simulations, we checked whether increasing the number of beads would restore the cubic cell.
Shown in Figure \ref{fig:beads} is how the largest cell angle $\alpha$ varies as the number of beads is increased.
In all cases the distortion is rhombohedral, so the remaining two angles are equidistant from 90, within error bars.
They are not shown here so that that error bars on the upper angle can be seen at this scale.
Notice that even with 64 beads it is still well above $90^\circ$.

Recall that upon lowering the temperature, the cubic structure is still stable at 100 GPa.
Since quantum effects favor the more symmetric cubic structure, the cell angles in these simulations are already converged.
In other words, increasing the number of beads would not make it somehow less cubic.
Given that the number of beads required increases with decreasing temperature, and our low temperature simulations appear to be converged, we believe this is further evidence that all of our path-integrals have a sufficient number of beads.

\section{\label{sec:substructure}H sublattice structure}

\begin{figure}
    \begin{subfigure}{0.49\linewidth}
        \includegraphics[width=\textwidth]{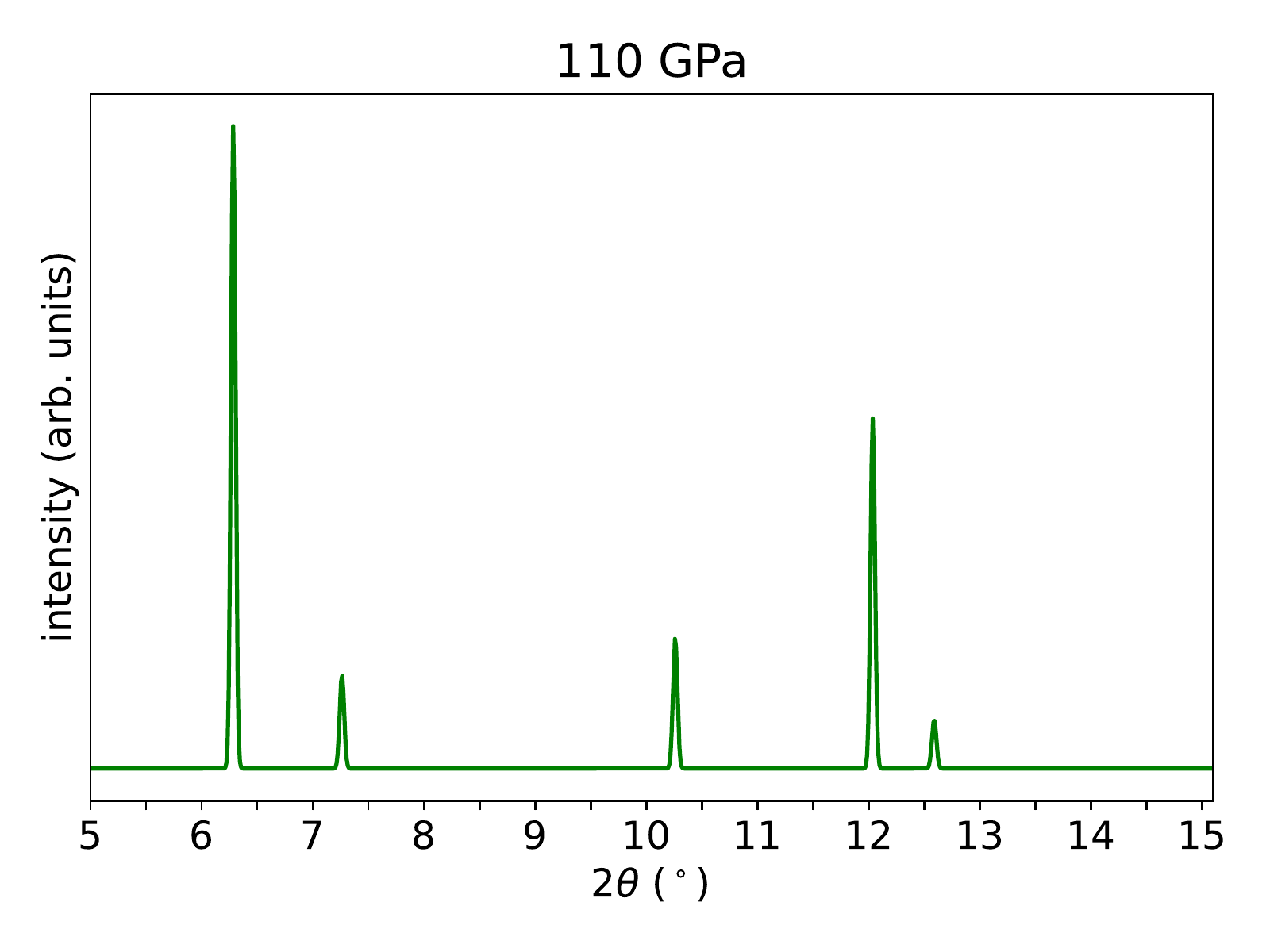}
        \caption{\label{fig:xrd_110}}
    \end{subfigure}
    \begin{subfigure}{0.49\linewidth}
        \includegraphics[width=\textwidth]{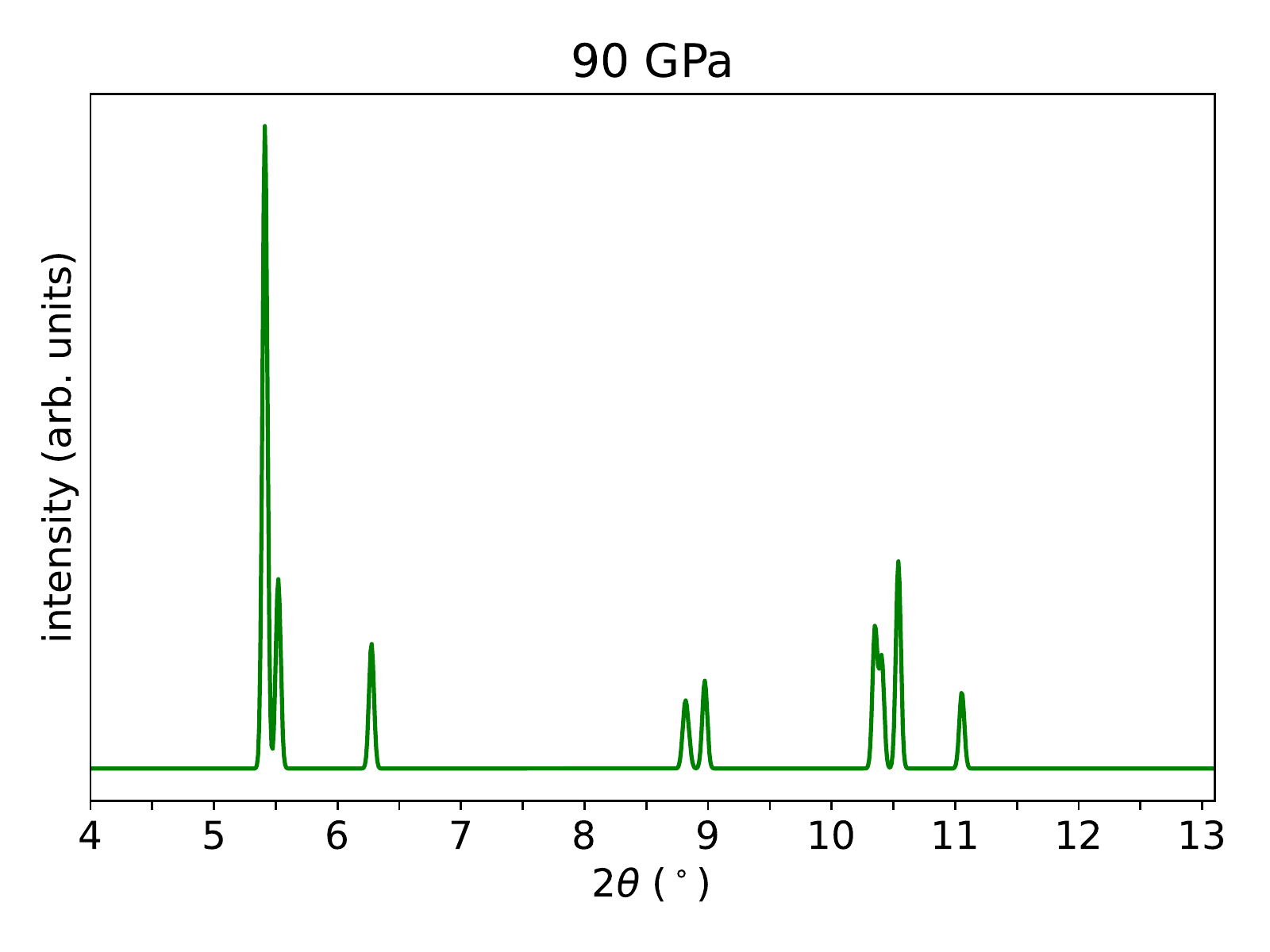}
        \caption{\label{fig:xrd_90}}
    \end{subfigure}
    \caption{\label{fig:xrd}}
\end{figure}

Shown in Figure \ref{fig:xrd} are simulated XRD patterns based on the hydrogen sublattice in our simulations at two different pressures.
These are calculated in exactly the same way as our La patterns, except now we consider here only the hydrogen sublattice.
While the hydrogen sublattice has not yet been resolved experimentally, we can attempt to determine whether or not there is a notable change here as the cell undergoes a distortion.
This does not appear to be the case: the reflections in the H patterns have the same locations as the La reflections, with significantly modulated intensities due to stronger fluctuations.
The splitting of the peaks as the pressure is lowered is similar to that seen in the La peaks.
In other words, the change in the hydrogen sublattice appears to be entirely commensurate with the overall rhombohedral distortion, with no discernible qualitative difference otherwise.

\bibliography{bibliography}